\documentclass[preprint, 12pt]{aastex}
\newcommand\T{\rule{0pt}{2.6ex}}

\usepackage{color}

\begin{document}

\title{Luminous Satellites versus Dark Subhaloes: Clustering in the Milky Way}

\author{Brandon Bozek\altaffilmark{1}, Rosemary F. G. Wyse\altaffilmark{1}, Gerard Gilmore\altaffilmark{2}}
\altaffiltext{1}{Johns Hopkins University, Department of Physics and Astronomy, 3400 N. Charles Street, Baltimore, MD 21218, USA; bbozek@pha.jhu.edu}
\altaffiltext{2}{Institute of Astronomy, University of Cambridge, Madingley Road, Cambridge CB3 0HA, UK}

\begin{abstract} 

The observed population of the Milky Way satellite galaxies offer a unique testing ground for galaxy formation theory on small-scales. Our novel approach was to investigate the clustering of the known Milky Way satellite galaxies and to quantify the amount of substructure within their distribution using a two-point correlation function statistic in each of three spaces: configuration space, line-of-sight velocity space, and four-dimensional phase-space. These results were compared to those for three sets of subhaloes in the Via Lactea II Cold Dark Matter simulation defined to represent the luminous dwarfs. We found no evidence at a significance level above $2\sigma$ of substructure within the distribution of the Milky Way satellite galaxies in any of the three spaces. The ``luminous" subhalo sets are more strongly clustered than are the Milky Way satellites in all three spaces and over a broader range of scales in four-dimensional phase-space. Each of the ``luminous" subhalo sets are clustered as a result of substructure within their line-of-sight velocity space distributions at greater than $3\sigma$ significance, whereas the Milky Way satellite galaxies are randomly distributed in line-of-sight velocity space. While our comparison is with only one Cold Dark Matter simulation, the inconsistencies between the Milky Way satellite galaxies and the Via Lactea II subhalo sets for all clustering methods suggest a potential new `small-scale' tension between Cold Dark Matter theory and the observed Milky Way satellites. Future work will obtain a more robust comparison between the observed Milky Way satellites and Cold Dark Matter theory by studying additional simulations.

\end{abstract}


\section{\label{sec:intro} Introduction}

Structure formation within the concordance model of cosmology, Cold Dark Matter (CDM) with a cosmological constant ($\Lambda CDM$), is a hierarchical process where large structures are built ``bottom up" via merging \citep{1978MNRAS.183..341W, 1984Natur.311..517B}. Small dark-matter haloes collapse first, early in the universe, and then merge to form larger structures like the Milky Way. Smaller galaxies that are accreted by larger hosts are tidally disrupted, the degree of which depends on relative densities of the host and the satellite and on the orbital parameters. The resultant set of satellite galaxies of the Milky Way that survives to be observed today provides an ideal testing ground for theories of galaxy formation. 

The focus of this paper is, first, to quantify the amount of substructure in the distribution of the known satellite galaxies of the Milky Way and, second, to compare these results with the amount of substructure found in the distributions of several sets of subhaloes from the Via Lactea II Cold Dark Matter N-body simulation \citep{2008Natur.454..735D}. Our method of quantifying the amount of substructure for each data set is as follows: First, we measured the degree of clustering in the distribution of each data set using a two-point correlation function statistic in each of configuration space, line-of-sight velocity space, and four-dimensional phase-space. Next, we construct a mock set of  ``dwarf galaxies" with the same radial density profile and line-of-sight velocity distribution as each data set. The clustering of these mock sets are then measured using the two-point correlation function statistics mentioned above and compared to the two-point correlation function of the corresponding data sets, in order to determine the degree the smooth underlying density profile and line-of-sight velocity distribution contributes to each data sets' clustering signal. Substructure is defined to exist on any scale where the clustering of the data set is greater than the clustering due to the smooth underlying density profile and line-of-sight velocity distribution. 

We should note the limitations of our approach. The suite of two-point correlation functions used in this analysis are global statistics that are suited for quantifying the amount of substructure in the distribution of satellites, and are therefore insufficiently sensitive to singular features within the satellite distribution. The known Milky Way satellite galaxy population has been reported to have a planar distribution dubbed the ``disk of satellites" \citep{2005A&A...431..517K, 2009MNRAS.394.2223M}, has been argued to host numerous ``ghostly streams" consisting of different sets of galaxies and globular clusters \citep{1995MNRAS.275..429L, 2008ApJ...686L..61D}, and has been proposed to have associated pairs such as Leo IV and Leo V \citep{2008ApJ...686L..83B}.  In order to identify a particular substructure within a larger set of satellites using a two-point correlation function, the pairs of satellites in the substructure need to pick out a particular scale of galaxy separations. A single stream, consisting of a few members, or a pair of galaxies within an otherwise randomly distributed set of satellites, or a larger structure, like a disk of satellites, with many members that possibly do not have a common scale of galaxy separations, could be missed in this analysis. We explored the relationship of the Milky Way satellite galaxies clustering results to the disk of satellites, ``ghostly streams", and associated pairs of galaxies in the discussion section. 

The paper is organized as follows. The data sets of Milky Way satellite galaxies and the data sets of Via Lactea II subhaloes are detailed in Section 2. In Section 3, we describe the two-point correlation function statistics that are used in the analysis. The results for the various data sets are given in Section 4. A discussion of these results is presented in Section 5 and a summary of the paper is provided in Section 6.

\section{\label{sec:Data}Data Sets}

\subsection{Milky Way Satellites}

The Milky Way is orbited by more than two dozen known satellite galaxies. The population of identified satellites has recently experienced a rapid increase, from the eleven bright satellites,§ with the slew of discoveries of low luminosity ``Ultra Faint Dwarfs" enabled by the excellent uniform photometry of the Sloan Digital Sky Survey (SDSS) \citep{ 2006ApJ...647L.111B, 2007ApJ...654..897B, 2008ApJ...686L..83B, 2005AJ....129.2692W, 2005ApJ...626L..85W, 2006ApJ...653L..29S, 2006ApJ...650L..41Z, 2006ApJ...643L.103Z, 2007ApJ...662L..83W}. Some of these discoveries have been confirmed to be dark-matter dominated dwarf galaxies through follow-up spectroscopy of candidate member stars (e.g. \citet{2007MNRAS.380..281M, 2007ApJ...670..313S, 2009ApJ...692.1464G, 2011ApJ...736..146K}), while others remain only tentatively categorized as a galaxy (e.g. \citet{2008ApJ...686L..83B}). Nevertheless, for present purposes here, we considered them all to be galaxies, i.e. to have dark-matter haloes. The candidate Milky Way satellite galaxies include Leo T and Leo A at the large distances of 420~kpc and 800~kpc respectively. Leo T and Leo A are not confirmed to be bound satellites of the Milky Way \citep{1999A&ARv...9..273V, 2007ApJ...656L..13I}, therefore the most distant galaxy we considered is Leo I at 260~kpc from the Galactic center. We also excluded Pisces II \citep{2010ApJ...712L.103B} because it lacks the line-of-sight-velocity data our analysis required for the 4D phase-space and line-of-sight velocity two-point correlation function statistics. The resultant set of twenty four galaxies, that makes up the set designated ``MW-A", is detailed in Table \ref{tab:Gallist}. 

We additionally considered a set of galaxies, denoted ``MW-B", consisting of only those that are found within the SDSS DR7 ``Northern Contiguous" footprint \citep{2000AJ....120.1579Y,2009ApJS..182..543A}, where ``Northern Contiguous" refers to SDSS stripes 9 through 39 that are in the Northern Galactic Hemisphere and are simply connected. Table \ref{tab:Gallist} also indicates the sixteen galaxies that are in the ``MW-B" set. We excluded non-contiguous stripes to avoid possible introduction of unknown biases into the clustering results. While further limiting the number of an already small galaxy sample, the purpose of the ``MW-B" set was to enable the study of a region where the selection biases controlling detectability are reasonably well understood. 

\subsection{Via Lactea II Subhaloes}

The full Via Lactea II (VL-II) public data set\footnote{See VL-2 Data, http://www.ucolick.org/~diemand/vl/data.html} consists of the zero-redshift positions and velocities of subhaloes of a Milky Way-sized Cold Dark Matter halo out to its virial radius (defined as the radius where the host halo density is 200 times the mean density of the universe) that are well above the simulation's resolution limit (i.e. they have a peak circular velocity of at least 4~$km s^{-1}$ at some point in their evolution). We obtained the first set, denoted hereafter as the ``VL-A" set, by selecting all of these subhaloes that are within the Galctocentric radial boundaries of the observed satellite galaxies, namely 4~kpc and 260~kpc. It should be noted that subhaloes with low Galactic latitudes that are within the unobservable regions of the SDSS near the Galactic disk are included in this set.

Our analysis of the clustering in the line-of-sight velocity space required us to define an observer position, which we chose to be in the x-y plane at 8~kpc from the ``Milky Way" center along the x-axis of the host halo, $R_{obs} = $ (8~kpc, 0, 0). This ``observer" position is arbitrary, so a set of subhaloes (VL-Arot) was defined as ``observed" from a  ``rotated" position in the same x-y plane but along the y-axis at $R_{rot} = $ (0, 8~kpc, 0). We additionally constructed two subsets of the dark subhaloes that are ``observed" to lie within the region defined by the SDSS footprint (described above) for each ``observer" position. The subhaloes that make up these sets are denoted respectively as ``VL-B" and ``VL-Brot". The number of subhaloes in each of the four VL sets are listed in Table \ref{tab:Cluster_nums}.

\subsection{ The ``Luminous" Subhalo Sets}

\subsubsection{The Largest Before Accretion Model}

We selected three sets of subhaloes from each of the dark subhalo sets, using semi-analytic models that have been proposed as possible resolutions of the ``Missing Satellites Problem", to represent the population of ``luminous" subhaloes. These semi-analytic models prescribe a set of constraints that limit star formation to only the most massive and/or earliest forming subhaloes. The Largest Before Accretion (LBA) model identifies proposed ``luminous" subhaloes by defining a circular velocity threshold below which stars cannot form \citep{2004ApJ...609..482K, 2007ApJ...667..859D}. The velocity threshold is defined in terms of the largest value of the maximum circular velocity (assuming spherical symmetry) for a dark-matter subhalo, evaluated over the complete history for that halo and denoted with the subscript ``p", $V_{max,p}$.  These criteria select a set of subhaloes that are the most massive prior to accretion into the Milky Way dark-matter halo, with the total number dependent on the choice of the velocity threshold.

It has been proposed that the expected number of satellite galaxies for the Milky Way can be estimated by taking into account selection effects of the SDSS survey and extrapolating the known number of satellite galaxies in the SDSS footprint to a value for the full volume \citep{2008ApJ...686..279K, 2007ApJ...670..313S}. The total number of satellites in this estimation depends on the assumed minimum of the luminosity function of all Milky Way satellites. In their analysis of the first Via Lactea simulation, VL-I, \citet{2008ApJ...679.1260M} defined their LBA set using a velocity threshold of $V_{max,p} = 21.9 km s^{-1}$, giving 65 total ``luminous" subhaloes, in order to match these estimates. The VL-II simulation effectively used different cosmological parameters and has a slightly larger host Dark Matter halo mass than the VL-I simulation, and so achieving a set of 65 subhaloes requires a larger velocity cut, in this case $V_{max,p} \geq 24 km s^{-1}$. We used these criteria to define our ``LBA-A" set.

\subsubsection{The Early Formation Model}

The Early Formation (EF) model was constructed to model the expected suppression of star formation in low $V_{circ}$ systems due to reionization and associated heating of the intergalatic medium above the virial temperature of these small systems \citep{1992MNRAS.256P..43E}. Prior to the epoch of reionization, stars are assumed to form in all subhaloes above the atomic cooling limit ($T \ge 10^4$K), which corresponds to the threshold of  $V_{max}(z_{reion}) > 16.2 km s^{-1}$. After the epoch of reionization, only large mass subhaloes are still able to accrete the warmed intergalactic gas and form stars. The minimum sufficient mass of a subhalo that would be able to form stars after reionization is an open question, so we followed \citet{2008ApJ...679.1260M} and defined this threshold to be $V_{max,p} \geq 38 km s^{-1}$. Following \citet{2007ApJ...667..859D}, we adopted $z_{reion} = 9.1$ as the reionization redshift, which is the closest accessible snapshot to their choice of $z_{reion} = 9.6$. These selection criteria define the ``EF-A" set.

\subsubsection{The Modified Early Formation Model}

The Early Formation model builds upon the Largest Before Accretion model by providing a physical motivation for limiting the number of subhaloes that formed stars. The Early Formation model, however, fails at reproducing the luminosity function of Milky Way satellite galaxies (corrected for the selection effects of the SDSS survey) due to the sharp $V_{circ}$ cutoff in the EF model between the subhaloes that are defined to host galaxies and those that do not \citep{2009ApJ...696.2179K}. The modified Early Formation Model provides a better fit to this completeness-corrected satellite luminosity function by replacing this sharp divide with a gradual transition. The modified Early Formation Model, ``EFmod", subhalo sets we used here were constructed following the \citet{2009ApJ...696.2179K} ``Model 3b" prescription for assigning stellar mass to each dark-matter halo (see \citet{2009ApJ...696.2179K} for details).

We constructed four sets of subhaloes for each of the three semi-analytic models following the same procedure described above for the dark subhalo sets. The same naming convention that was used for the dark subhalo sets was also used for the ``luminous" subhalo sets for the ``observer" positions in both volumes. The number of subhaloes in each semi-analytic model set is listed in Table \ref{tab:Cluster_nums}. It is important to note that while the each of these models alleviate the discrepancy at the heart of the ``Missing Satellites Problem", they fail to account accurately for the mass function of the Milky Way satellites. For example, over twenty of the subhaloes in the full-area LBA set have a $V_{max,p}  \geq 35 km s^{-1}$, whereas only Draco and Ursa Minor have an estimated median value of $V_{max,p}$ that is consistent at $2\sigma$ with a circular velocity as large as $V_{max,p}  = 35 km s^{-1}$ \citep{2012MNRAS.422.1203B}. However, this mass function discrepancy would be alleviated somewhat if the Milky Way dark matter halo is less massive than in the Via Lactea II simulation \citep{2012MNRAS.422.1203B}.  

\section{\label{sec:Methods}Methods to Measure Clustering}

\subsection{Spatial Two-Point Correlation Function}

The spatial two-point correlation function, $\xi(r)$, is the most commonly used statistic to measure the spatial clustering of a distribution of galaxies. This statistic measures the probability, $dP$, in excess of a random distribution, of
finding two galaxies separated by a distance $r$ within a volume $dV$, where $r = \sqrt{(x_i - x_j)^2 + (y_i - y_j)^2 + (z_i - z_j)^2}$. This excess probability is defined by: 
\begin{equation}
dP = n(1 + \xi(r))dV,
\end{equation}
where $n$ is the average galaxy number density. 

Various estimators are used in practice but the variance of the Landy and Szalay (LS) estimator is essentially Poissonian, making the LS estimator best suited for sparse data sets \citep{1993ApJ...412...64L}. For each data set we determined the spatial two-point correlation function using the LS estimator:
\begin{equation}
\label{eq:LS}
\xi(r) = \frac{N_{DD}(r) - 2N_{DR}(r) + N_{RR}(r)}{N_{RR}(r)}.
\end{equation}
The quantities $N_{DD}$, $N_{RR}$, $N_{DR}$ in equation (\ref{eq:LS}) are, respectively, the number of data-data, random-random, and data-random galaxy pairs (all appropriately normalized) within a separation of $r \pm dr/2$, where $dr$ is the bin width. For the two-point correlation function, $\xi(r)> 0$ indicates that the population of galaxies is clustered on scales of length $r$, $\xi(r) = 0$ signifies a randomly distributed population, and $\xi(r) < 0$ indicates an under-populated region with length scale $r$. We measured the correlation length, $r_0$, where $\xi(r_0) = 1$, and the critical length, $r_c$, where $\xi(r_c) = 0$ for each data set (for multiple crossings of zero the first instance was reported). The spatial two-point correlation function was calculated in 20 equally spaced bins within the separations of 0 to 250~kpc. The errors were modeled assuming that the data-data pair counts were Poissonian, $\sigma_{\xi(r)} = \frac{1 + \xi(r)}{\sqrt{N_{DD}}}$. We note that the Poissonian method for error estimation, while frequently used in clustering studies, can potentially underestimate the errors for two-point correlation functions compared to other more robust methods, so the errors reported here may be smaller than the true error. For each two-point correlation function statistic, we report all clustering signals that have a minimum significance of $2.6\sigma$ (or $99\%$ likelihood).

The clustering measure made by the two-point correlation function estimator is strongly dependent on the random set of ``galaxies". This random set is required to be uniformly distributed within an ``observational window" that is defined by the volume that contains the galaxies in the data set. We used two observational windows: the first was defined by the full halo of the Milky Way within the radial range of 4~kpc to 260~kpc, while the second window was defined by the subsection of the full volume constrained by the SDSS DR7 Northern Contiguous footprint. The subhalo data sets of the Via Lactea simulation (described above) were selected using the same observational windows. The random sets each contain 50,000 ``galaxies''. This is $\sim10 - 1000$ times larger in number than each data set, in order to achieve an unbiased estimation of the clustering of each actual data set. The errors in the spatial two-point correlation functions due to random seed variations are $\sim0.5\%$ for a random set of this size. Random seed variations between the two random sets used to determine the spatial two-point correlation functions of the two full-area subhalo sets with different ``observer" locations (i.e. VL-A and VL-Arot) caused differences on this scale in what would otherwise be identical spatial clustering results. These small differences did not affect our final conclusions, although we report all differences in the spatial clustering results between different ``observer" locations that arise from these variations below.

\begin{figure*}
\begin{center}
\includegraphics[trim = 0mm 0mm 0mm 13mm, clip,width=0.99\textwidth]{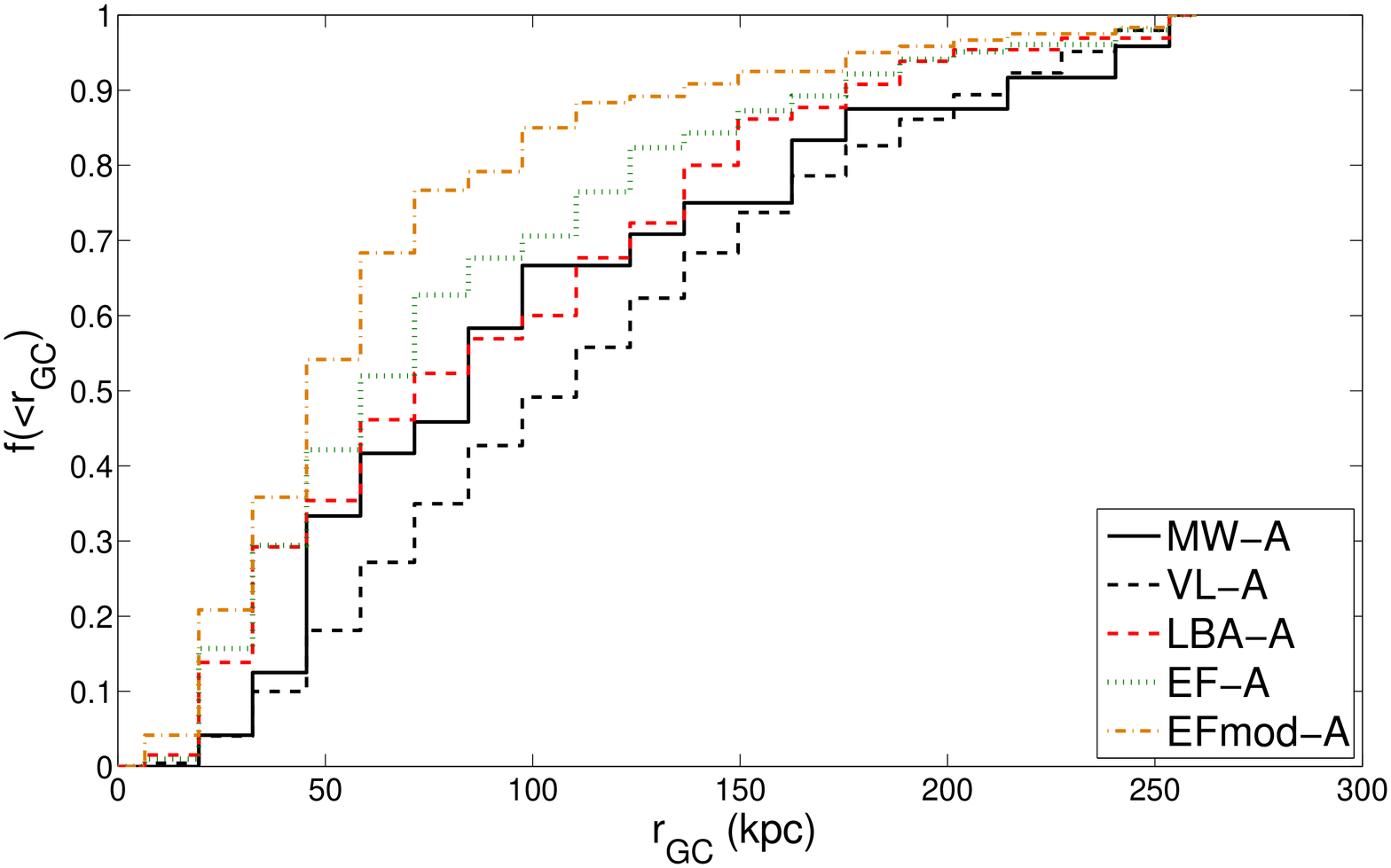} 
\caption{\scriptsize Cumulative radial distributions of the MW-A (solid, black), VL-A (thin-dashed, black), LBA-A (thick-dashed, red), EF-A (dotted, green), and the EFmod-A (dot-dashed, orange) sets. The three ``luminous" subhalo sets are all more centrally concentrated than the VL-A dark subhalo set and the Milky Way satellite galaxies.}
\label{fig:Radial_main}
\end{center}
\end{figure*}

We explored the impact of the radial density profile of each data set on the spatial clustering results by constructing a mock set of 10,000 ``galaxies" that are randomly distributed in $(\theta,\phi)$, in a spherically symmetric configuration, with the same mean radial number density profile as the corresponding data set. Each of the two-point correlation functions of these sets was denoted with a ``-sm" suffix, i.e. VL-A-sm. The radial density profile was determined by finding the lowest nth-order polynomial that fit the cumulative radial distribution of each set to better than $90\%$ confidence, according to a Kolmogorov-Smirnov test. The radial density profiles of the various data sets are shown in Fig.~\ref{fig:Radial_main} and the polynomial degree of each set is listed in Table \ref{tab:Cluster_nums}.  Random seed variations in the mock set results between different ``observer" locations are on the same scale as the random sets mentioned above and, therefore, did not affect the final conclusions.

\subsection{Line-of-sight Velocity Two-Point Correlation Function}

The Galactocentric line-of-sight velocity two-point correlation function measures  galaxy-galaxy clustering in terms of their separation in line-of-sight velocity space, defined as the magnitude of the line-of-sight velocity vector difference\footnote{The line-of-sight velocity vector difference includes the angular separation of the galaxies on the sky.} : $|\delta \overrightarrow{v_{los}}| = |\overrightarrow{v_{los,i}} - \overrightarrow{v_{los,j}}|$. We used the LS estimator:
\begin{equation}
\label{eq:LS_2}
\xi(|\delta \overrightarrow{v_{los}}|) = \frac{N_{DD}(|\delta \overrightarrow{v_{los}}|) - 2N_{DR}(|\delta \overrightarrow{v_{los}}|) + N_{RR}(|\delta \overrightarrow{v_{los}}|)}{N_{RR}(|\delta \overrightarrow{v_{los}}|)},
\end{equation}
in a manner similar to that for the spatial two-point correlation function. We binned the data from $|\delta \overrightarrow{v_{los}}| = 0 km s^{-1}$ to $|\delta \overrightarrow{v_{los}}| = 250 km s^{-1}$ using a total of 20 equally spaced bins. 

The ``galaxies" of the random data sets were each assigned a line-of-sight velocity magnitude using a uniform probability distribution, as required by the two-point correlation estimator, within the same line-of-sight velocity space observational window as the respective data set to which they are compared. The line-of-sight velocity vector of each random set ``galaxy" is then defined by the position of each ``galaxy" and a hypothetical ``observer" chosen to be 8~kpc from the center of the volume. 

We found that a single random set of 50,000 ``galaxies" produced errors due to random seed variations that could be as large as $5\%$. This is up to ten times larger than the random seed error for the spatial and 4D phase-space analyses. We limited this variation by calculating a final line-of-sight velocity two-point correlation function that is the average over 50 line-of-sight velocity two-point correlation functions, each calculated using a random set of 10,000 ``galaxies". Our final result effectively used a random set of 500,000 galaxies and the error associated with a random seed for the final result was found to be no larger than $1-2\%$. This method has been adopted previously in other two-point correlation function analyses and is computationally more efficient than using a single set of 500,000 points \citep{1994ApJ...425L...5G}. The total error for the final line-of-sight two-point correlation function is given by: 
\begin{equation}
\sigma_{\xi(|\delta \overrightarrow{v_{los}}|)} = \sqrt{\sigma_{mean}^2 + \langle \sigma_i \rangle^2},
\end{equation}
where 
\begin{equation}
\begin{array} {c c}
\sigma_i = \frac{1 + \xi(|\delta \overrightarrow{v_{los}}|)}{\sqrt{N_{DD}}},  \\
\\
\sigma_{mean} = \sqrt{\Sigma(\frac{1}{(N)(N-1)})(\xi_{mean} - \xi_i)^2},
\end{array}
\end{equation}
and $\langle \rangle$ brackets denote the average over the 50 sets.

While the single dimension of the line-of-sight velocity limits the utility of this statistic, the physical interpretation of the statistic is aided by the angular separation of the galaxy pairs included in the vector difference. For example, galaxy pairs with a small $|\delta \overrightarrow{v_{los}}|$ likely have a similar speed and either the same line-of-sight velocity sign coupled with a relatively small angular separation on the sky or the opposite line-of-sight velocity sign combined with a nearly maximum angular separation. An excess in clustering signal strength for small $|\delta \overrightarrow{v_{los}}|$ beyond that found for the underlying line-of-sight velocity distribution would indicate that pairs of either type are not rare within their host. While the two-point correlation function signal of a particular bin is not straightforward to interpret, the differences in the clustering signal between the observed Milky Way satellites and the various subhalo sets of the Via Lactea II simulation are robust. Although it is beyond the scope of this paper, differences in clustering strengths between sets of satellites or subhaloes on particular scales can be further studied to identify the combinations that are responsible for a given detected two-point correlation function signal. 

A line-of-sight velocity was also randomly assigned to the ``galaxies" of the mock sets according to the line-of-sight velocity distribution of each corresponding data set. This allowed a study of the influence of the line-of-sight velocity distribution on the clustering results of the data sets. A Gaussian distribution was used if it fit the line-of-sight velocity distribution to better than $85\%$ confidence (according to a Kolmogorov-Smirnov test). If a Gaussian did not satisfy this criterion then an nth-order polynomial was fit to the whole line-of-sight velocity distribution. If an nth-order polynomial failed to fit the whole line-of-sight velocity distribution, then the positive and negative velocity distributions were fit separately. The line-of-sight velocity distribution models used to generate each of the line-of-sight velocity distributions for the mock sets are listed in Table \ref{tab:Cluster_nums}.

\subsection{Four-dimensional Phase-Space Two-Point Correlation Function} 

A statistical analysis of clustering in phase-space would provide the strongest determination of substructure by simultaneously evaluating both spatial and kinematic associations. However, only a small fraction of the Milky Way satellite galaxies have even mean proper motion data. The maximum information that is available for all the known Milky Way satellite galaxies is limited to the estimates of the three-dimensional position and the line-of-sight velocity, which can be combined to define a four-dimensional (4D) phase-space coordinate for each object. The clustering in 4D phase-space can then be determined with a two-point correlation function of separations in this space. \citet{2009ApJ...698..567S} and \citet{2010arXiv1011.1926C} conducted similar analyses of clustering in 4D phase-space of stars in each of the Milky Way stellar halo and simulated stellar haloes. 

The scalar difference of line-of-sight speed was used, rather than the vector difference that was used in the line-of-sight velocity analysis described above, since the vector difference includes redundant information on the angular separation of the two objects. The 4D phase-space length is then defined to be: 
\begin{equation}
\label{eq:theta}
\theta = \sqrt{w_r|\overrightarrow{r_i} - \overrightarrow{r_j}|^2 + w_v(v_{los,i} - v_{los,j})^2},
\end{equation}
where the $w_r$ and $w_v$ terms are weights that are defined by:
\begin{equation}
\begin{array} {c c}
w_r = (\frac{\displaystyle 1}{\displaystyle \Delta r_{max}})^2 \frac{\displaystyle (\frac{r_{err,i}}{r_i})^2 + (\frac{r_{err,j}}{r_j})^2}{\displaystyle 2 \langle \frac{r_{err}}{r} \rangle^2}, \\
\\
w_v = (\frac{\displaystyle 1}{\displaystyle \Delta v_{max}})^2 \frac{\displaystyle v_{err,i}^2 + v_{err,j}^2}{\displaystyle 2 \langle v_{err} \rangle^2}.
\end{array}
\end{equation}
The symbols $r_{err}$ and $v_{err}$ respectively represent the error for the distance and line-of-sight velocity measurements (ignored in the other clustering analyses). The angle brackets in the denominator denote an average over the whole sample. 

The normalization terms $\Delta r_{max}$ and $\Delta v_{max}$ set the physical scale for the 4D phase-space length. \citet{2010arXiv1011.1926C} determined, after exploring many different weighting options for their definition of 4D phase-space distance, that there was not a physically well-motivated optimal choice of normalization; happily different choices for normalization did not affect their final result. We experimented with different normalizations and also found that our conclusions remain unchanged regardless of the choice of normalization. Without a clear physical choice, we set their values to the largest observed values in our largest Milky Way satellite galaxy data set ``MW-A": $\Delta r_{max} = 353$~kpc and $\Delta v_{max} = 291 km s^{-1}$. We used the same normalization for every set of galaxies or subhaloes in order to have a one-to-one comparison between different sets. A different choice of this global normalization would result in a simple scaling in both $\theta$ and $\xi_{4D}(\theta)$ that would maintain the relative strength of clustering between different data sets.

The second component of each weighting term is the error for each object in terms of the average error of the entire set.  The inclusion of errors into the weighting pushes small 4D phase-space lengths that are poorly measured to larger, less physically meaningful values. Pairs of galaxies that are tightly bound in 4D phase-space, such as the Leo IV and Leo V dwarf galaxies, have been considered more likely to be related \citep{2008ApJ...686L..83B}. A small 4D phase-space length is not independently sufficient for determining the authenticity of associations since any proximity in 4D phase-space could be coincidental. This motivates the use of a two-point correlation statistic in 4D phase-space.

We determined the clustering in 4D phase-space by calculating the two-point correlation function with the LS estimator: 
\begin{equation}
\xi_{4D}(\theta) =  \frac{N_{DD}(\theta) - 2N_{DR}(\theta) + N_{RR}(\theta)}{N_{RR}(\theta)}
\end{equation}
The data were binned from $\theta = 0$ to $\theta = 1$ using a total of 20 equally spaced bins (the scale of $\theta$ ranging from zero to one is due to the choice of weights discussed above) and the errors were modeled assuming that the data-data pair counts are Poissonian, $\sigma_{\xi(\theta)} = \frac{1 + \xi(\theta)}{\sqrt{N_{DD}}}$. 

The weighting terms of the 4D phase-space metric require that position and velocity errors are assigned to each of : the ``galaxies" of the random sets; mock sets of ``galaxies" modeled to represent the smooth underlying density profile and line-of-sight velocity distribution; and the sets of subhaloes from the VL-II simulation. A ``galaxy" or subhalo in each of these sets was randomly assigned, using a uniform probability distribution, a line-of-sight velocity error ($\sigma_{v,los}$) and a fractional distance error ($\sigma_d/d$) from the range of values measured for the full-area set of Milky Way satellite galaxies, MW-A.

\section{Results}
\subsection{Spatial Clustering of the Milky Way Satellite Galaxies}

The spatial clustering results of the Milky Way satellite galaxies are shown in the top two panels of Fig.~\ref{fig:dwarf_galaxies_spatial}. The two-point correlation function of the set of all known Milky satellite galaxies (MW-A), represented by the solid black curve in the top left panel, shows that the satellite galaxies are clustered, at a significance of greater than $2.6\sigma$, for bin 2 and bins 5 through 10. The contribution of the smooth underlying density profile to this spatial clustering signal was determined by comparing the Milky Way satellites' spatial two-point correlation function to the two-point correlation function of a mock set of  ``dwarf galaxies" that are randomly distributed with the same mean radial number density profile as the Milky Way satellites (represented by the dotted black curve shown in same panel). The two curves are within 2$\sigma$ of each other on all scales, indicating that the clustering signal of the set of all known Milky Way satellite galaxies is the result of a steep density profile, rather than the result of substructure within the set.

Our analyses finds no evidence of spatial substructure within the Milky Way satellite distribution; however, as noted above, the Milky Way satellite galaxies have been reported to be oriented in a disk-like structure \citep{2005A&A...431..517K, 2007MNRAS.374.1125M, 2009MNRAS.394.2223M}. Although there is not a $3\sigma$ detection of substructure in the MW-A spatial two-point correlation function, there is a hint of a disk of satellites. There is a small oscillation, on the $1\sigma$ level, in the amplitude of bins two through six of the MW-A spatial two-point correlation function around the two-point correlation function of the corresponding mock set that can be attributed to the disk-like distribution of the satellite galaxies. For example, the clustering signal of the second bin is primarily due to satellites near the center of the disk that are clustered on scales smaller than the thickness of the disk of satellites\footnote{Defined here as twice the root-mean-square height of the disk \citep{2009MNRAS.394.2223M}: $2 \times \Delta = 57$~kpc.}, and over half the pairs that contribute to the clustering signal of the sixth bin are separated on scales larger than the disk thickness but oriented along the plane of the disk. The dip in the two-point correlation function amplitude between these two bins is due to a lack of pairs intermediate to these types of associations, as would be expected from a disk-like structure.

The two-point correlation function of the Milky Way satellite galaxy footprint set (MW-B) is represented by the dashed red curve in the top right panel of Fig.~\ref{fig:dwarf_galaxies_spatial}. The amplitude of this two-point correlation function is diminished compared to that for the full-area set, such that, on all scales, the MW-B set is consistent with a random distribution to within a significance level of $2.6\sigma$.

\subsection{Line-of-Sight Velocity Space and Four-Dimensional Phase-Space Clustering of the Milky Way Satellite Galaxies}

\begin{figure*}
\begin{center}
\begin{tabular}{lll}
\includegraphics[trim = 10mm 2.5mm 20mm 10mm, clip,width=0.49\textwidth]{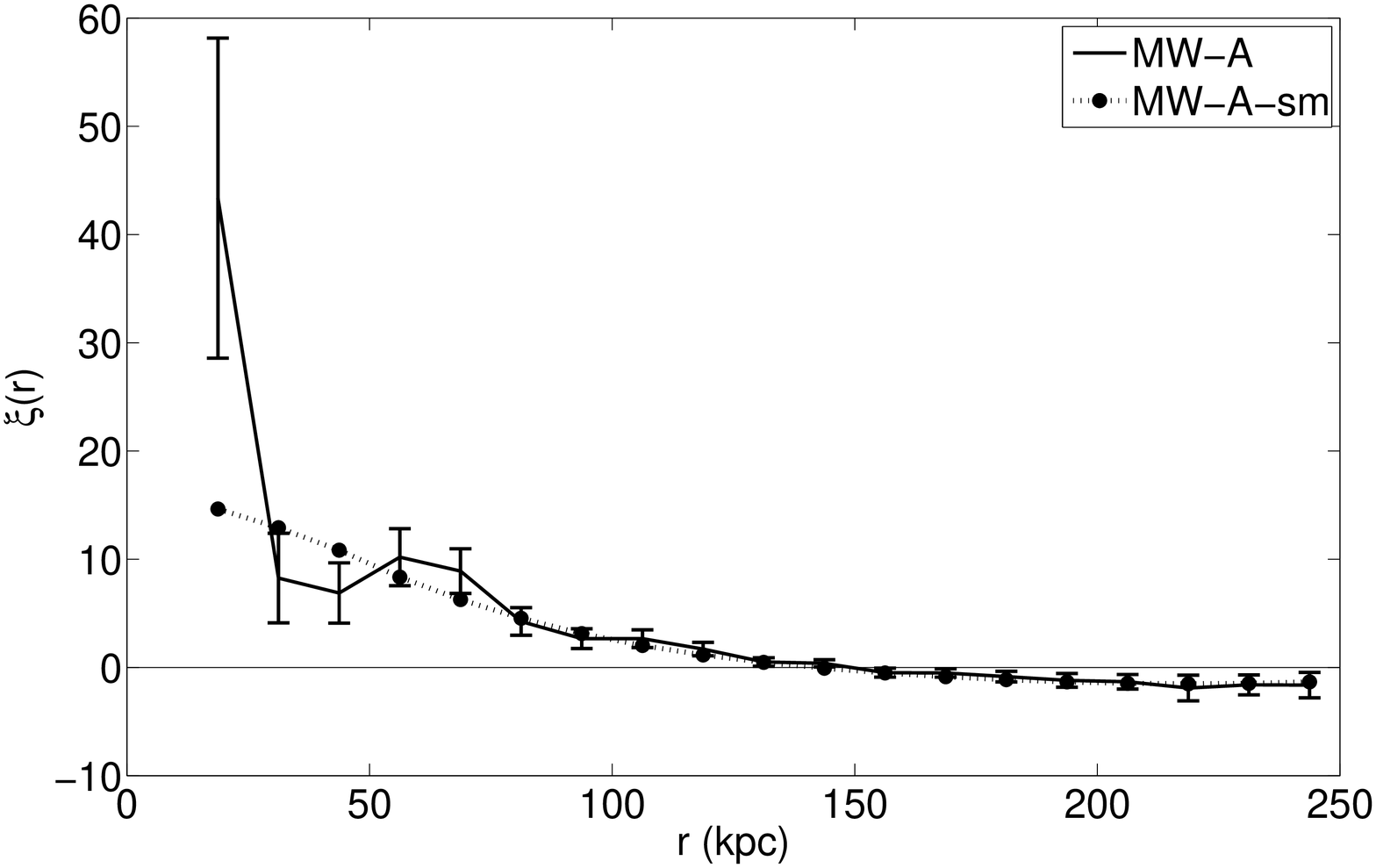} &\includegraphics[trim = 10mm 2.5mm 20mm 10mm, clip,width=0.49\textwidth]{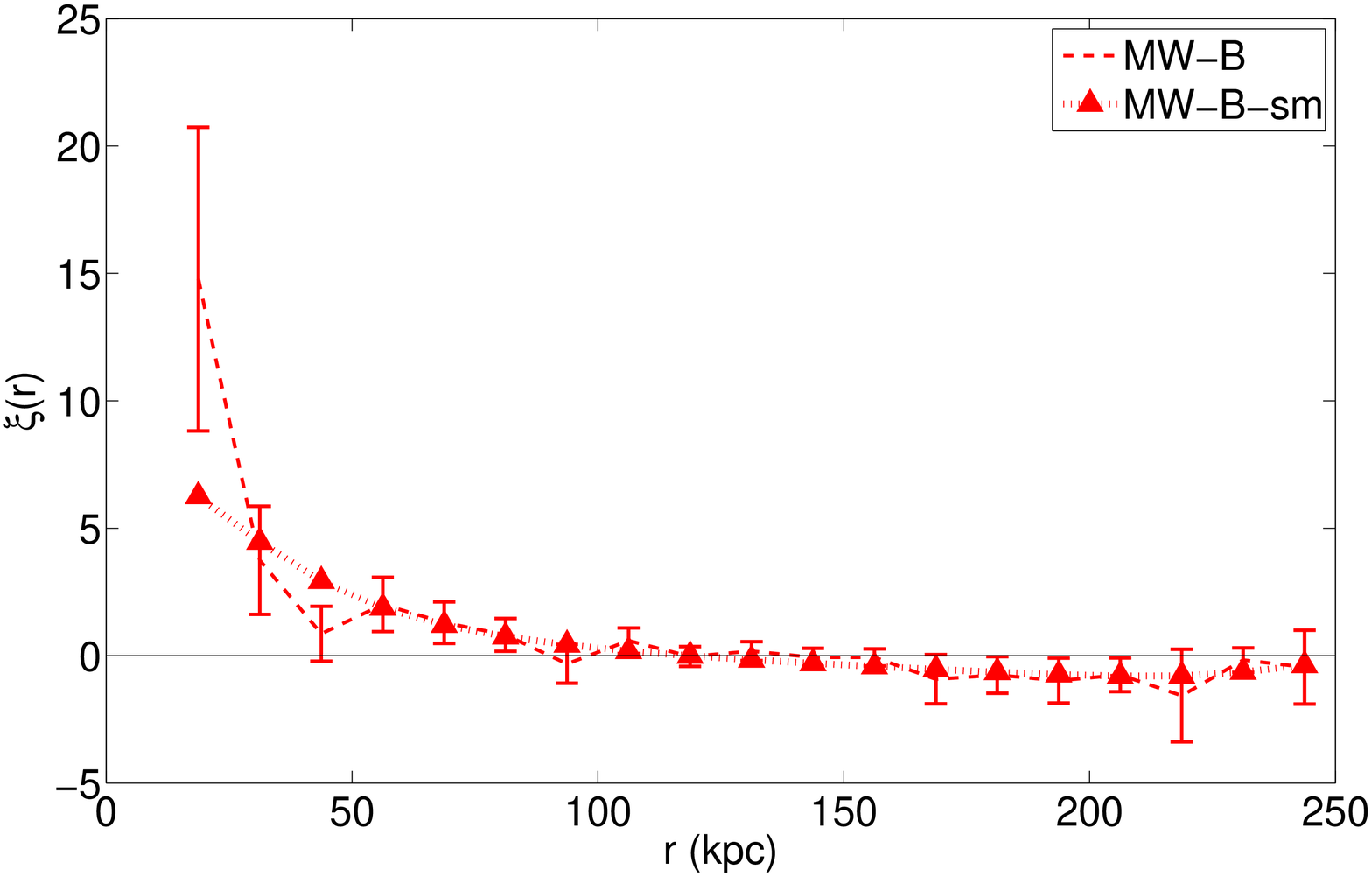} \\
\includegraphics[trim = 5mm 2.5mm 20mm 10mm, clip,width=0.49\textwidth]{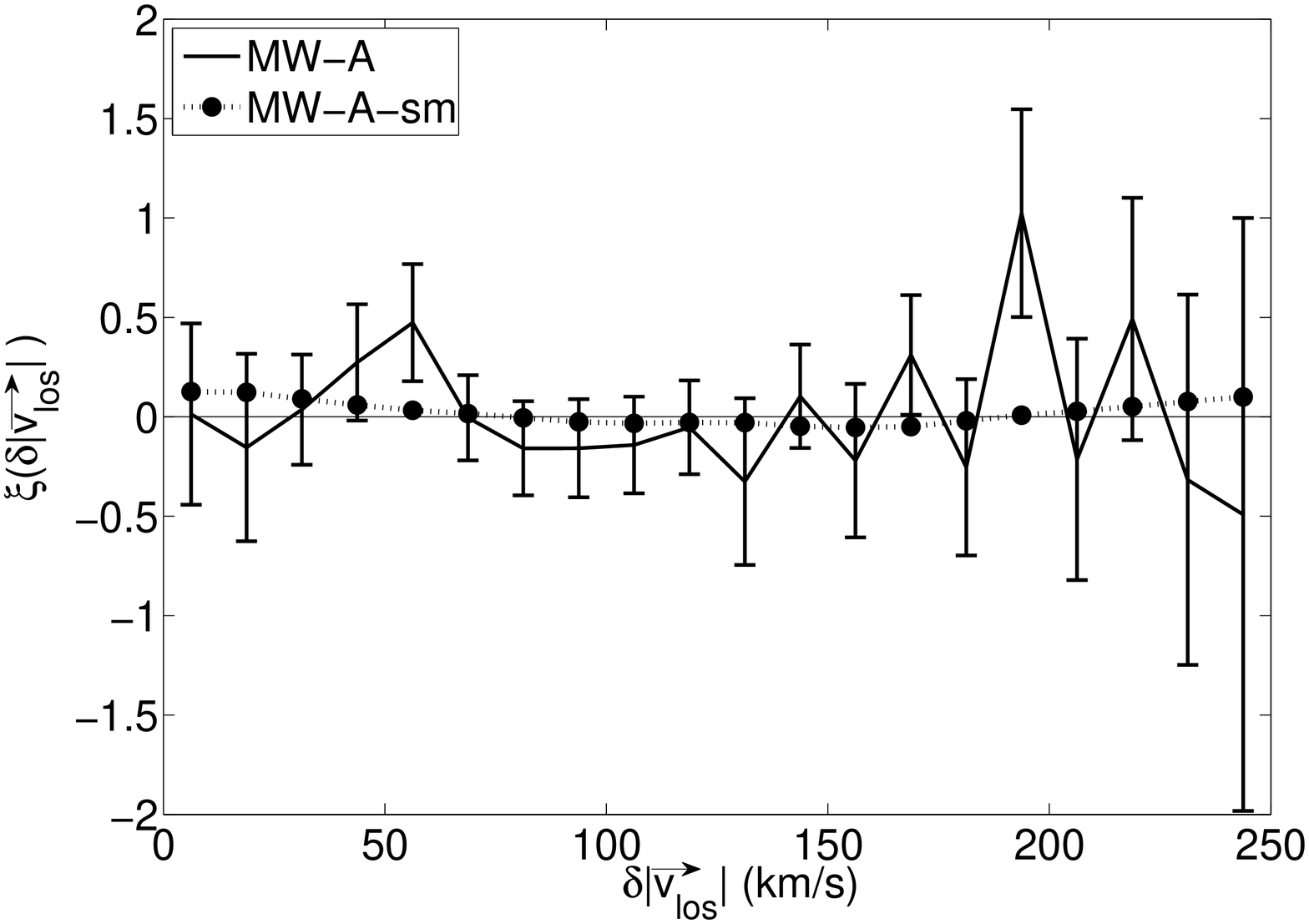} & \includegraphics[trim = 5mm 2.5mm 20mm 10mm, clip,width=0.49\textwidth]{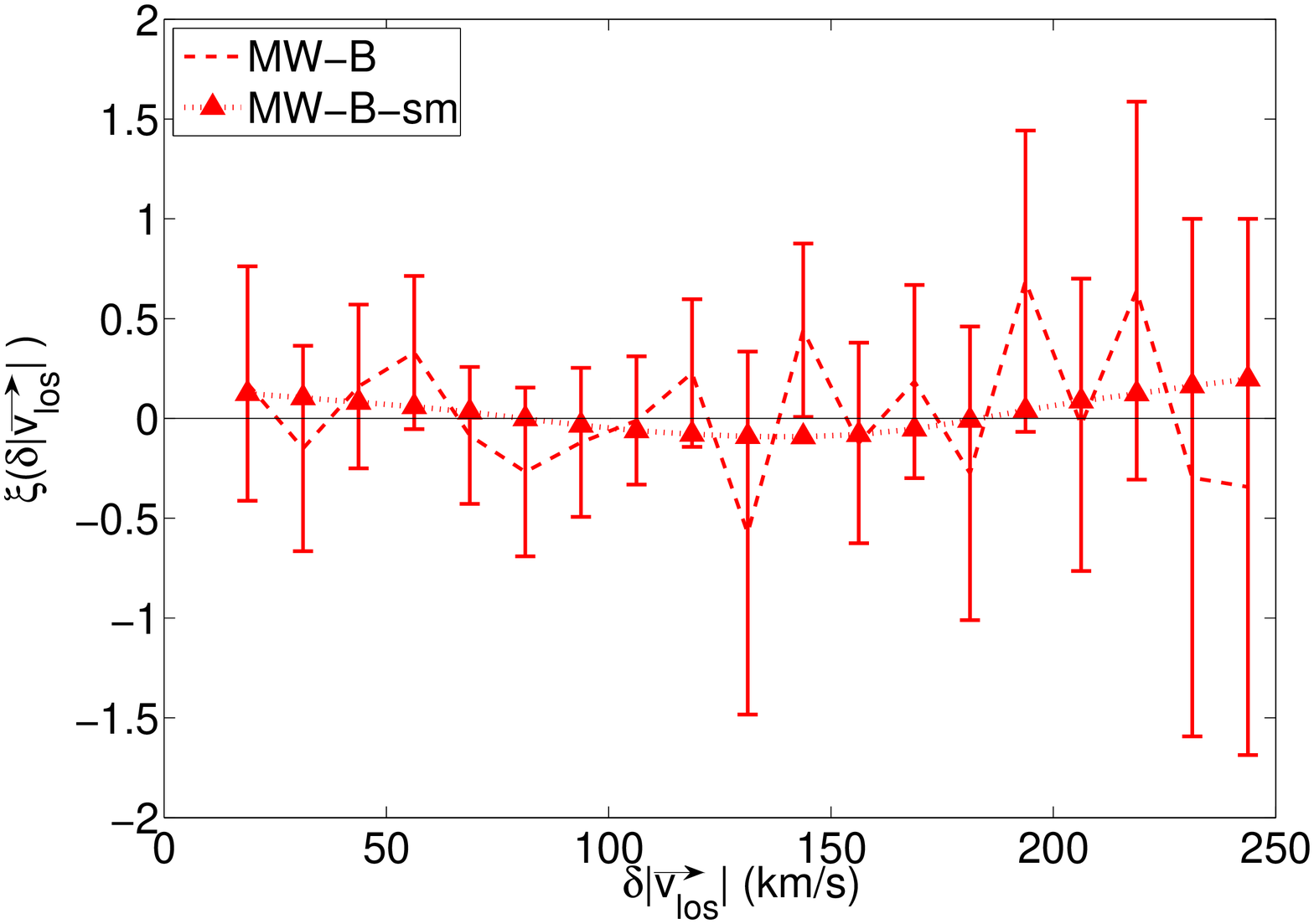} \\
\includegraphics[trim = 10mm 2.5mm 20mm 10mm, clip,width=0.49\textwidth]{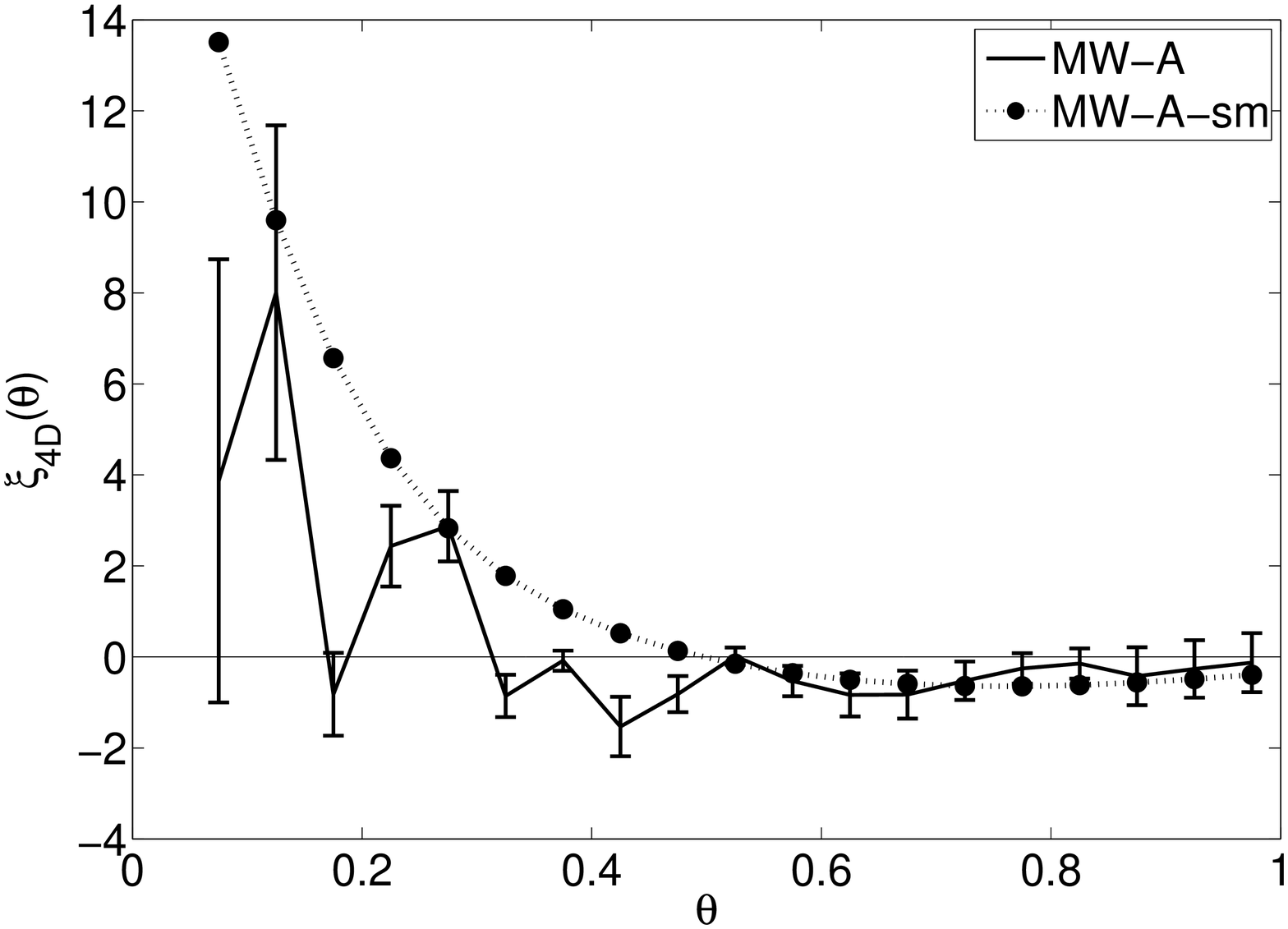} & \includegraphics[trim = 10mm 2.5mm 20mm 10mm, clip,width=0.49\textwidth]{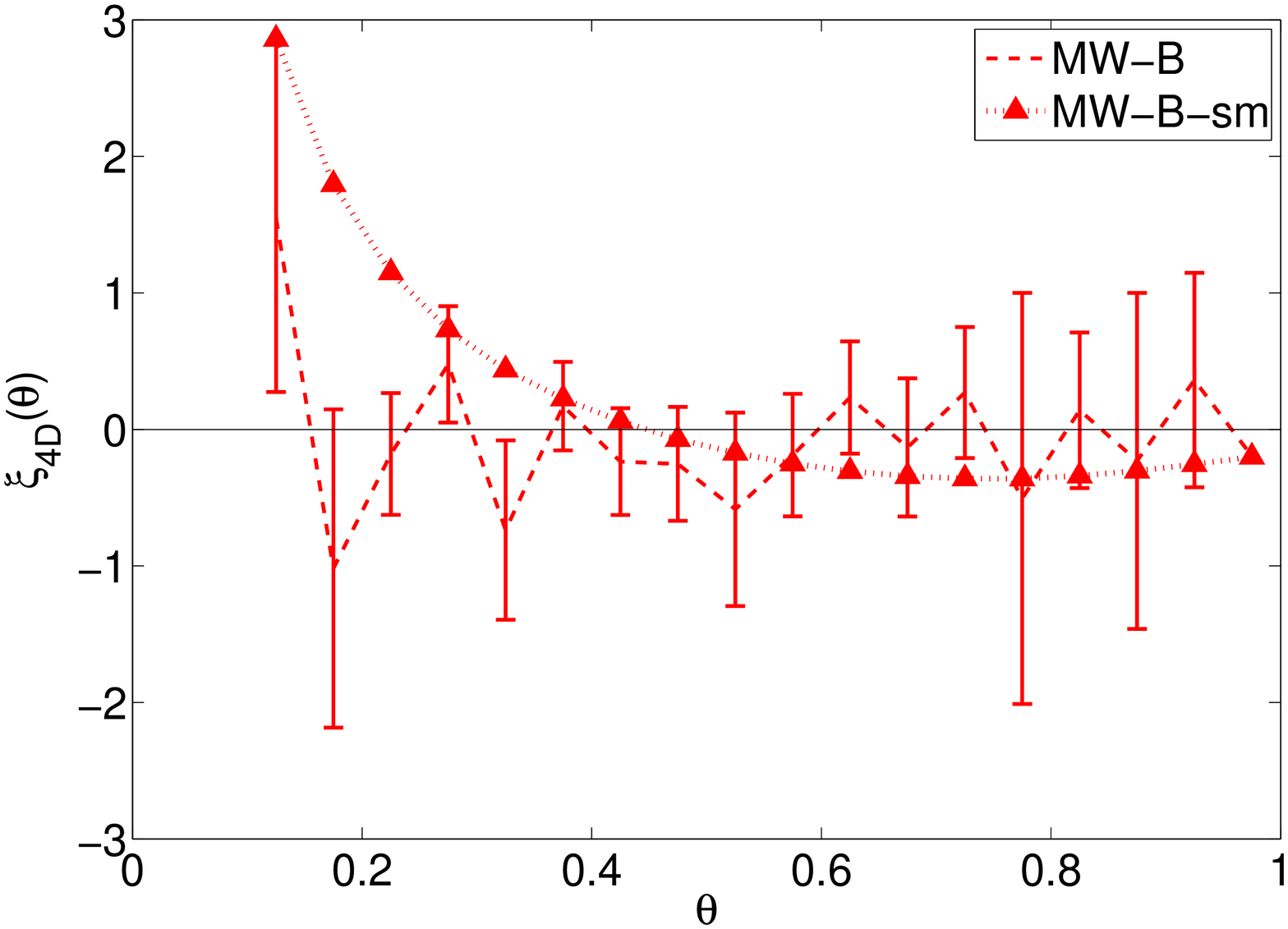} \\
\end{tabular}
\caption{\scriptsize The spatial (top row), line-of-sight velocity (center row), and 4D phase-space (bottom row) two-point correlation functions for the Milky Way satellite galaxies: MW-A (solid black, left panels), and the Milky Way satellite galaxies that are contained within the boundaries of the SDSS DR7 Northern Contiguous Footprint, MW-B (dashed red, right panels). The mock sets of ``galaxies", MW-A-sm (dotted, black, circle markers) and MW-B-sm (dotted, red, triangles), that represent the clustering signal due to the smooth underling density profile and line-of-sight velocity distribution of the two data sets. The MW-A and MW-B are clustered in configuration space and four-dimensional phase-space. Both sets are randomly distributed in line-of-sight velocity space. There is no evidence of substructure within the distribution of either set in any space. The black horizontal line in all panels shows the value for the two-point correlation function that corresponds to a uniform random distribution. The following bins of the noted space's two-point correlation functions have no data-data pairs and are, therefore, not shown: bin 1(MW-A and MW-B spatial), bin 1 (MW-A 4D phase-space), bin 1 and 2 (MW-B 4D phase-space).} 
\label{fig:dwarf_galaxies_spatial}
\end{center}
\end{figure*}

Both the set of all known Milky Way satellites galaxies and the footprint set are randomly distributed in line-of-sight velocity space as shown in the central panels of Fig.~\ref{fig:dwarf_galaxies_spatial}. The 4D phase-space two-point correlation function of the MW-A set, represented by the solid black curve in the bottom left panel of Fig.~\ref{fig:dwarf_galaxies_spatial}, shows the MW-A set is clustered in 4D phase-space for bins 5 and 6 at a significance level above $2.6 \sigma$. The clustering signal of these bins is consistent to within $2\sigma$ of the 4D phase-space two-point correlation function of the corresponding mock set of ``galaxies" (MW-A-sm), represented by the dotted black curve. The consistency of these two curves indicates the clustering signal of the MW-A set is due to the underlying 4D phase-space distribution of the Milky Way satellite galaxies, rather than substructure within this distribution. The overall amplitude of the 4D phase-space two-point correlation function for the MW-B set, represented by the solid red curve in the bottom right panel of Fig.~\ref{fig:dwarf_galaxies_spatial}, is diminished with respect to the MW-A two-point correlation function and, as a result, the MW-B set is consistent with a random distribution to $2\sigma$.

\subsection{Via Lactea II Subhalo Spatial Clustering Results} 

The spatial two-point correlation functions of the ``luminous" subhalo sets of the Via Lactea II simulation are shown in the top panels of Figs. \ref{fig:LBA results} - \ref{fig:KOP results}. The solid curves, in the top left panels, represent the full-area sets, while the dashed and dot-dashed curves in the top right panel represent the two footprint sets. The dotted curves in each panel represent the two-point correlation functions of the mock sets of ``subhaloes" indicating the contribution of each subhalo set's smooth underlying density profile to its clustering signal. Substructure within the spatial distribution of the subhalo set is defined to exist on any scale where the clustering of the data set is greater than the clustering due to the smooth underlying density profile. The ``rotated'' ``observer" location results for the full-area sets are not shown in the top left panels for clarity, since they are nearly identical to the first. The scales where the clustering signal of each set exceeds the contribution from the smooth underlying density profile by $2.6\sigma$ or more is listed in Table \ref{tab:Cluster_nums} for each set. 

The full-area ``luminous" subhalo sets are each clustered, with a two-point correlation function amplitude greater than zero at a significance level above $4.3\sigma$ on nearly all scales. We found no evidence of substructure within the spatial distributions of any of the full-area ``luminous" subhalo sets. The amplitude of the LBA-A set's two-point correlation function in the twelfth bin is in excess of the two-point correlation function amplitude of the corresponding mock set of ``subhaloes", LBA-A-sm, by $3.4\sigma$. However, at a two-point correlation function amplitude of $\sim0.5$, the LBA-A clustering strength is modest and, therefore, not a clear detection of substructure. The spatial two-point correlation function of the full-area ``dark" subhalo set, VL-A, represented by the black solid curve in the top left panel of Fig.~\ref{fig:VL results}, shows that this set is spatially clustered at a significance level above $7\sigma$ for the first 13 bins . Evidence of substructure is detected at a significance of greater than $6.5\sigma$ in the first 8 bins of the VL-A spatial two-point correlation function.

\begin{figure*}
\begin{center}
\begin{tabular}{lll}
\includegraphics[trim = 10mm 2.5mm 20mm 10mm, clip,width=0.49\textwidth]{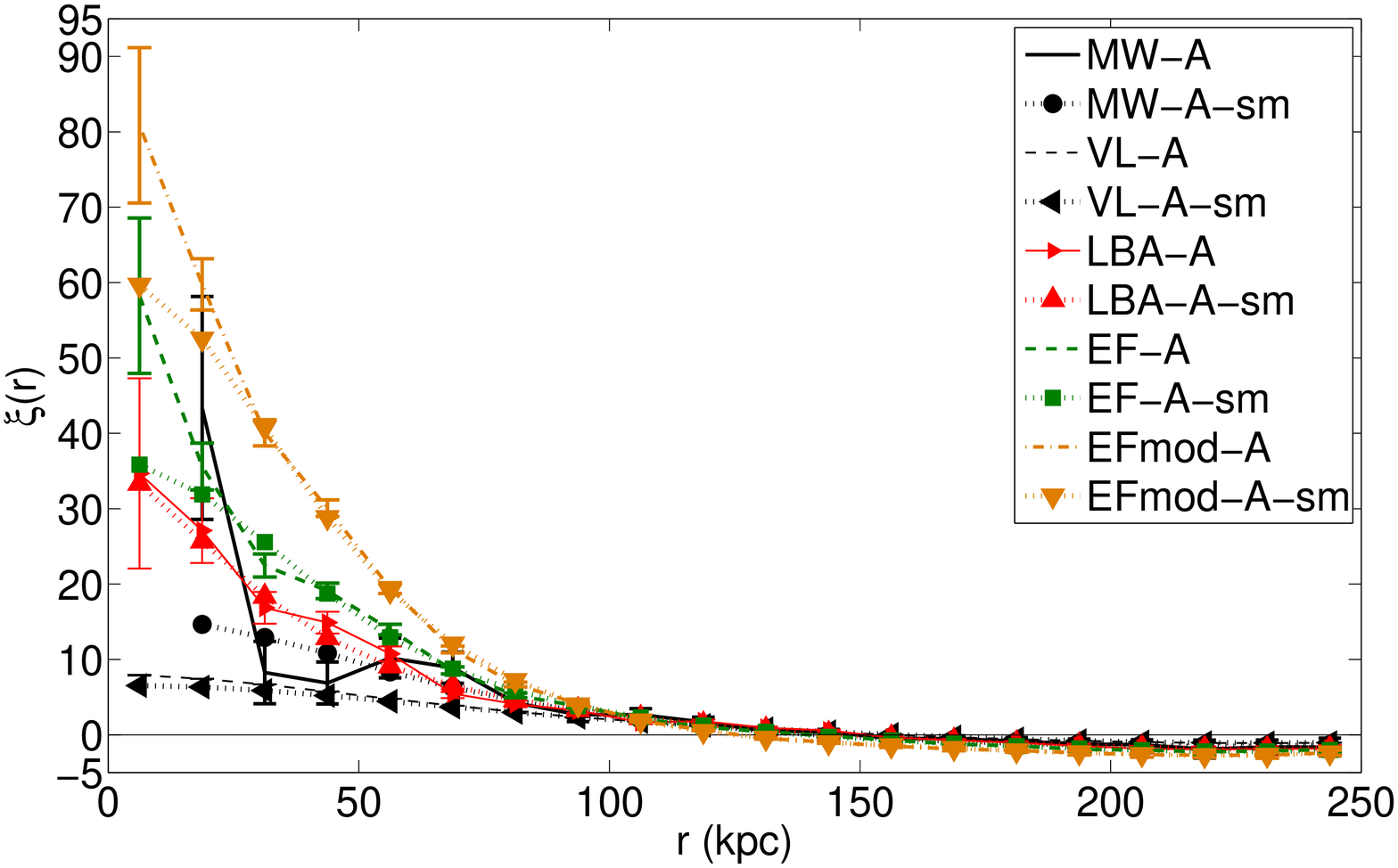} &\includegraphics[trim = 10mm 2.5mm 20mm 10mm, clip,width=0.49\textwidth]{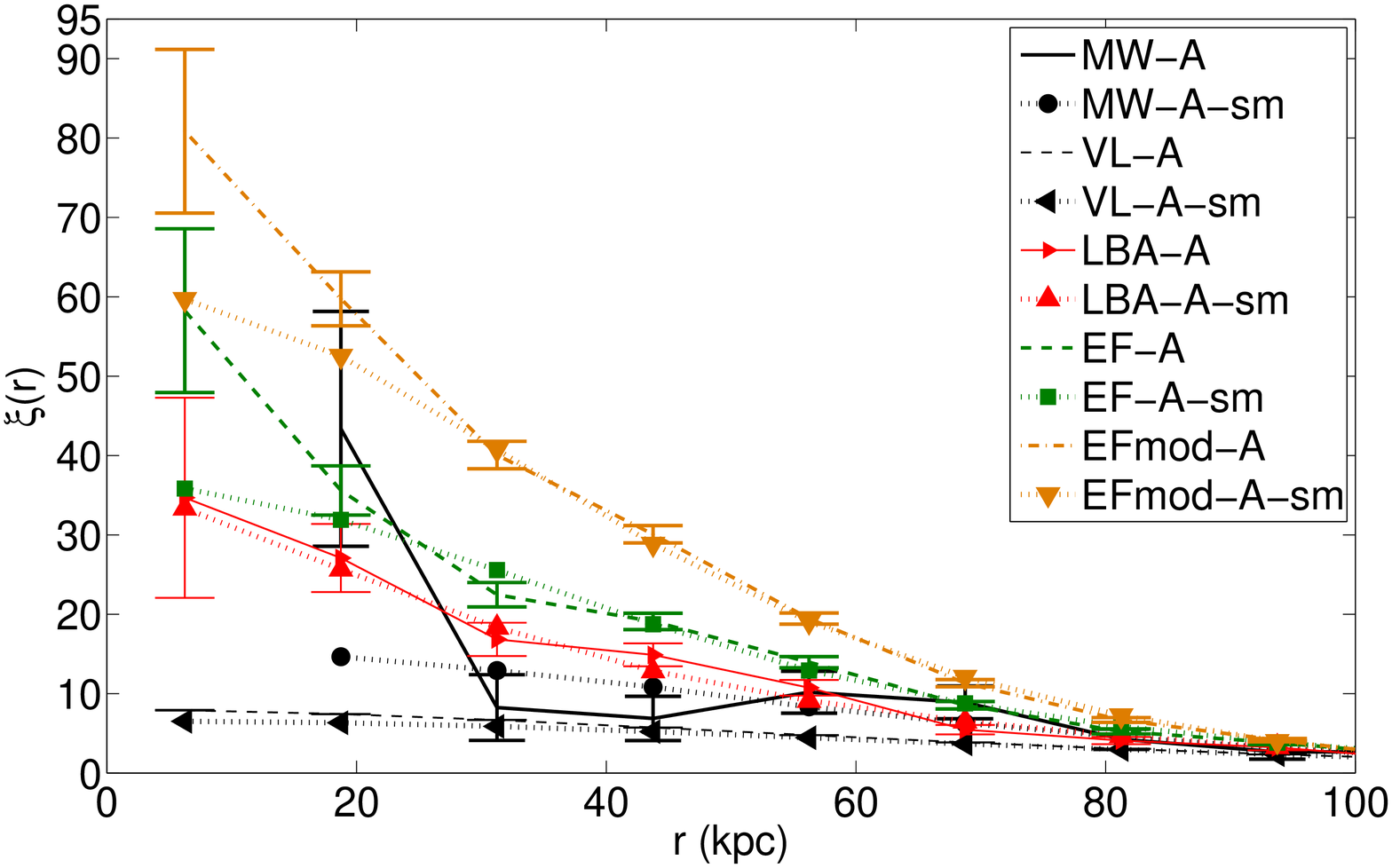} \\
\includegraphics[trim = 5mm 2.5mm 20mm 10mm, clip,width=0.49\textwidth]{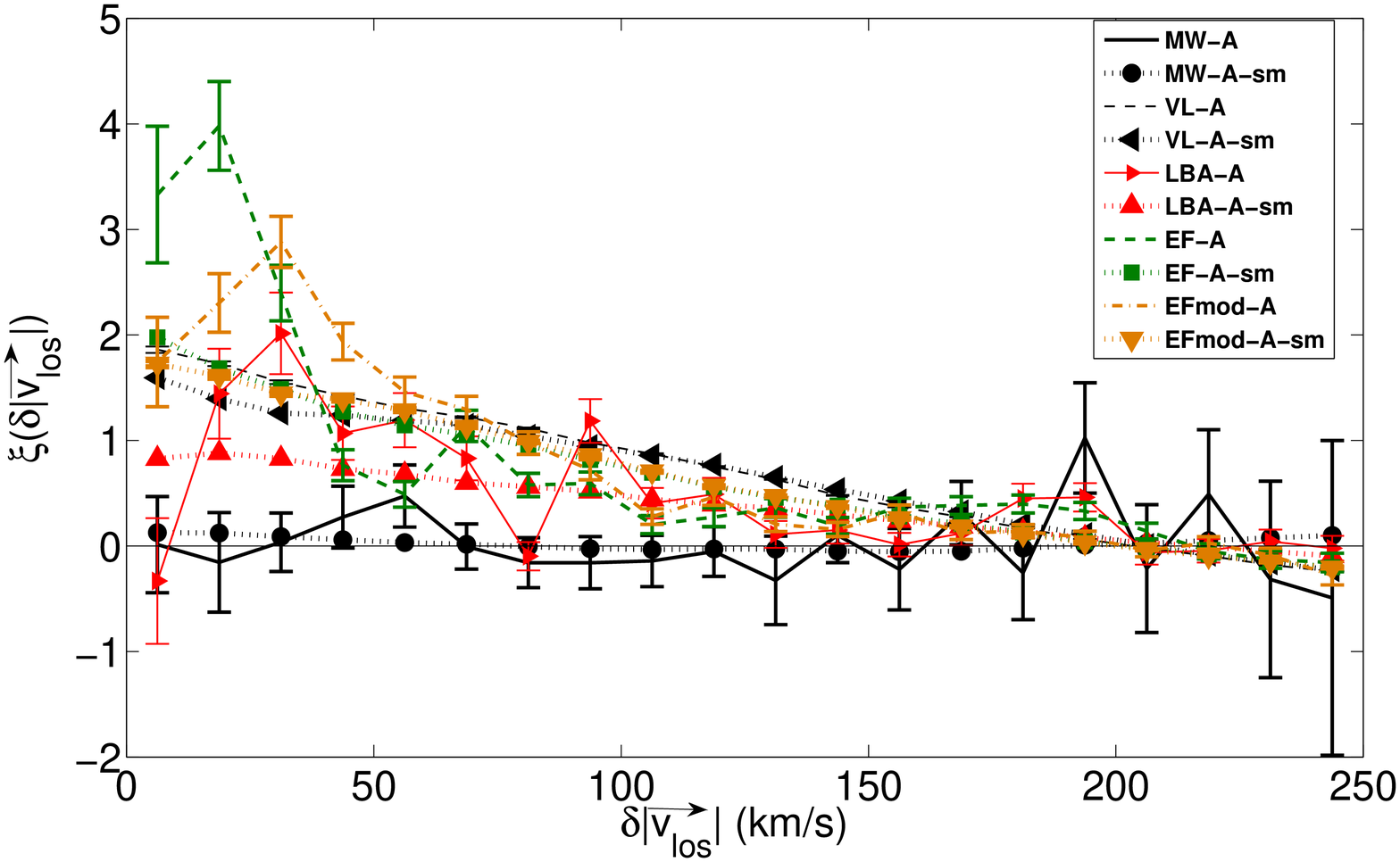} & \includegraphics[trim = 5mm 2.5mm 20mm 10mm, clip,width=0.49\textwidth]{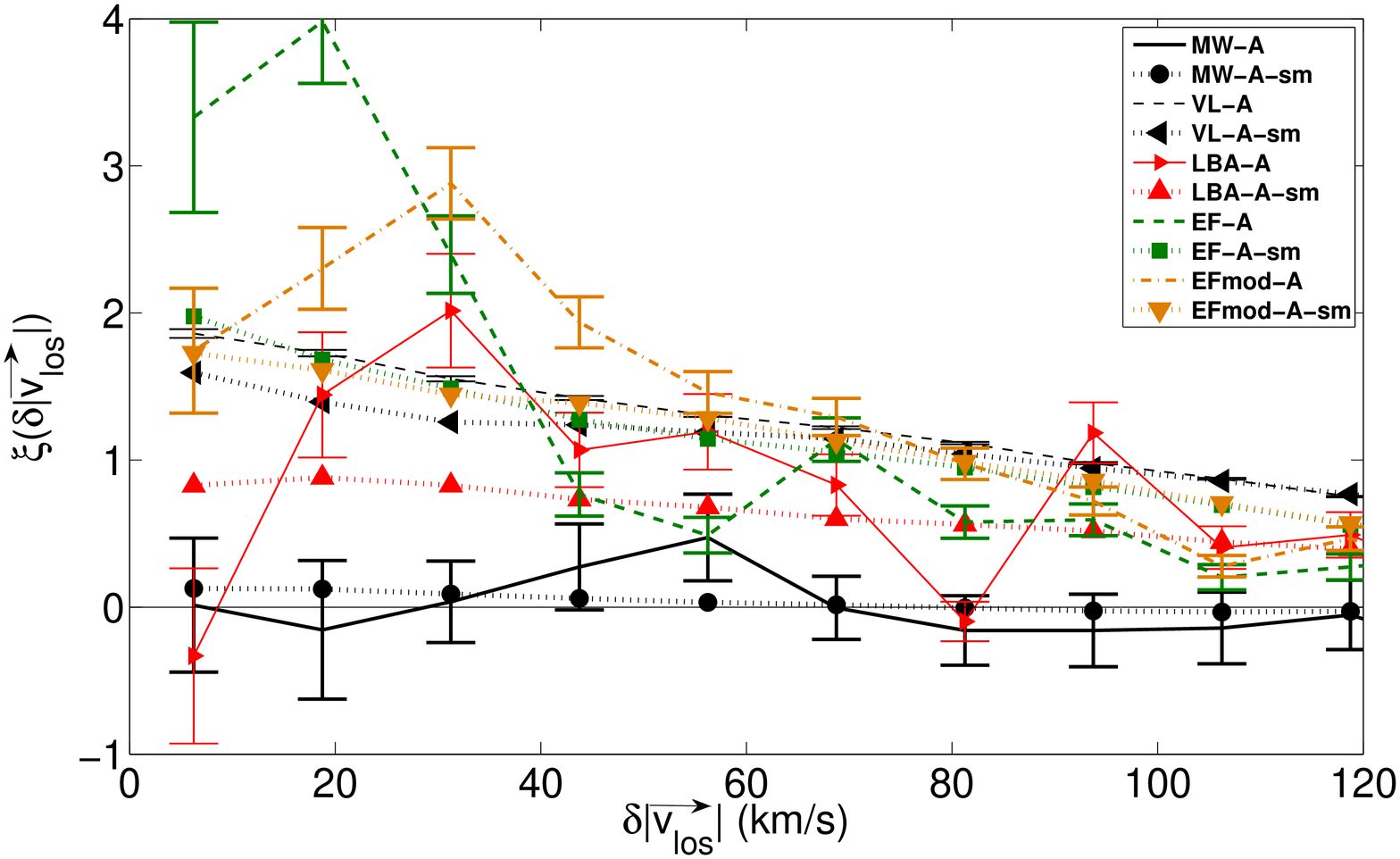} \\
\includegraphics[trim = 5mm 2.5mm 20mm 10mm, clip,width=0.49\textwidth]{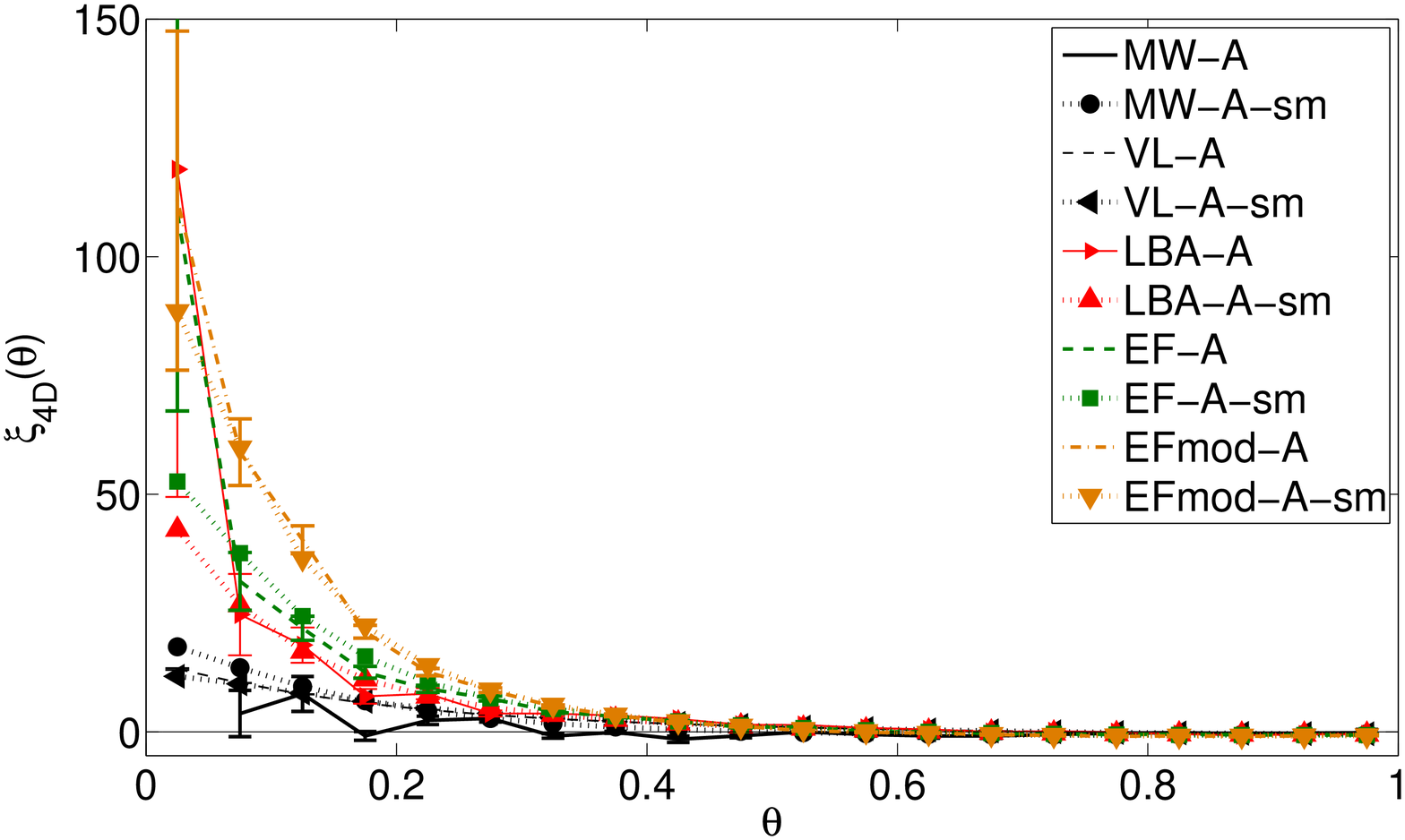} & \includegraphics[trim = 5mm 2.5mm 20mm 10mm, clip,width=0.49\textwidth]{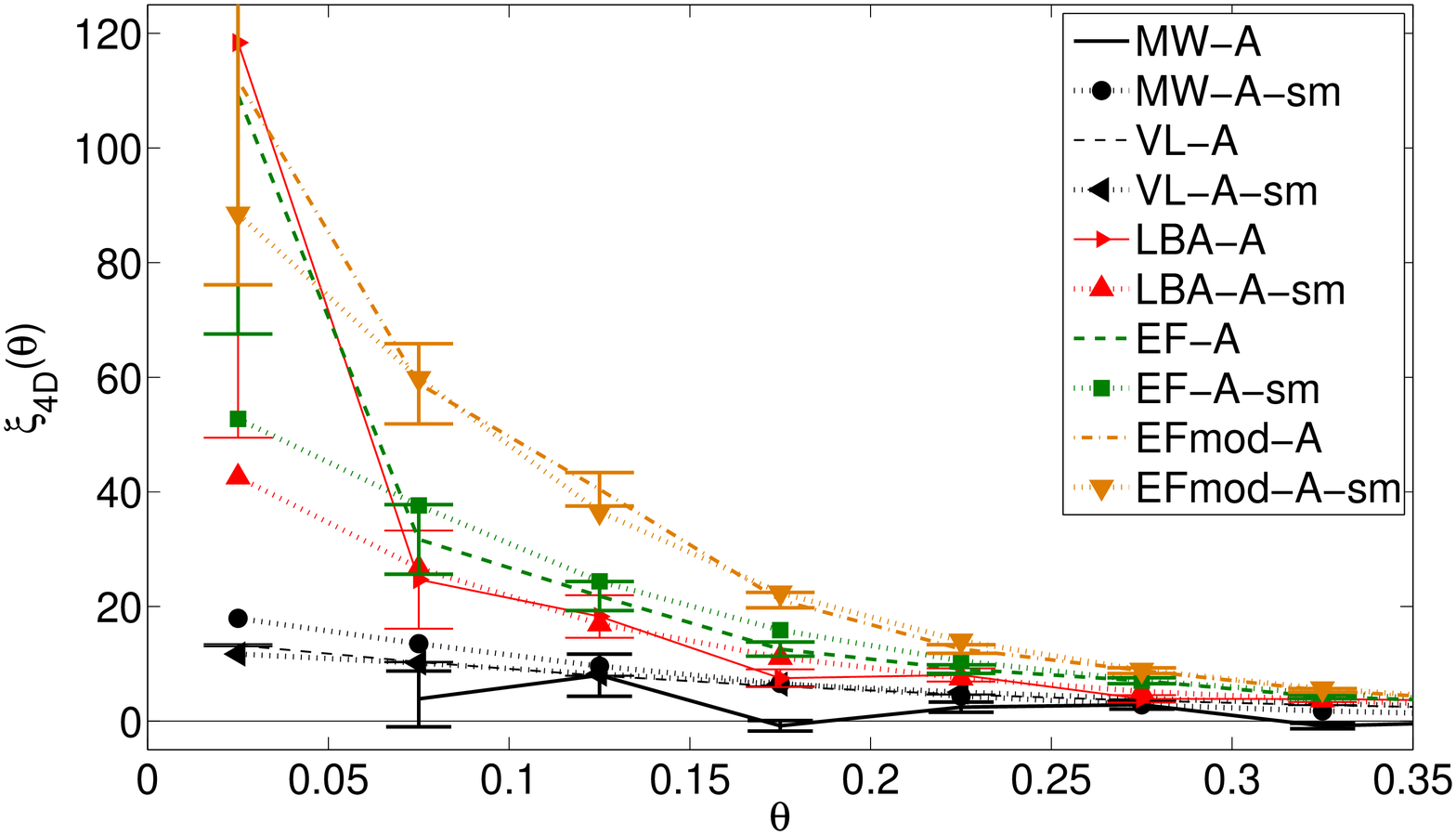} \\
\end{tabular}
\caption{\scriptsize The spatial (top row), line-of-sight velocity (center row), and 4D phase-space (bottom row) two-point correlation functions for the full-area sets: MW-A (solid black) VL-A (thin dashed back), LBA-A (thin solid red, right triangle), EF-A (thick dashed green), and EFmod-A (dot-dashed orange). The mock sets that represent the clustering signal due to the smooth underling density profile and line-of-sight velocity distribution of the corresponding data sets: MW-A-sm (dotted black, circle) VL-A-sm (dotted black, left triangle), LBA-A-sm (dotted red, up triangle), EF-A (dotted green, square), and EFmod-A-sm (dotted orange, down triangle) are also shown. The ``luminous" subhalo sets are more strongly clustered than the Milky Way satellite galaxies in all three spaces and over a broader range of scales in four-dimensional phase-space. The right panel in each row is a ``zoomed in" view of the left panel. The central right panel shows the scales where the ``luminous" subhalo sets are clustered in line-of-sight velocity space as a result of substructure, while the Milky Way satellites are randomly distributed. The error bars for the mock sets are not shown since they are smaller than each of the points in these panels. The black horizontal line in all panels shows the value for the two-point correlation function that corresponds to a uniform random distribution.The first bin of the MW-A spatial and 4D phase-space two-point correlation functions have no data-data pairs and are not shown.}
\label{fig:main_pts}
\end{center}
\end{figure*}

All three of the ``luminous" subhalo sets are more strongly clustered than the full-area dark subhalo set for scales smaller than $112.5$~kpc, where the spatial two-point correlation functions of all four sets (shown together in the top panels of Fig.~\ref{fig:main_pts}) have an amplitude greater than one. This stronger small-scale spatial clustering of the ``luminous" subhalo sets compared to the dark VL-A set is mainly due to the fact that the ``luminous" subhalo sets are more centrally concentrated, demonstrated by the relative amplitudes of the two-point correlation functions of the corresponding mock ``subhalo" sets, shown in Fig.~\ref{fig:main_pts}. The radial density distributions of all the VL-II subhalo sets of Fig.~\ref{fig:Radial_main} show the stronger central concentrations of the ``luminous" subhalo set compared to the dark subhalo set. 

The spatial two-point correlation functions of the ``luminous" subhalo sets that are constrained to lie within the SDSS DR7 Northern Contiguous footprint region are shown in the top right panels of Figs. \ref{fig:LBA results} - \ref{fig:KOP results}. The dashed and dot-dashed curves respectively show the footprint set results for the two ``observer" locations. The LBA sets have several scales where the spatial clustering signal is more significant than $2.6\sigma$. The first seven bins of the EF-B and EFmod-B spatial two-point correlation functions and the first eight bins of the EF-Brot and EFmod-Brot spatial two-point correlation functions (with the exception of bin six for the EFmod-Brot set) indicate, at a significance level above $2.7\sigma$, that these sets are spatially clustered. There is no evidence of spatial substructure in any of the ``luminous" subhalo footprint sets' two-point correlation functions at a significance level above $2.6\sigma$, with the exception of the second bin of the EFmod-Brot set at $2.7\sigma$ significance. The first seven bins of the spatial two-point correlation functions of the dark subhalo footprint sets, VL-B and VL-Brot, shown in the top right panel of the Fig.~\ref{fig:VL results}, indicate, at greater than $6.9\sigma$ significance, the presence of substructure in the spatial distribution of these two sets on those scales.

\subsection{Via Lactea II Subhalo Line-of-Sight Velocity Space Clustering Results}

\begin{figure*}
\begin{center}
\begin{tabular}{lll}
\includegraphics[trim = 10mm 2.5mm 20mm 10mm, clip,width=0.49\textwidth]{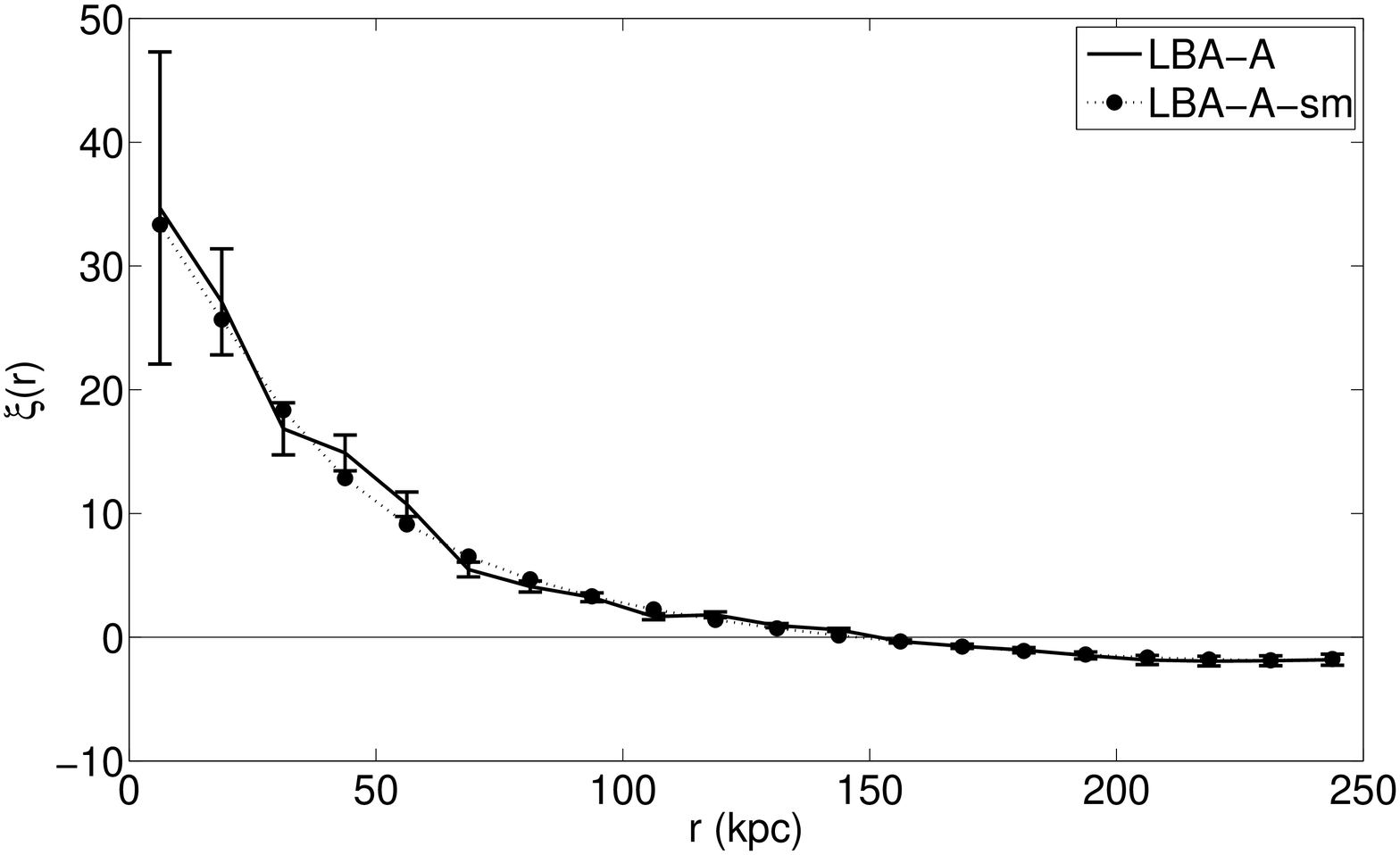} &\includegraphics[trim = 10mm 2.5mm 20mm 10mm, clip,width=0.49\textwidth]{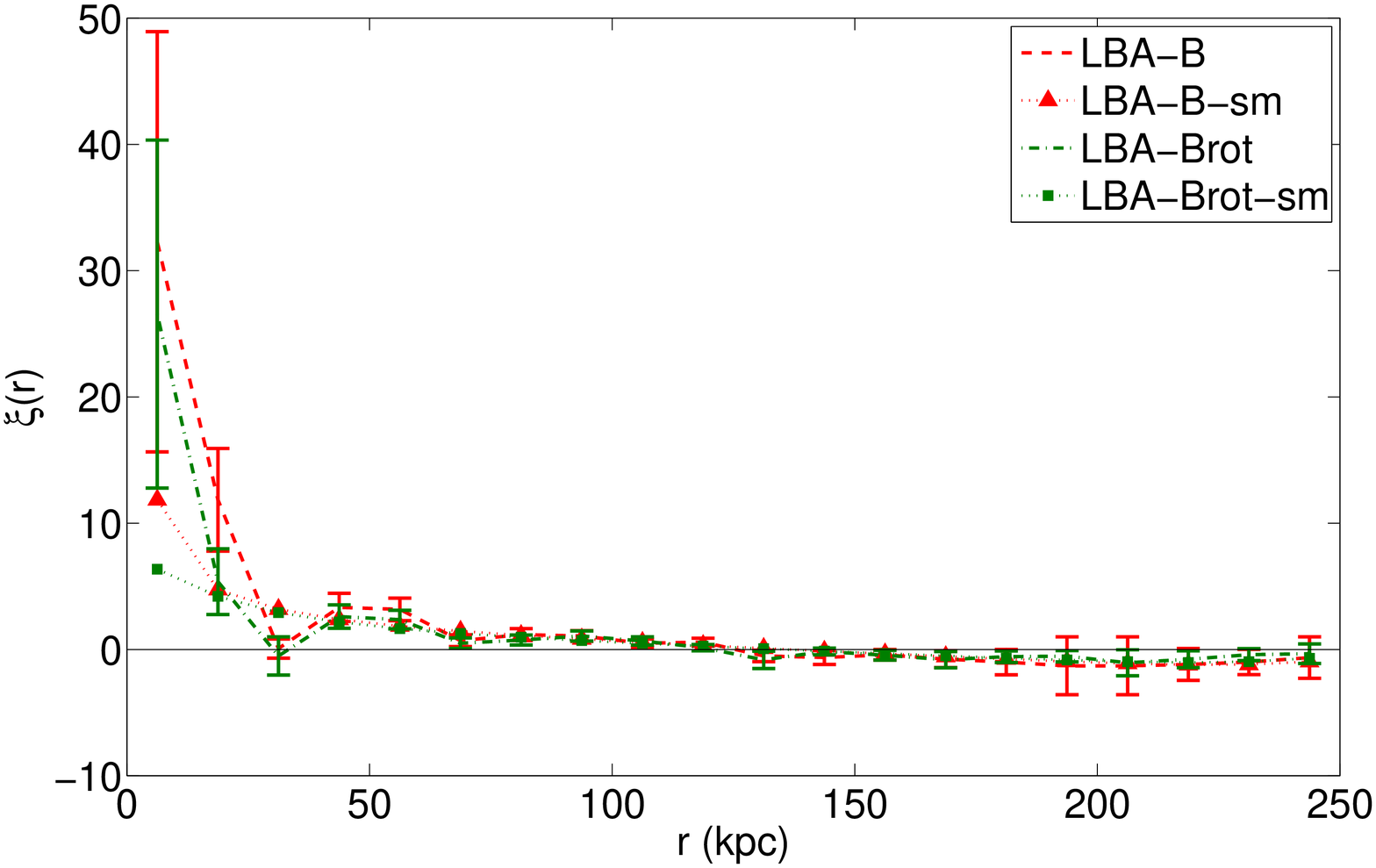} \\
\includegraphics[trim = 5mm 2.5mm 20mm 10mm, clip,width=0.49\textwidth]{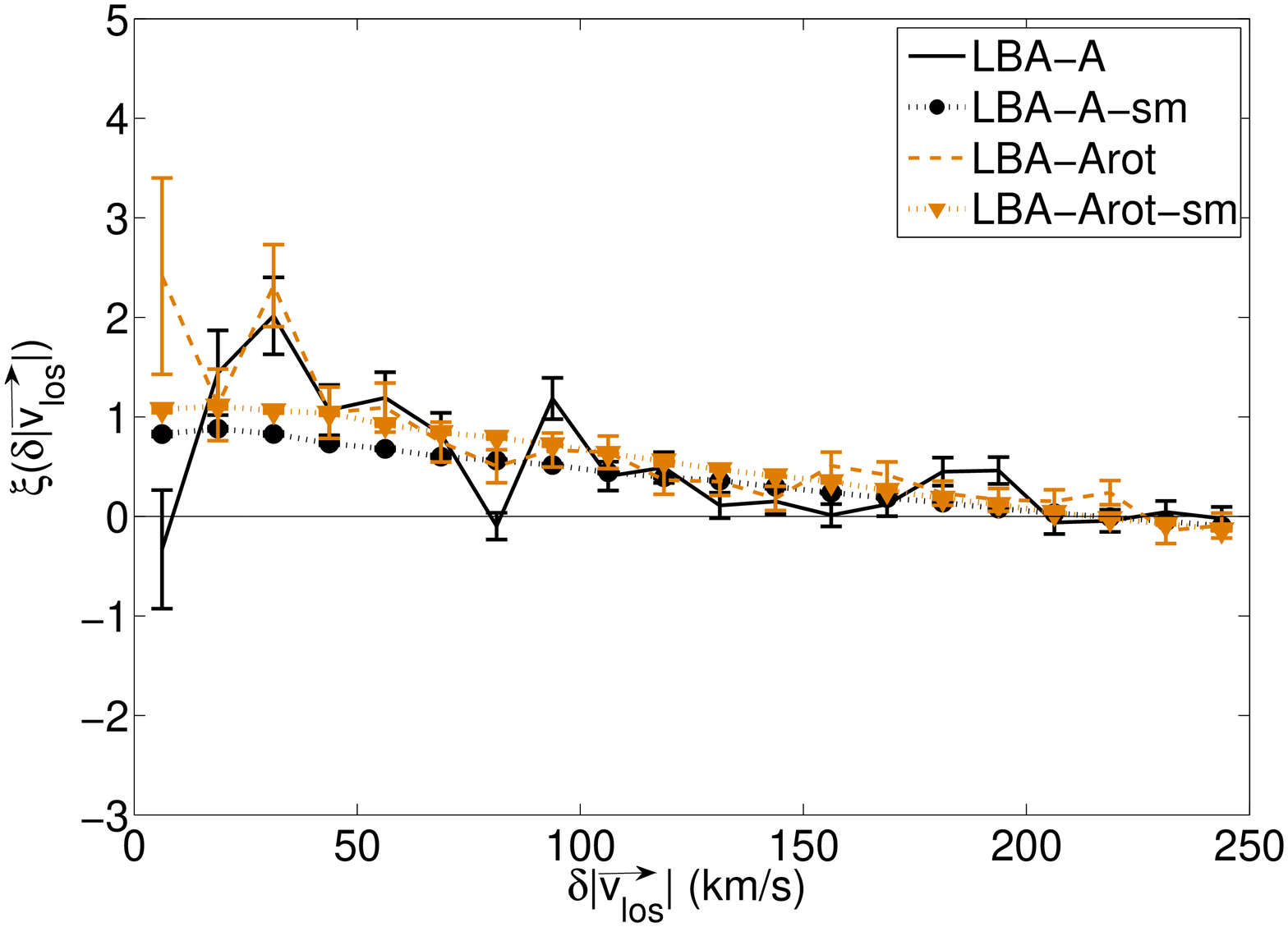} & \includegraphics[trim = 5mm 3mm 20mm 10mm, clip,width=0.48\textwidth]{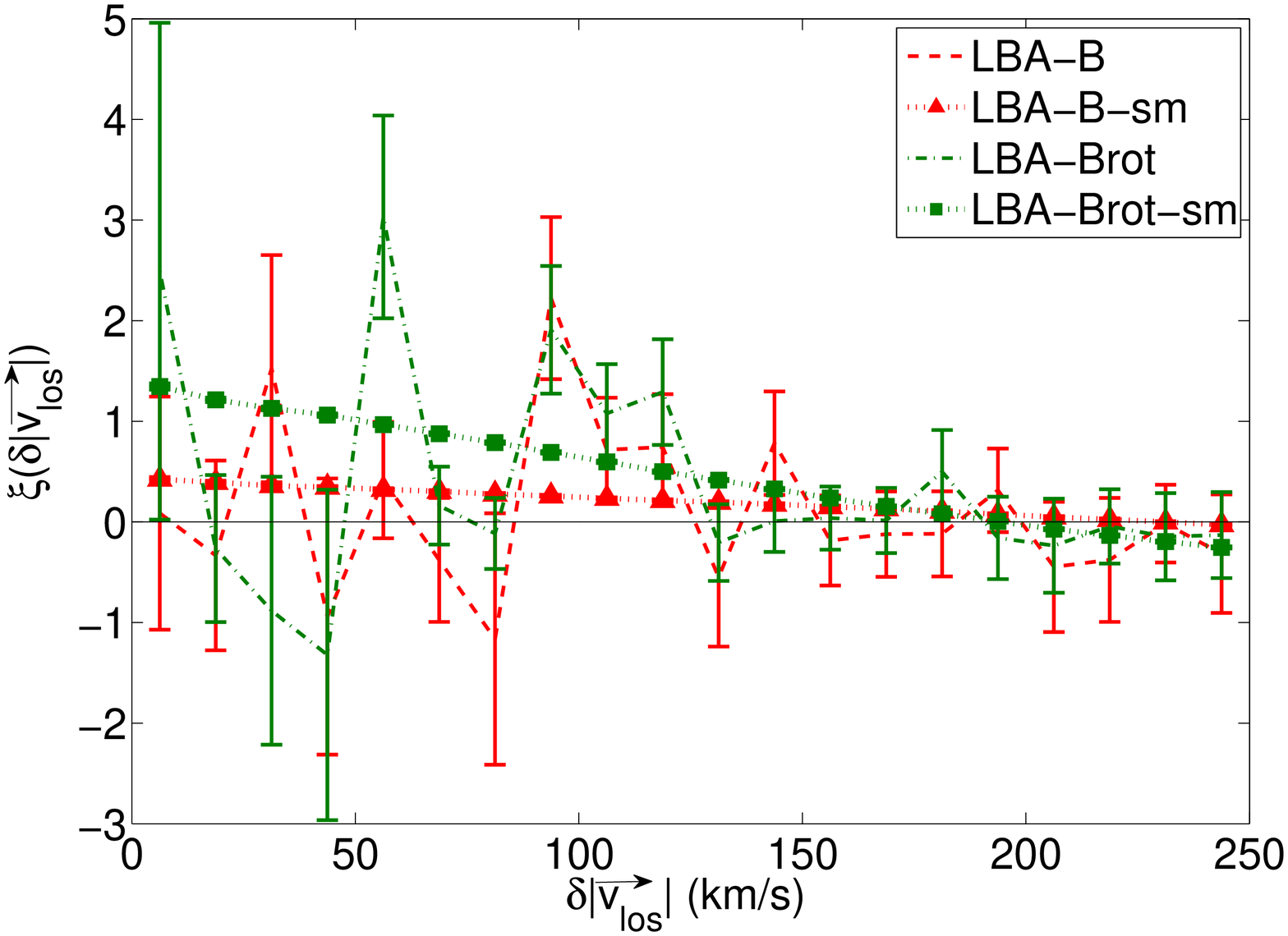} \\
\includegraphics[trim = 5mm 10mm 20mm 10mm, clip,width=0.49\textwidth]{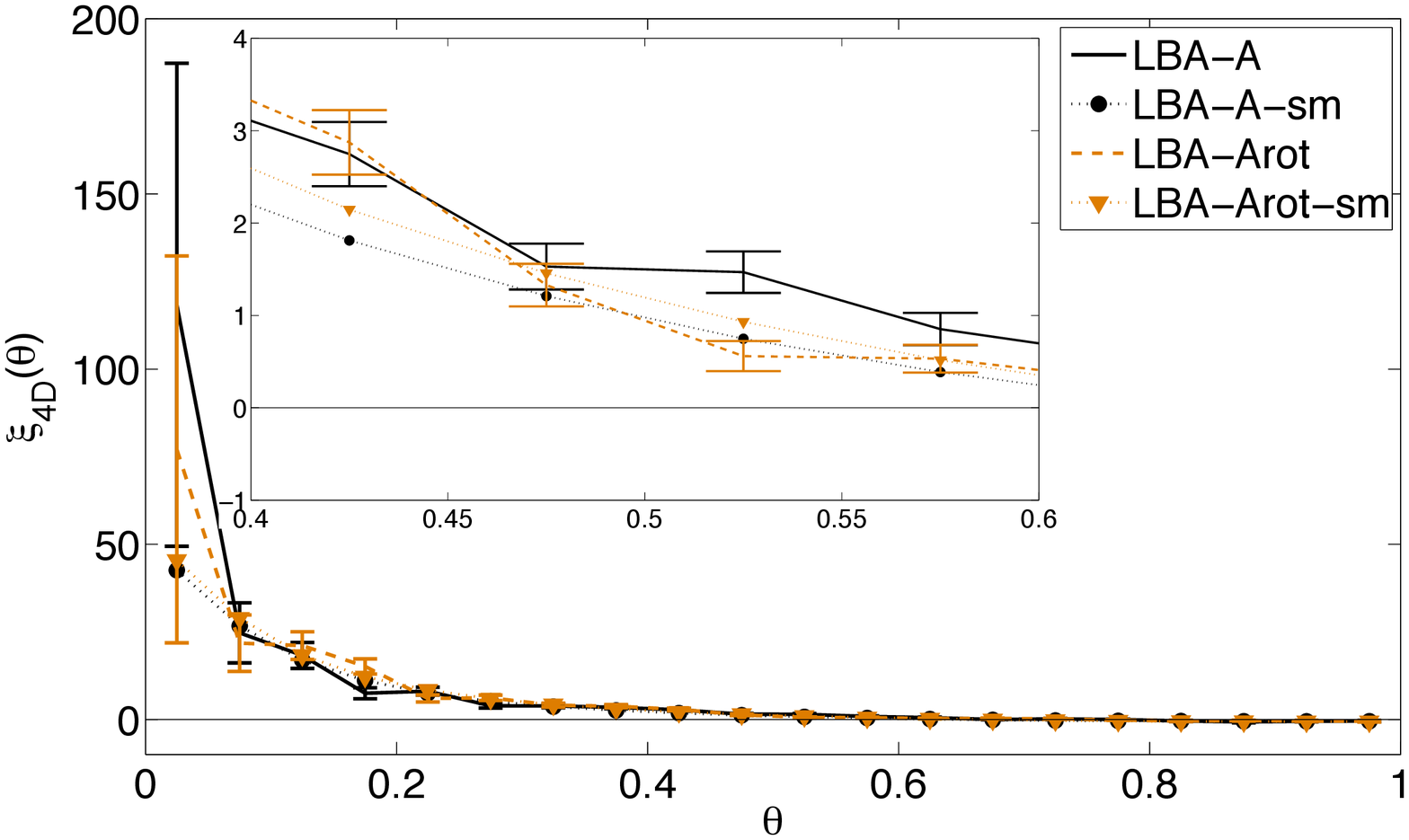} & \includegraphics[trim = 10mm 8mm 20mm 13mm, clip,width=0.47\textwidth]{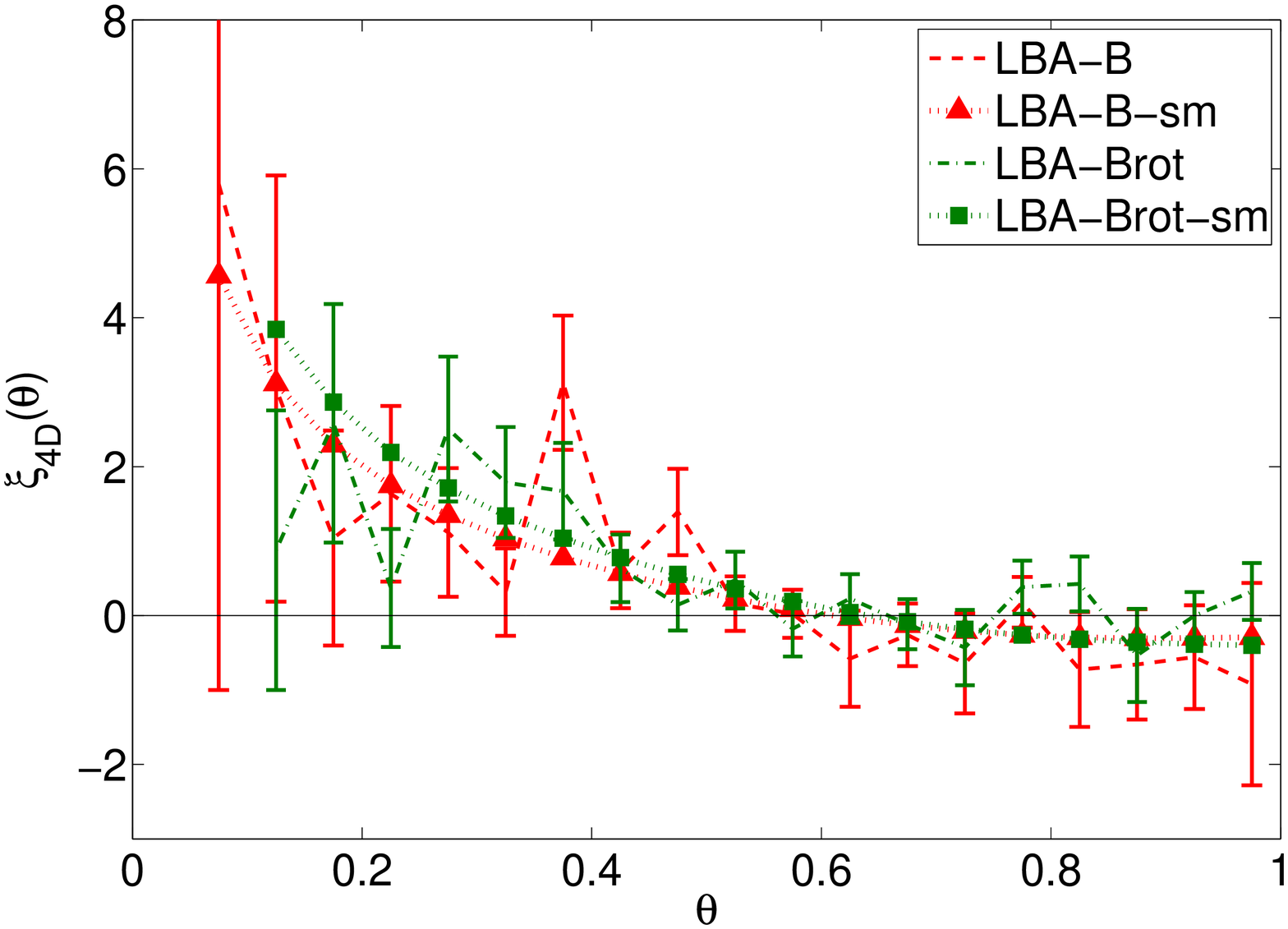} \\
\end{tabular}
\caption{\scriptsize The spatial (top row), line-of-sight velocity (center row), and the 4D phase-space (bottom row) two-point correlation functions for the Largest Before Accretion model (LBA) sets. Left panels: LBA-A (solid black), LBA-Arot (dashed orange). Right Panels: LBA-B (dashed red), LBA-Brot (dashed-dotted green). The mock sets of ``subhaloes" that represent the clustering signal due to the smooth underling density profile and line-of-sight velocity distribution of the corresponding data sets:  LBA-A-sm (dotted black, circle), LBA-Arot-sm (dotted orange, down triangle), LBA-B-sm (dotted red, up triangle), LBA-Brot-sm (dotted green, square). The LBA-A and LBA-Arot sets each contain substructure in their line-of-sight velocity space distribution. Evidence of substructure is also found in the 4D phase-space two-point correlation functions of the LBA-A and LBA-B set. The inset in the bottom left panel shows a ``zoomed in" region of the LBA-A 4D phase-space two-point correlation function where the clustering signal is in excess of the two-point correlation function of the LBA-A-sm set. The LBA-Arot spatial two-point correlation functions are not shown in the top panels as they are nearly identical to the LBA-A set. The error bars for the mock ``subhalo" sets are not shown in the panels of the top and bottom rows since they are smaller than each of the points in these panels. The black horizontal line in all panels shows the value for the two-point correlation function that corresponds to a uniform random distribution. The first bin of the LBA-B and the first two bins of the LBA-Brot 4D phase-space two-point correlation functions have no data-data pairs and are, therefore, not shown.}
\label{fig:LBA results}
\end{center}
\end{figure*}

\begin{figure*}
\begin{center}
\begin{tabular}{lll}
\includegraphics[trim = 10mm 2.5mm 20mm 10mm, clip,width=0.49\textwidth]{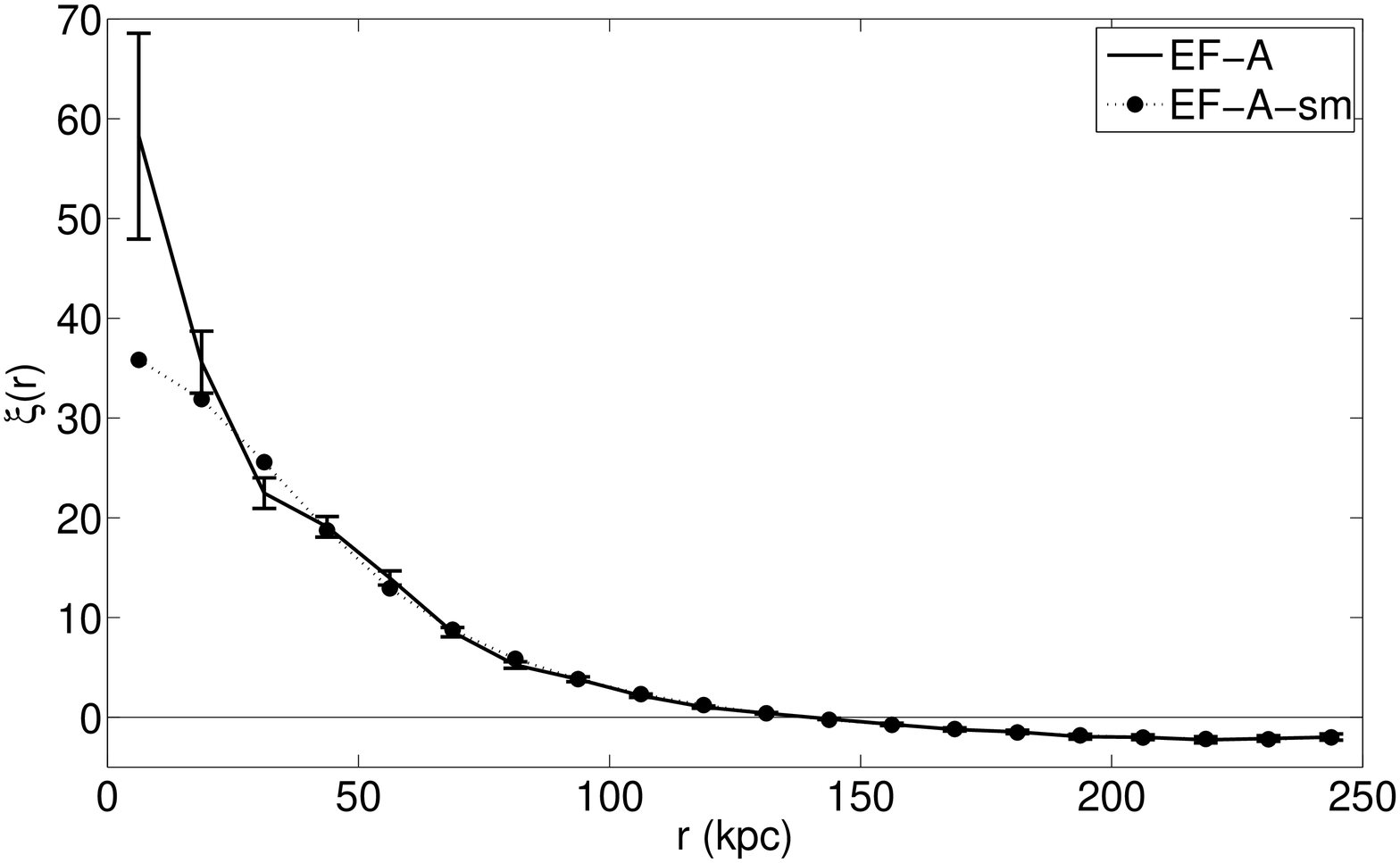} &\includegraphics[trim = 10mm 2.5mm 20mm 10mm, clip,width=0.49\textwidth]{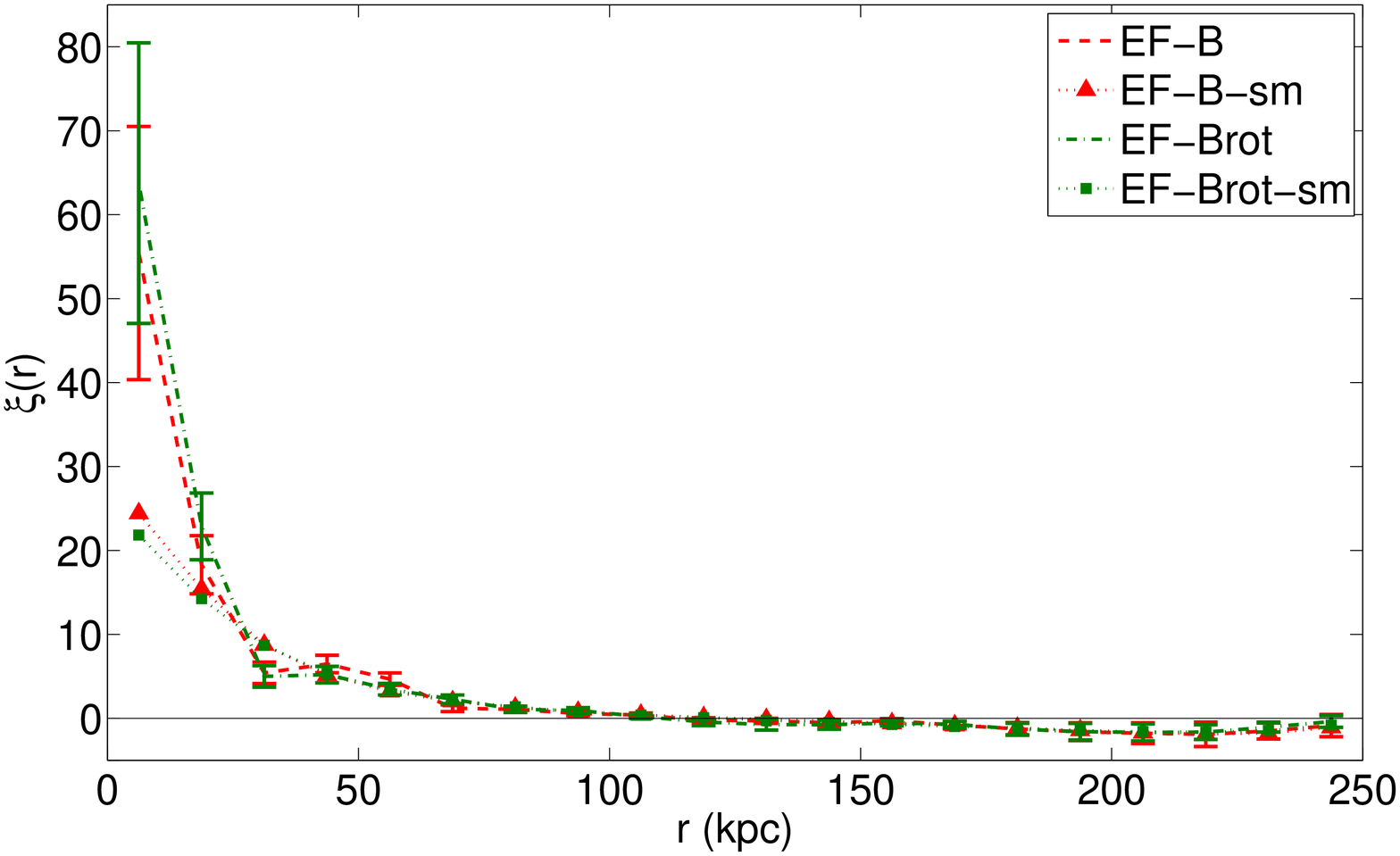} \\
\includegraphics[trim = 5mm 2.5mm 20mm 10mm, clip,width=0.49\textwidth]{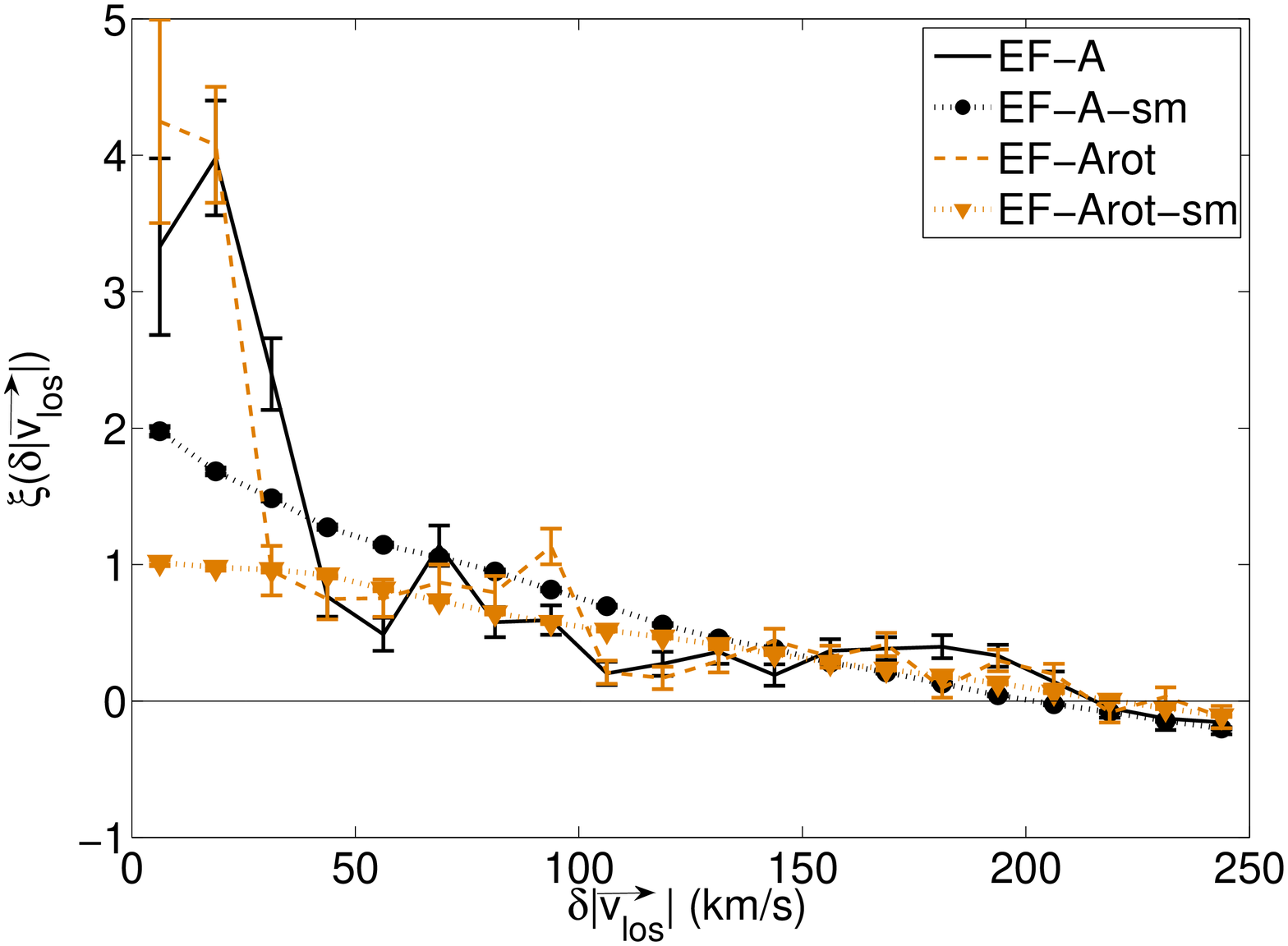} & \includegraphics[trim = 5mm 2.5mm 20mm 5mm, clip,width=0.49\textwidth]{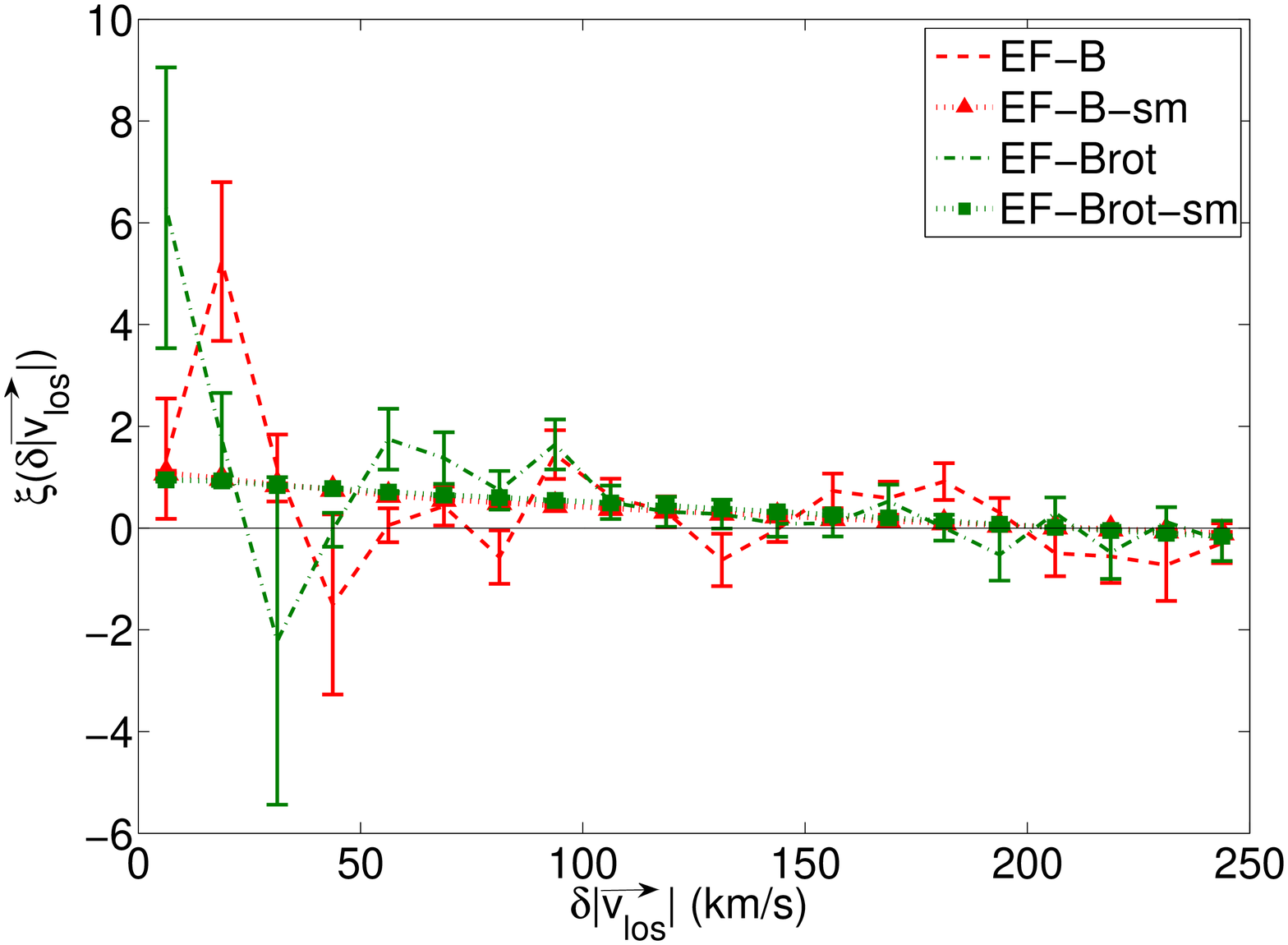} \\
\includegraphics[trim = 5mm 2.5mm 20mm 10mm, clip,width=0.49\textwidth]{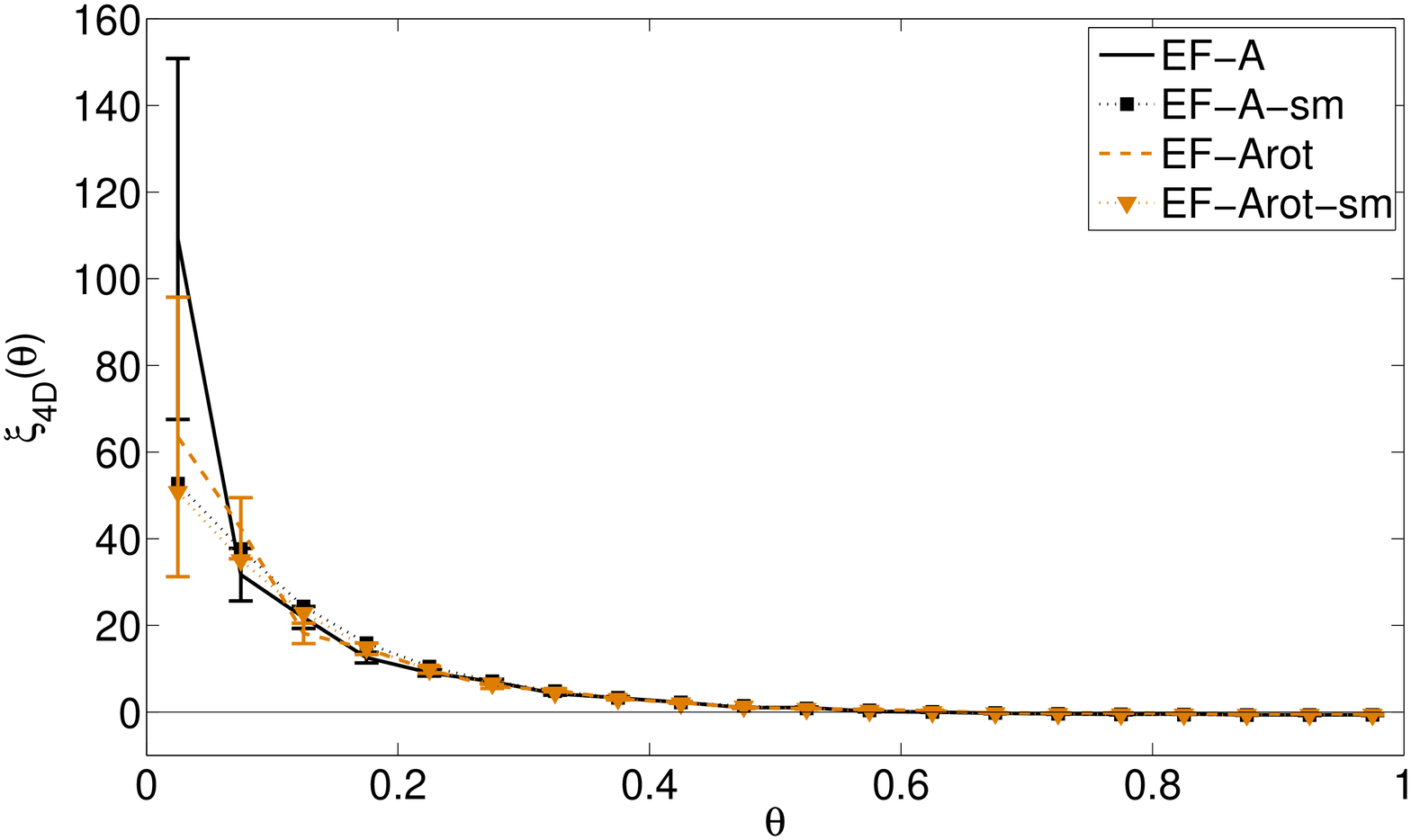} & \includegraphics[trim = 5mm 2.5mm 20mm 10mm, clip,width=0.49\textwidth]{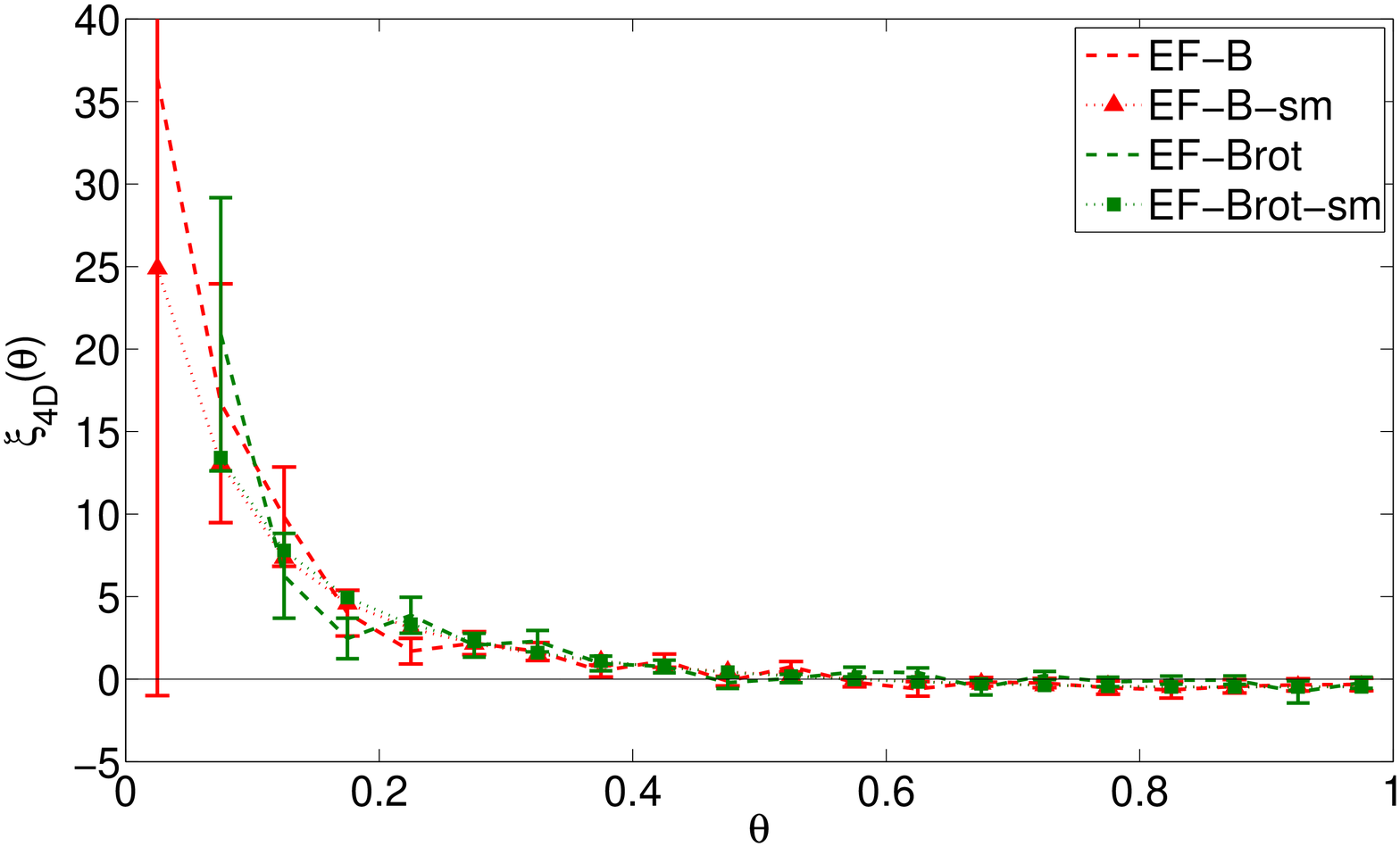} \\
\end{tabular}
\caption{\scriptsize The spatial (top row), line-of-sight velocity (center row), and the 4D phase-space (bottom row) two-point correlation functions for the Early Formation model (EF) sets. Left Panels: EF-A (solid black), EF-Arot (dashed orange). Right Panels: EF-B (dashed red), EF-Brot (dashed-dotted green). The mock sets of ``subhaloes" that represent the clustering signal due to the smooth underling density profile and line-of-sight velocity distribution of the corresponding data sets: EF-A-sm (dotted black, circle), EF-Arot-sm (dotted orange, down triangle), EF-B-sm (dotted red, up triangle), EF-Brot-sm (dotted green, square). Each of the EF-A, EF-Arot, and the EF-B sets contain substructure in their respective line-of-sight velocity distributions. The EF-Arot spatial two-point correlation functions are not shown in the top panels as they are nearly identical to the EF-A set. The error bars for the mock ``subhalo" sets are not shown in the panels of the top and bottom rows since they are smaller than each of the points in these panels.  The black horizontal line in all panels shows the value for the two-point correlation function that corresponds to a uniform random distribution. The first bin of the EF-Brot 4D phase-space two-point correlation function has no data-data pairs and is, therefore, not shown.}
\label{fig:EF results}
\end{center}
\end{figure*}

\begin{figure*}
\begin{center}
\begin{tabular}{lll}
\includegraphics[trim = 10mm 2.5mm 20mm 10mm, clip,width=0.49\textwidth]{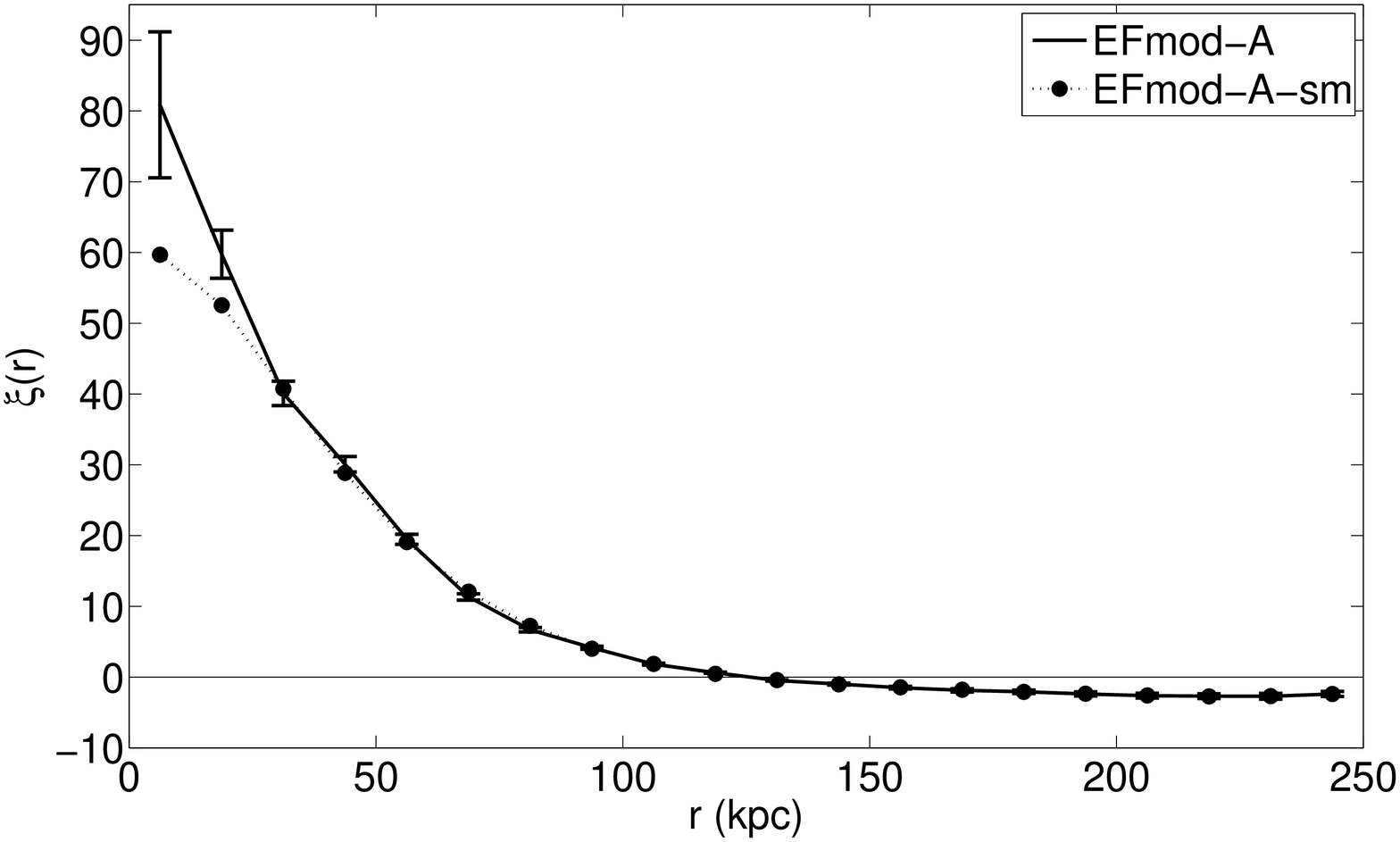} &\includegraphics[trim = 10mm 2.5mm 20mm 10mm, clip,width=0.49\textwidth]{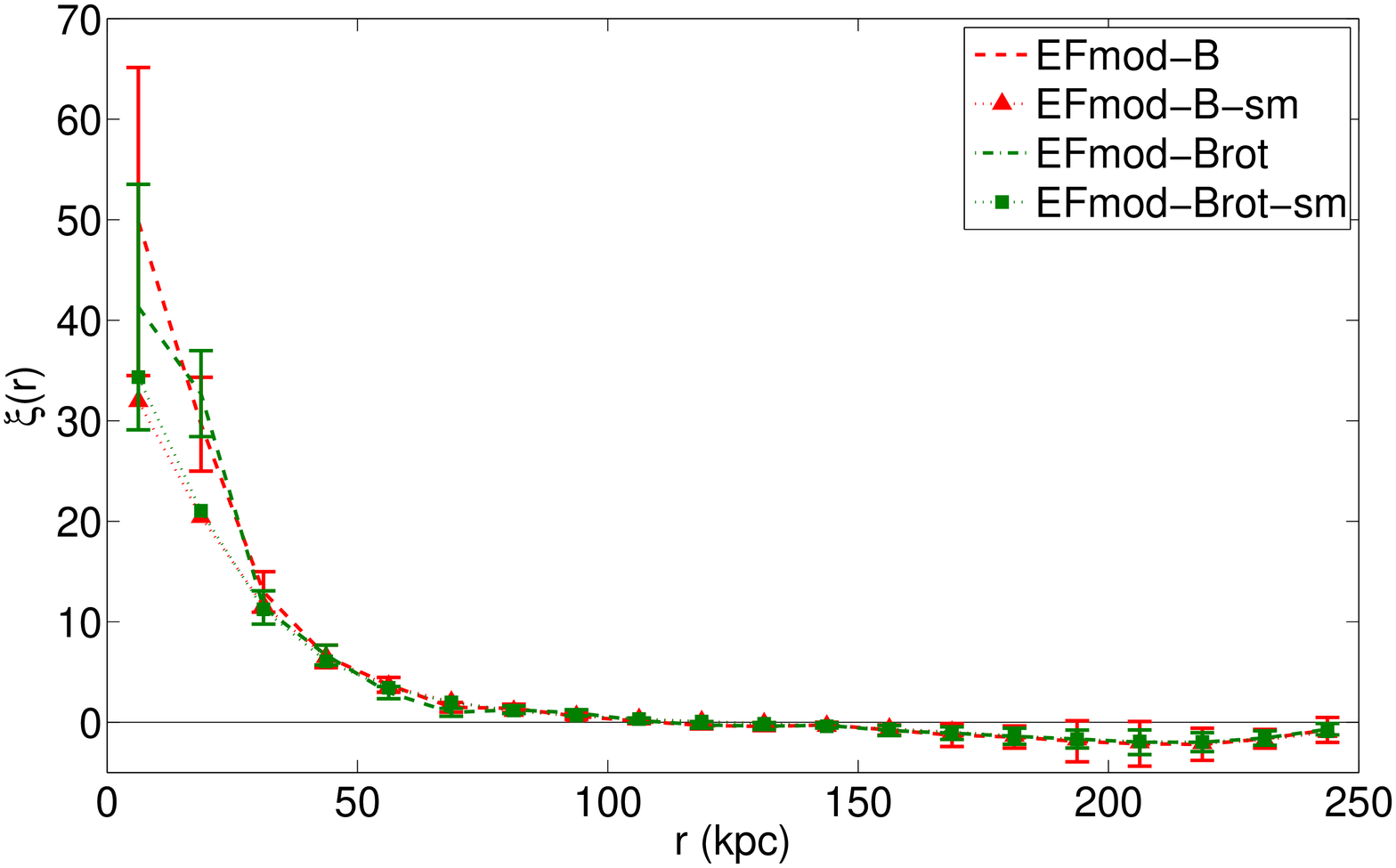} \\
\includegraphics[trim = 5mm 2.5mm 20mm 10mm, clip,width=0.49\textwidth]{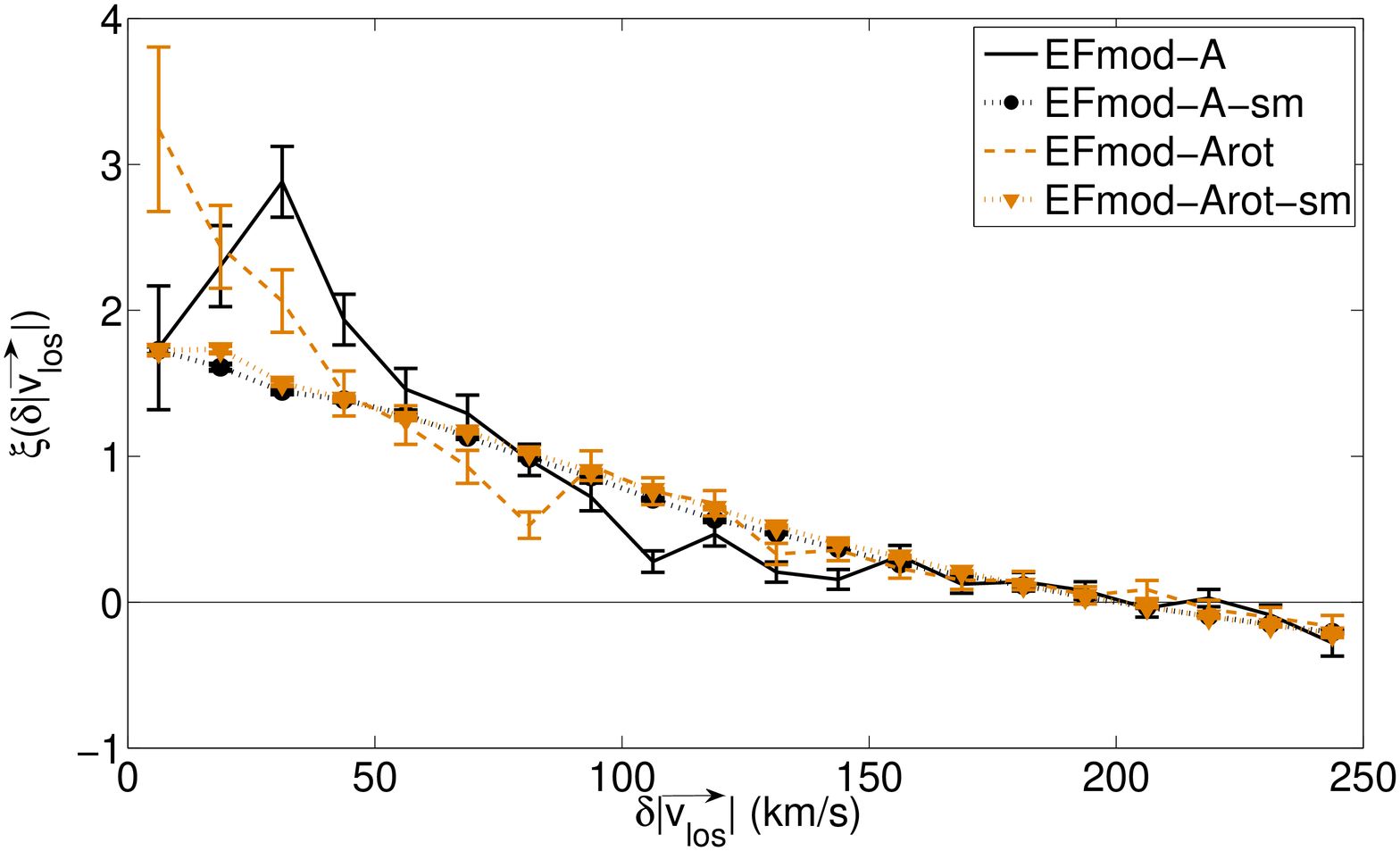} & \includegraphics[trim = 5mm 2.5mm 20mm 10mm, clip,width=0.49\textwidth]{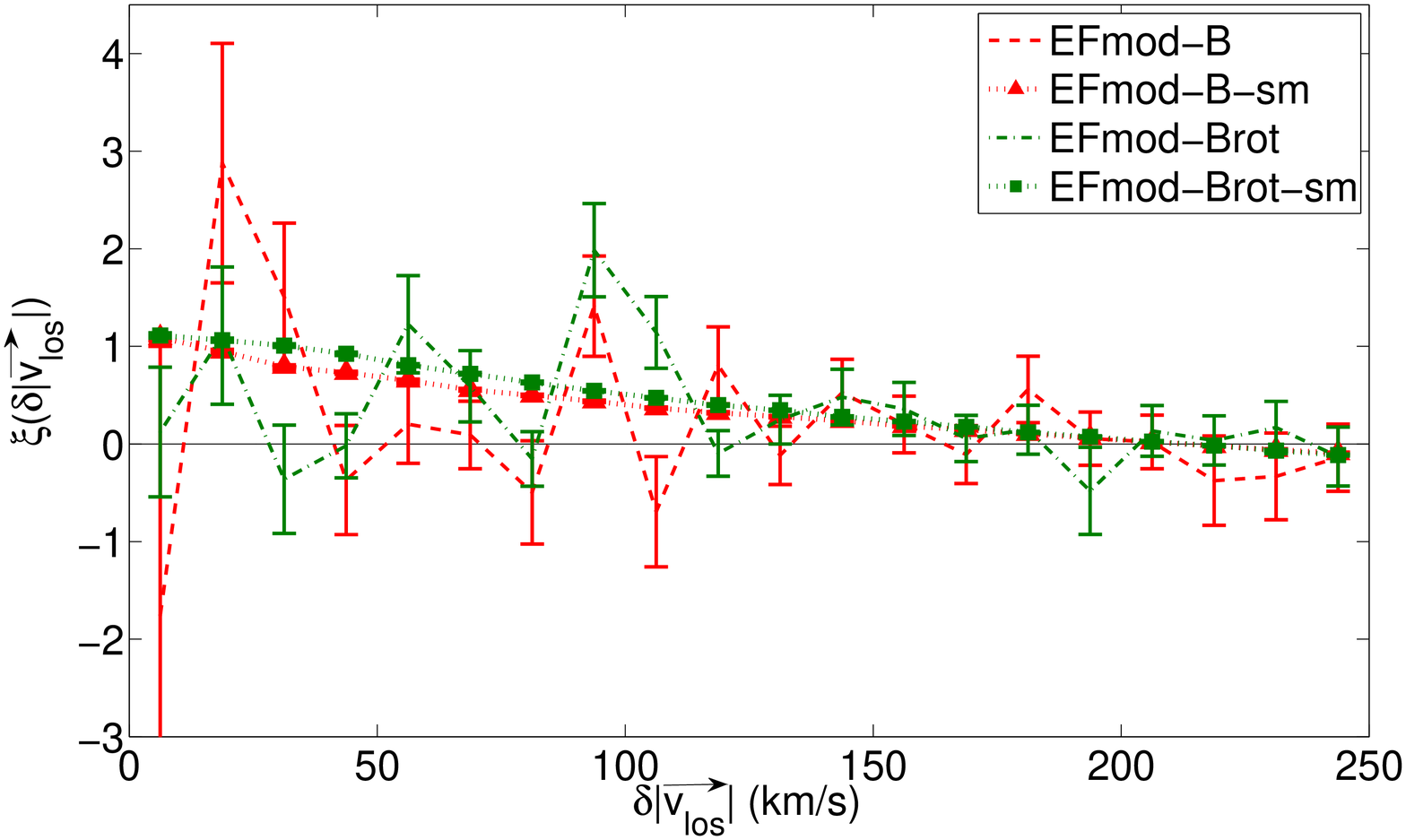} \\
\includegraphics[trim = 5mm 2.5mm 20mm 10mm, clip,width=0.51\textwidth]{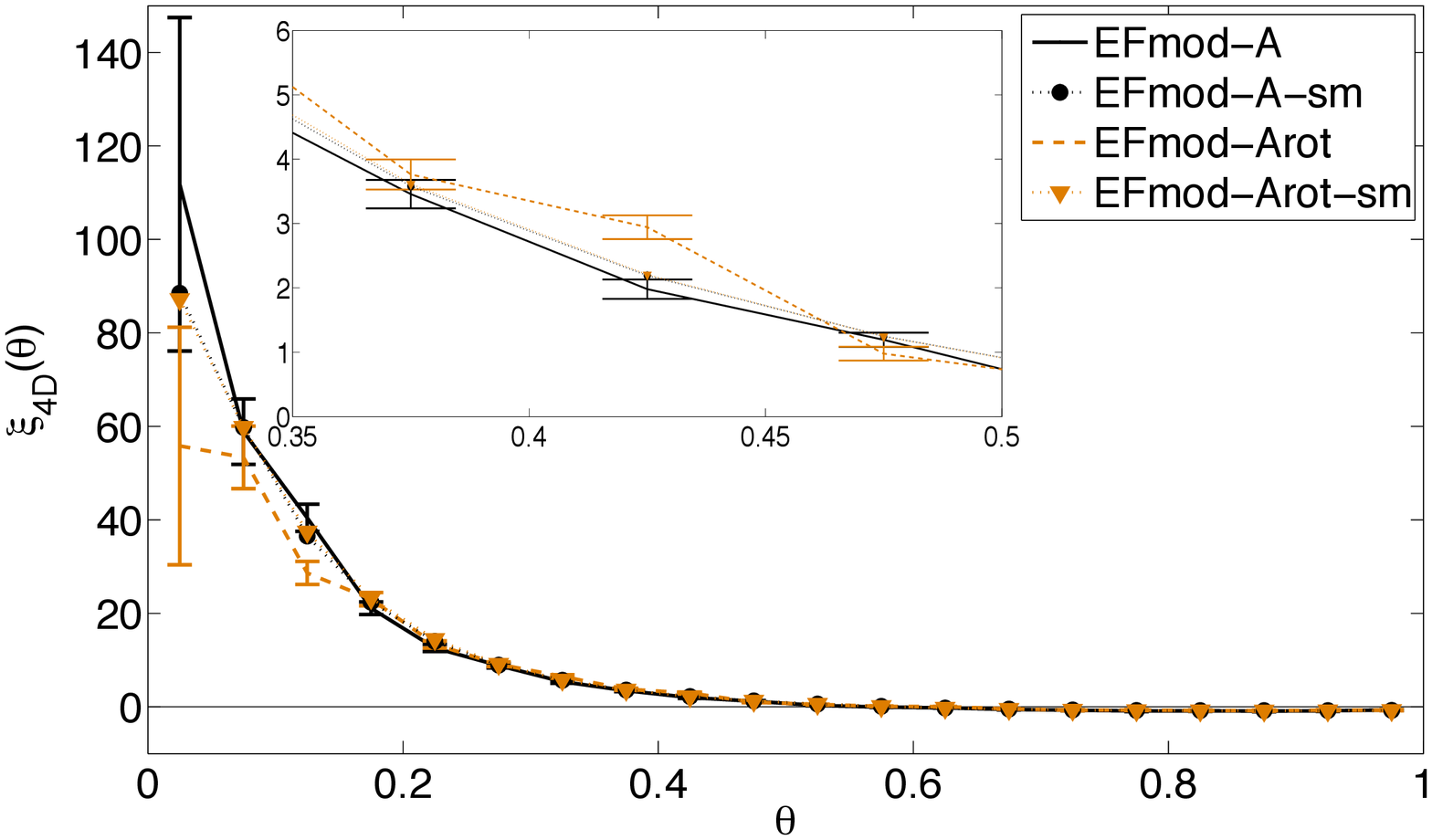} & \includegraphics[trim = 10mm 2.5mm 20mm 10mm, clip,width=0.51\textwidth]{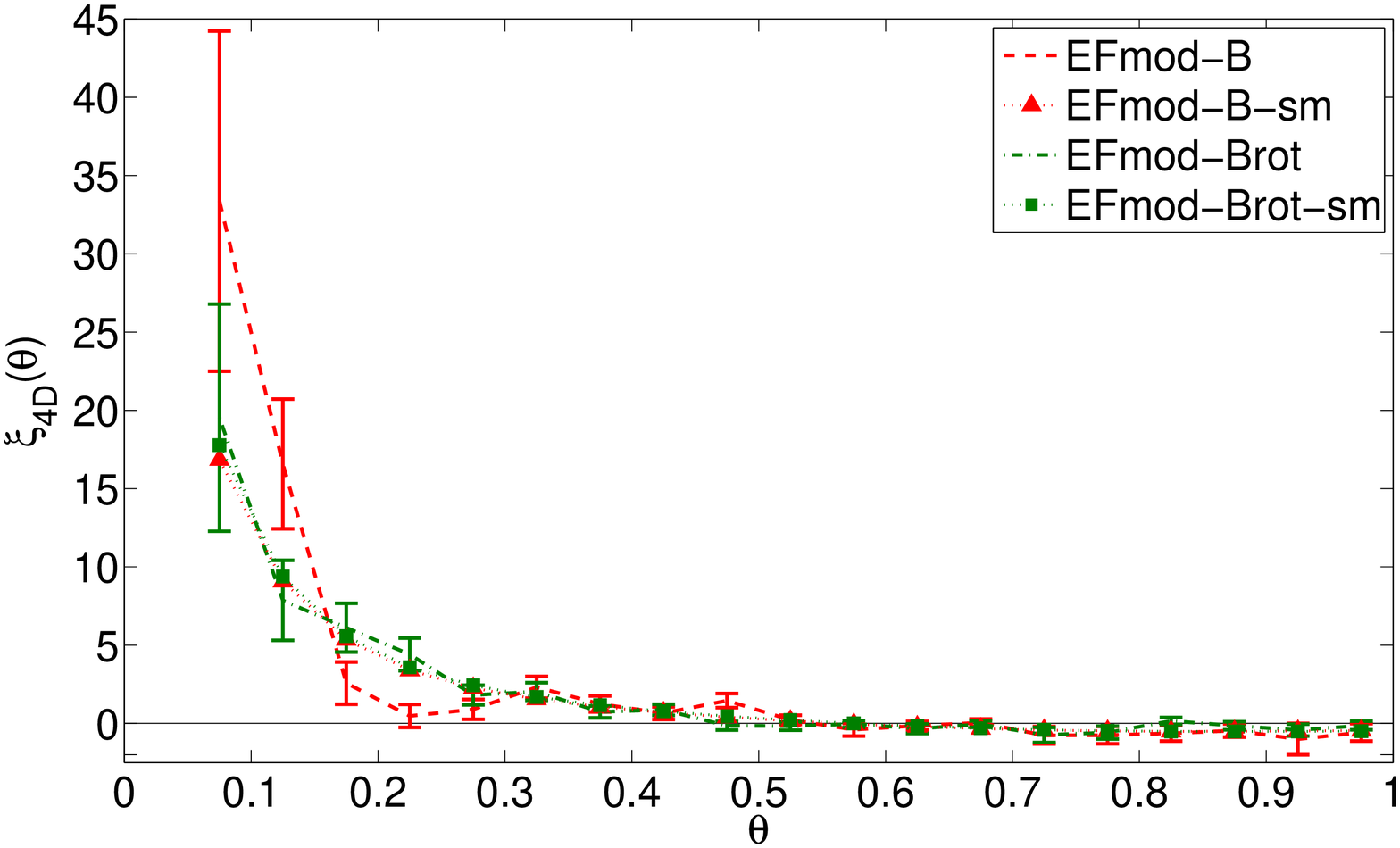} \\
\end{tabular}
\caption{\scriptsize The spatial (top row), line-of-sight velocity (center row), and the 4D phase-space (bottom row) two-point correlation function results for the Modified Early Formation model (EFmod) sets. Left Panels: EFmod-A (solid black), EFmod-Arot (dashed orange), EFmod-B (dashed red). Right Panels: EFmod-Brot (dashed-dotted green). The mock sets of ``subhaloes" that represent the clustering signal due to the smooth underling density profile and line-of-sight velocity distribution of the corresponding data sets: EFmod-A-sm (dotted black, circle), EFmod-Arot-sm (dotted orange, down triangle), EFmod-B-sm (dotted red, up traingle), EFmod-Brot-sm (dotted green, square). The EFmod-A and EFmod-Brot sets each contain substructure in their line-of-sight velocity distributions. The EFmod-Brot set also contains substructure in its configuration space distribution. The 4D phase-space distribution of the EFmod-Arot set contains substructure on one scale, as shown in the ``zoomed in" region in the inset of the bottom left panel. The EFmod-Arot spatial two-point correlation functions are not shown in the top left panel as they are nearly identical to the EFmod-A set. The error bars for the mock ``subhalo" sets are not shown in the panels of the top and bottom rows since they are smaller than each of the points in these panels. The black horizontal line in all panels shows the value for the two-point correlation function that corresponds to a uniform random distribution. The first bin of the EFmod-B and the EFmod-Brot 4D phase-space two-point correlation functions have no data-data pairs and are, therefore, not shown.}
\label{fig:KOP results}
\end{center}
\end{figure*}

The line-of-sight velocity two-point correlation functions of the Via Lactea subhalo sets are shown in the central panels of Figs. \ref{fig:LBA results} - \ref{fig:VL results}. The solid black curve and the dashed orange curve in the central left panel of each figure respectively represent the two-point correlation function for the full-area sets for each ``observer" location. There is a notable variation between the two-point correlation functions of each ``observer" location for each of the four full-area subhalo sets, but both sets follow a qualitatively similar trend. The two-point correlation functions of the mock ``subhalo" sets, that represent the clustering signal contribution due to each data set's smooth line-of-sight velocity distribution, is depicted by the dotted curves in each figure. The scales where substructure is detected, i.e. scales on which the clustering signal of each set exceeds the clustering signal due to the smooth line-of-sight velocity distribution at a significance of at least $2.6\sigma$, is listed in Table \ref{tab:Cluster_nums} for each set. 

For each of the ``luminous" subhalo sets, the two-point correlation function indicates the set is clustered in line-of-sight velocity space at a minimum significance of $2.6\sigma$, for a broad range of scales (and greater than $3\sigma$ significance for all but a few of these scales). This clustering signal, in at least one bin, for each of the ``luminous" subhalo models, is due to substructure, as shown in the central left panels of Figs. \ref{fig:LBA results} - \ref{fig:KOP results} by comparing the solid and dashed curves with the corresponding dotted curves. Evidence of substructure was detected in bin 3 of the two-point correlation function for both the LBA-A and LBA-Arot set respectively at $3\sigma$ and $2.9\sigma$ significance, and additionally in bin 8 for the LBA-A set at $3.1\sigma$ significance. No substructure was detected for the EFmod-Arot set, however there is evidence of substructure in the 3rd and 4th bin of the two-point correlation function of the EFmod-A set at $5.5\sigma$ and $2.9\sigma$ significance, respectively. The Early Formation Model sets, EF-A and EF-Arot, have the most significant detections of substructure amongst the ``luminous" subhalo sets. The clustering signal of the EF-A set indicates substructure on scales corresponding to bins 2 and 3 at a respective significance of $5.2\sigma$ and $3.2 \sigma$, while the EF-Arot two-point correlation function shows evidence of substructure in bins 1, 2, and 8 at a respective significance of $4.2\sigma$, $7\sigma$, and $4.1\sigma$. The line-of-sight velocity two-point correlation functions of the LBA-A and EF-A set each have at least one bin with a clustering signal that is in excess of the clustering signal of their respective line-of-sight velocity distribution by more than $2.6\sigma$, but with an amplitude that is too small ($\lesssim$ 0.5) to provide a clear detection of substructure.

The dark subhalo sets, VL-A and VL-Arot, are clustered on scales corresponding to the first 16 bins of each set's two-point correlation function at a significance of greater than $7\sigma$, as shown in the central left panel of Fig.~\ref{fig:VL results}. Evidence of substructure is detected in the first seven bins of the VL-A two-point correlation function at greater than $4\sigma$ significance, and in the first eight bins of the VL-Arot two-point correlation function at greater than $3\sigma$ significance. The first 16 bins of the line-of-sight velocity two-point correlation functions of the dark subhalo footprint sets, VL-B and VL-Brot (represented by the dashed green and red dot-dashed curves in the central right panel of the same figure), indicate each set is clustered with a significance greater than $4.2 \sigma$. Each of the dark subhalo footprint sets contain substructure within its line-of-sight velocity distribution, as indicated by the first eight bins of the VL-Brot two-point correlation function at greater than $2.8\sigma$ significance and bins 2-5 of the VL-B two-point correlation function at greater than $3.6\sigma$ significance.

The ``luminous" subhalo footprint sets show significant scatter in their line-of-sight velocity two-point correlation functions, represented by the dashed green and red dot-dashed curves in the central right panels of Figs. \ref{fig:LBA results} - \ref{fig:KOP results}, due to the smaller number of subhaloes in the footprint sets. Each set is clustered on some scale, with a line-of-sight velocity two-point correlation function amplitude greater than zero in at least one bin at a significance of $2.7\sigma$ or greater. The clustering signal in one bin of both the EFmod-Brot set and the EF-B set indicates the presence of substructure within each set's line-of-sight velocity distribution at a respective significance of $3\sigma$ and $2.7\sigma$.

\subsection{Via Lactea II Subhalo Four-Dimensional Phase-Space Clustering Results} 

\begin{figure*}
\begin{center}
\begin{tabular}{lll}
\includegraphics[trim = 10mm 2.5mm 20mm 10mm, clip,width=0.49\textwidth]{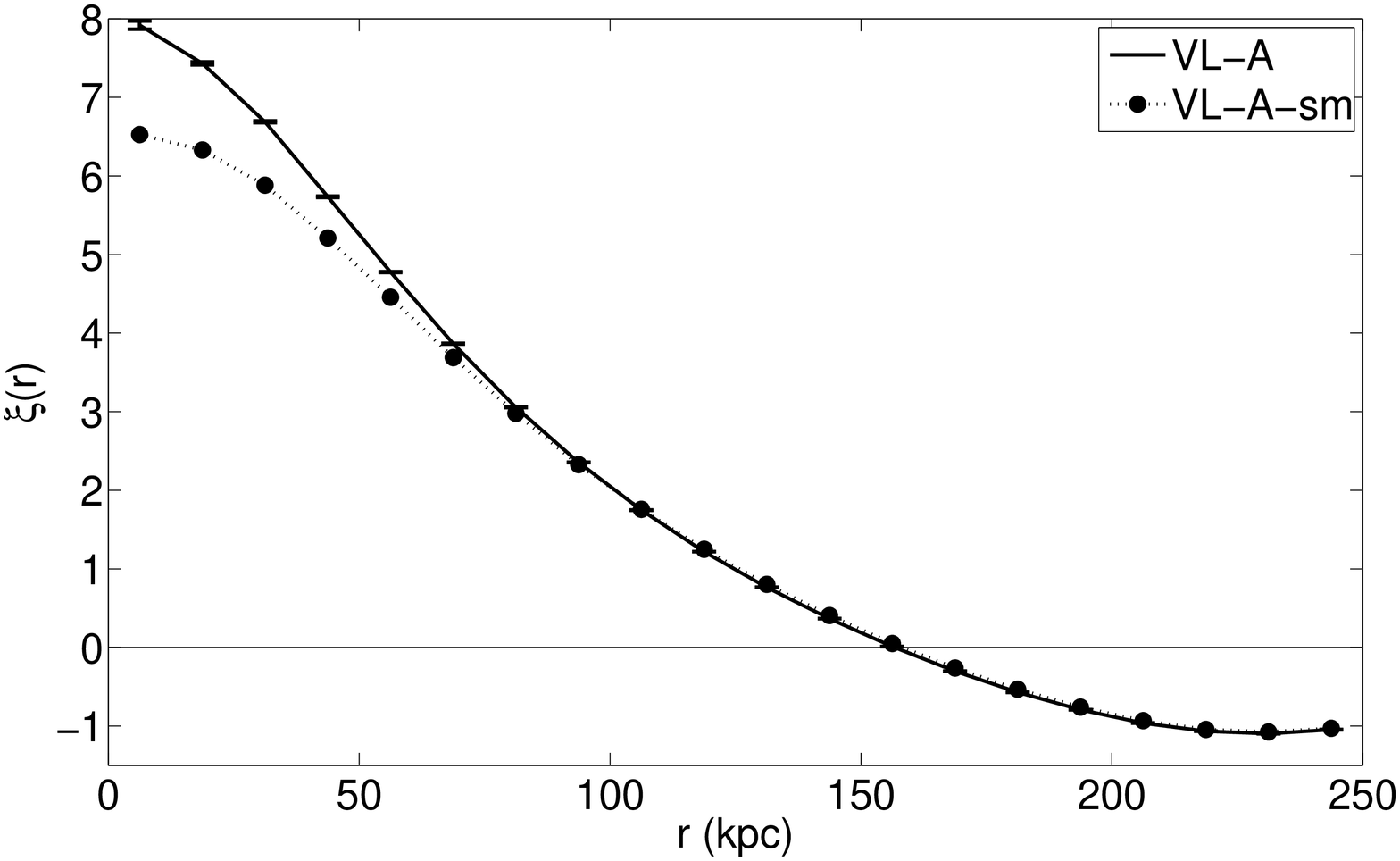} &\includegraphics[trim = 10mm 2.5mm 20mm 10mm, clip,width=0.49\textwidth]{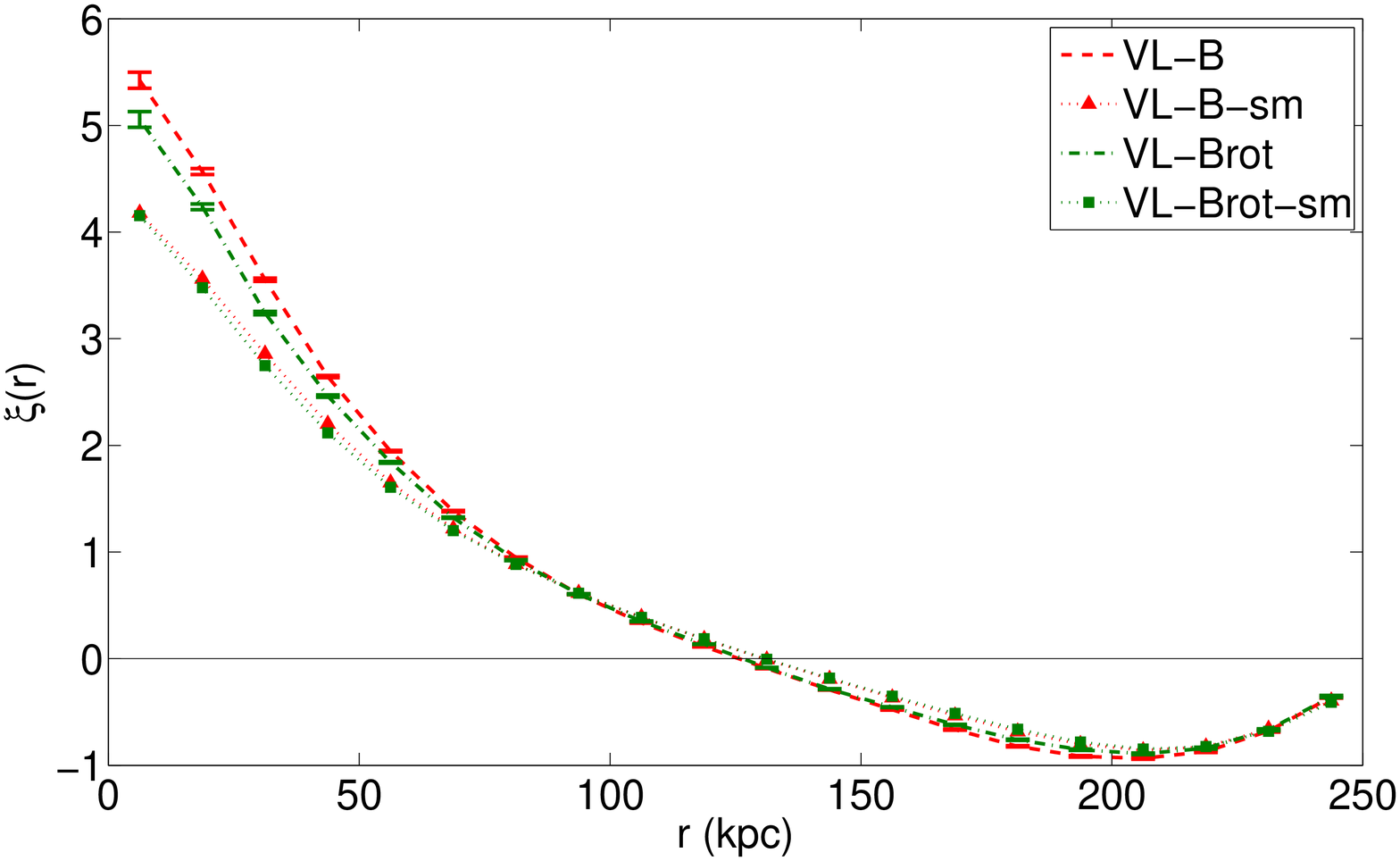} \\
\includegraphics[trim = 5mm 2.5mm 20mm 10mm, clip,width=0.49\textwidth]{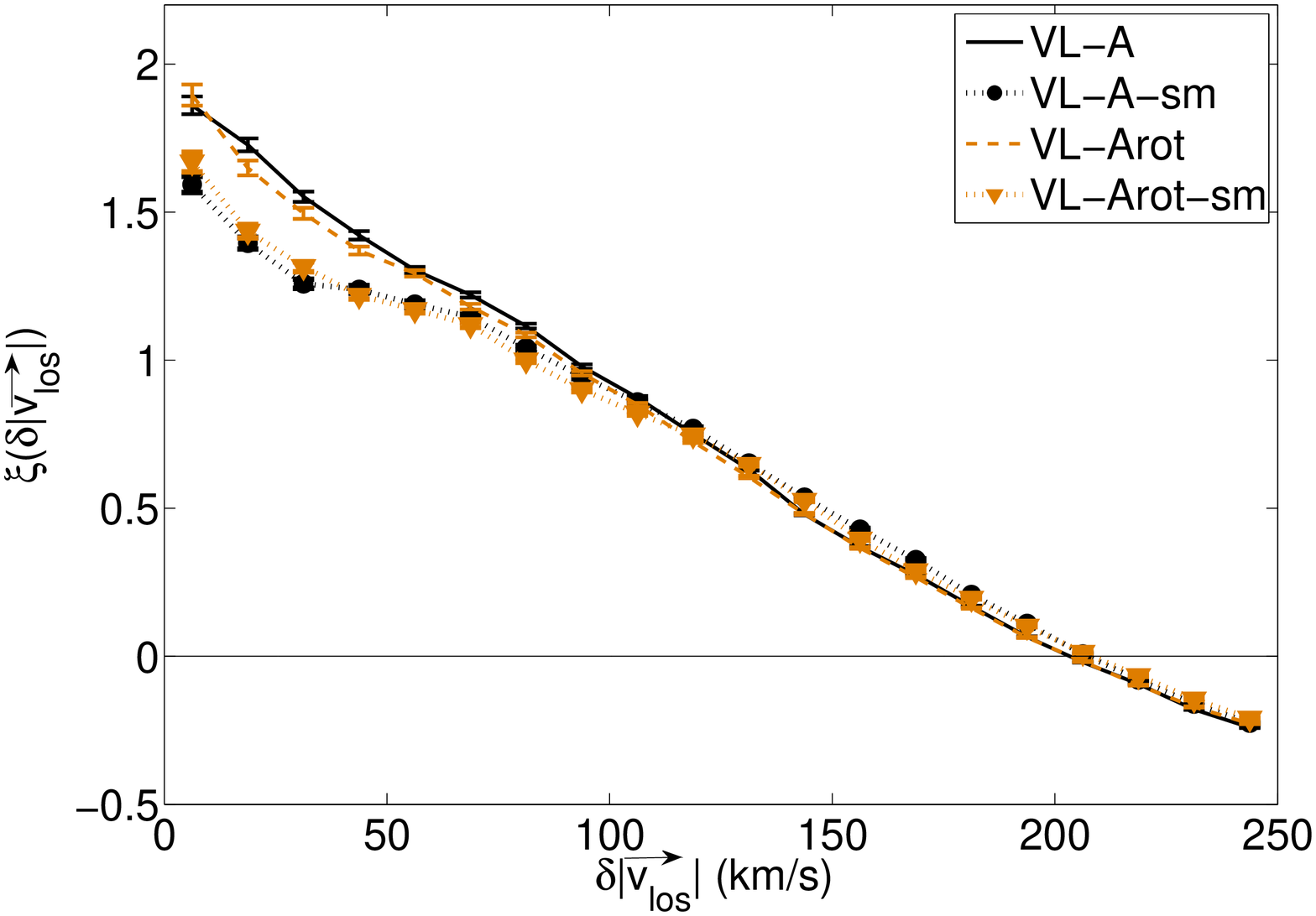} & \includegraphics[trim = 5mm 2.5mm 20mm 10mm, clip,width=0.49\textwidth]{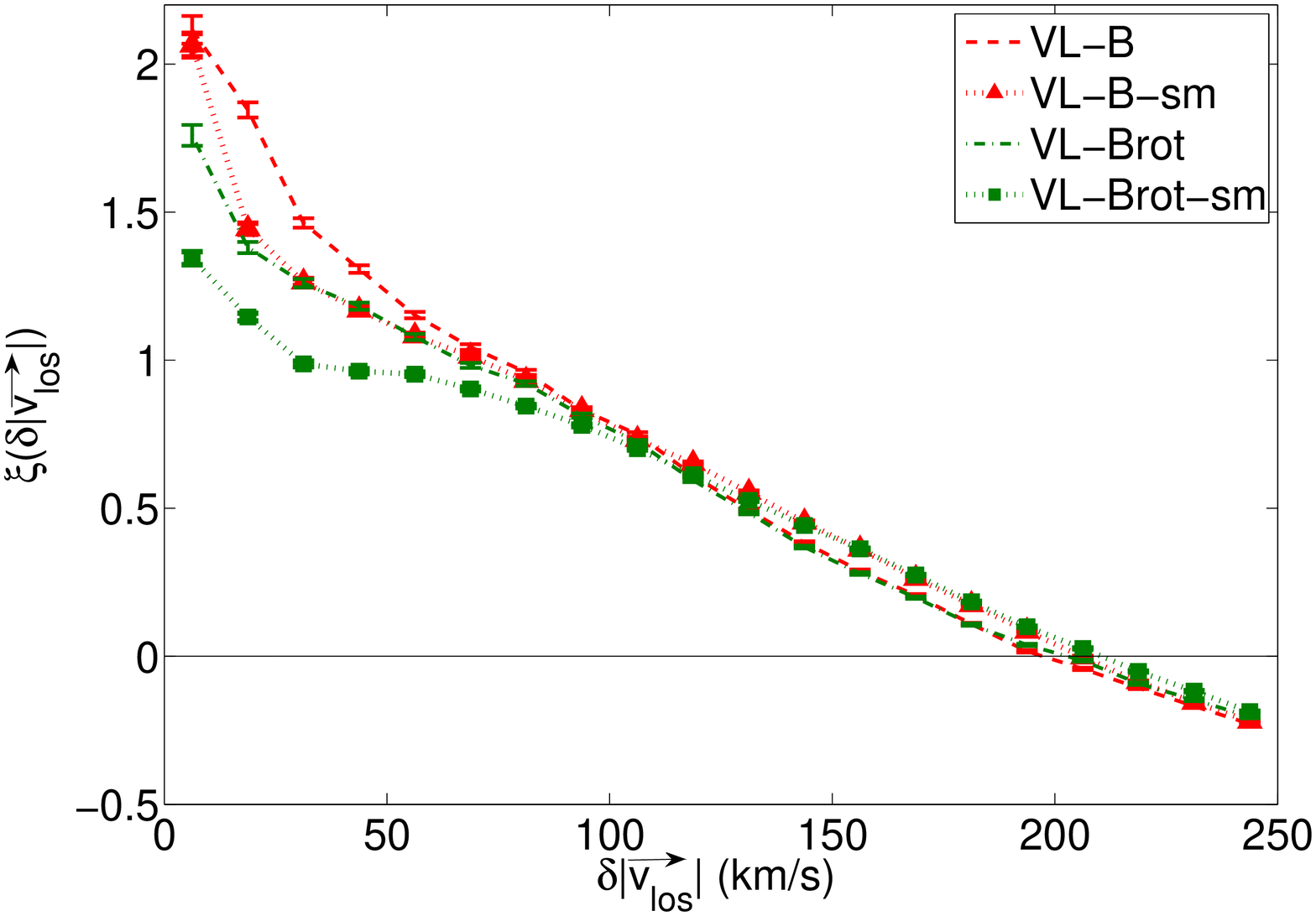} \\
\includegraphics[trim = 10mm 2.5mm 20mm 10mm, clip,width=0.49\textwidth]{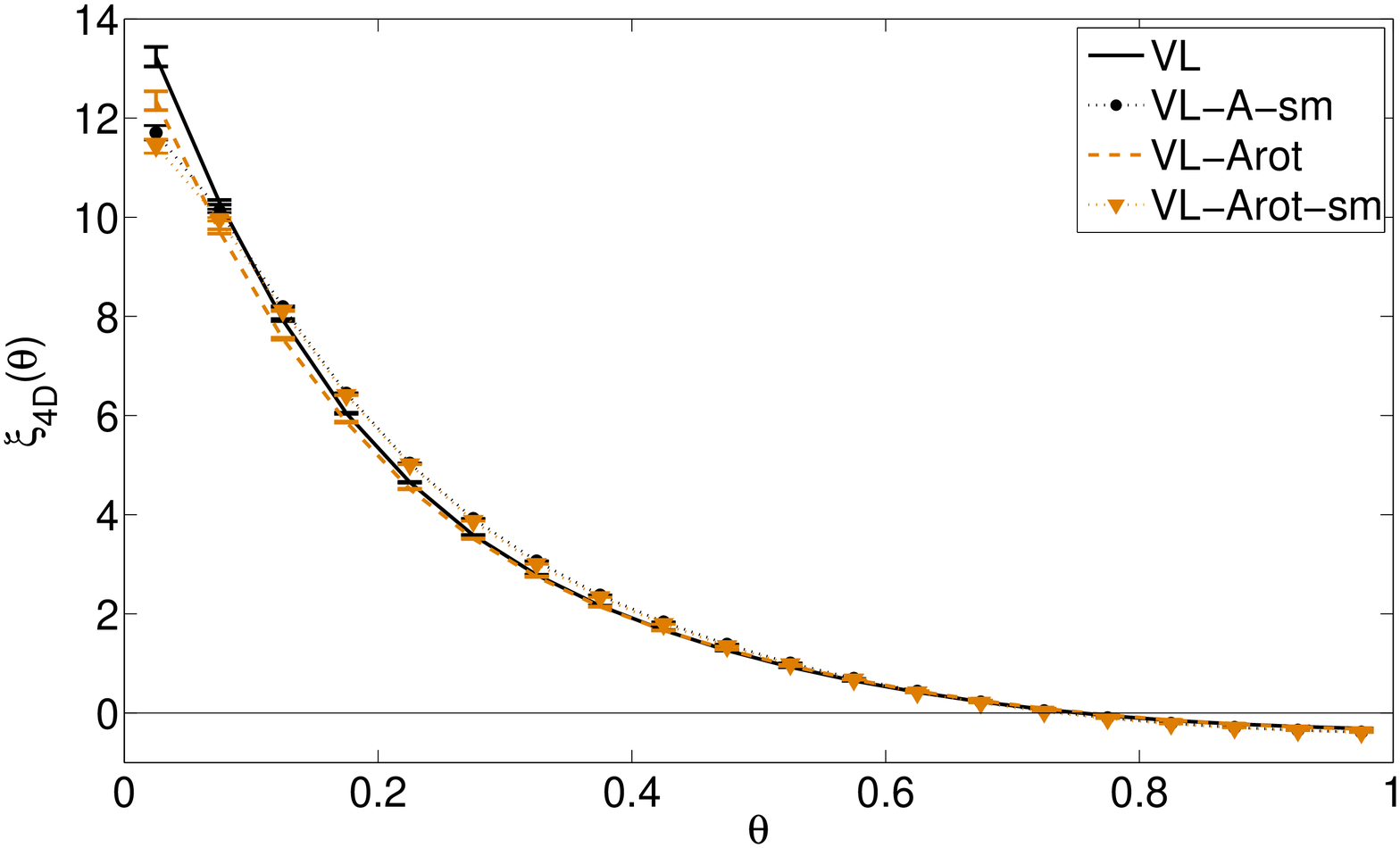} & \includegraphics[trim = 10mm 2.5mm 20mm 10mm, clip,width=0.49\textwidth]{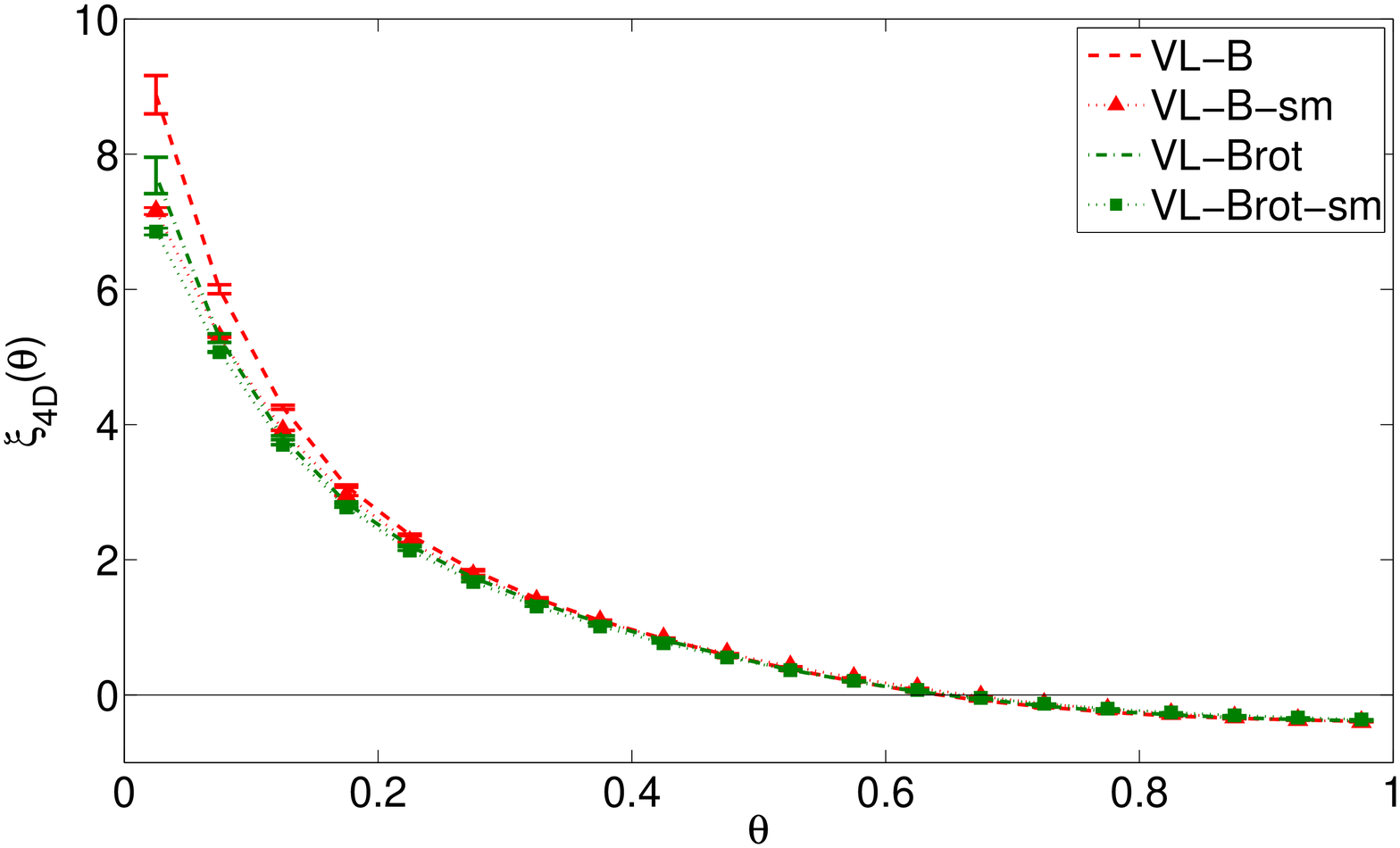} \\
\end{tabular}
\caption{\scriptsize The spatial (top row), line-of-sight velocity (center row), and the 4D phase-space (bottom row) two-point correlation functions for the four sets of dark Via Lactea II subhaloes. Left Panels: VL-A (solid black), VL-Arot (dashed orange). Right Panels: VL-B (dashed red), VL-Brot (dashed-dotted green). The mock sets of ``subhaloes" that represent the clustering signal due to the smooth underling density profile and line-of-sight velocity distribution of the corresponding data sets: VL-A-sm (dotted black, circle), VL-Arot-sm (dotted orange, down triangle), VL-B-sm (dotted red, up triangle), VL-Brot-sm (dotted green, square). Each of the dark subhalo sets are clustered due to substructure in each space on multiple scales, with the exception of the VL-A and VL-Arot sets that only contain substructure in their four-dimensional phase-space distributions on one scale. The VL-Arot spatial two-point correlation functions are not shown in the top left panel as they are nearly identical to the VL-A set. The error bars for mock ``subhalo" sets are not shown in the top row panels since they are smaller than each of the points in these panels. The black horizontal line in all panels shows the value for the two-point correlation function that corresponds to a uniform random distribution.}
\label{fig:VL results}
\end{center}
\end{figure*}

The 4D phase-space two-point correlation functions for the full-area Via Lactea II subhalo sets are shown in the bottom left panels of Figs. \ref{fig:LBA results} - \ref{fig:VL results}. The ``luminous" subhalo sets' 4D phase-space two-point correlation functions indicate that all the sets are clustered at a minimum of $2.6\sigma$ significance for a range of scales (and above $3\sigma$ significance for nearly all of these scales), as shown by the solid black and dashed orange curves in Figs.  \ref{fig:LBA results} - \ref{fig:KOP results}. The dotted curves in each panel depict the 4D phase-space two-point correlation function of the mock set of ``subhaloes" that indicates the contribution of each subhalo set's underlying four-dimensional phase-space distribution to its overall clustering signal. Of the ``luminous" subhalo sets only the LBA-A and EFmod-Arot sets contain substructure within their 4D phase-space distribution. The clustering signal in bins 9, 11, and 12 of the LBA-A set and bin 9 of the EFmod-Arot set indicate the presence of substructure at a respective significance of $2.7\sigma$, $3.2\sigma$, $2.6\sigma$, and $4\sigma$. 

The full-area dark subhalo sets, shown in the bottom left panel of Fig.~\ref{fig:VL results}, are significantly clustered in 4D phase-space, with a two-point correlation function amplitude greater than zero in the first 15 bins at greater than $7\sigma$ significance. The clustering signal in the first bin of the VL-A set and the VL-Arot set indicates that there is substructure in these set's 4D phase-space distribution at a respective significance of $4.5\sigma$ and $2.8\sigma$. There are several bins where the amplitude of the 4D phase-space two-point correlation function of the VL-A, VL-Arot, LBA-A, and EF-Arot sets exceed the clustering signal of their underlying 4D phase-space distribution by more than $2.6\sigma$, but with an amplitude of $\lesssim 0.5$  these bins do not provide a clear detection of substructure. 

The 4D phase-space two-point correlation functions of the dark subhalo footprint sets, VL-B and VL-Brot, are shown in the bottom right panel of Fig.~\ref{fig:VL results}. The clustering signal in the first 15 bins indicate that each set is clustered at greater than $7\sigma$ significance. For the VL-B set, evidence of substructure is detected in the first seven bins of the 4D phase-space two-point correlation function at greater than $3.6\sigma$ significance, and in the first nine bins for the VL-Brot set at a minimum significance of $2.9\sigma$. Each of the 4D phase-space two-point correlation functions of the ``luminous" subhalo footprint sets, shown in the bottom right panels of Figs. \ref{fig:LBA results} - \ref{fig:KOP results}, have at least one bin where the sets are shown to be clustered at a minimum significance of $2.6\sigma$. The LBA-B set is the only ``luminous" subhalo footprint set with evidence in one bin of the two-point correlation function of substructure in the set's 4D phase-space distribution. 

\section{\label{sec:Discussions} Discussion}

The clustering results of the Milky Way satellite galaxies show no significant evidence of substructure in the distribution of satellite galaxies in any of the three spaces considered. The clustering signal found in configuration and four-dimensional phase-space is simply due to the smooth underlying density profile. We, therefore, found no evidence for `ghostly streams' or groups with any kinematic correlation that would be a signature of group infall \citep{1976MNRAS.174..695L, 1976RGOB..182..241K, 1995MNRAS.275..429L, 2008ApJ...686L..61D, 2008MNRAS.385.1365L}, nor did we find evidence that the Milky Way satellite galaxies are oriented in a disk-like structure \citep{2005A&A...431..517K, 2007MNRAS.374.1125M, 2009MNRAS.394.2223M}. There was also no evidence to support proposed associations of galaxy pairs like Leo IV and Leo V \citep{2008ApJ...686L..83B}.  However, such structures are not ruled out in this analysis. The suite of two-point correlation function statistics used in this analysis is not sufficiently sensitive to singular substructures within the distribution of satellites, since the only information used by the two-point correlation functions to measure clustering is the frequency of galaxy separations. In order to identify a particular substructure, such as a disk of satellites, within a larger set of satellites using a two-point correlation function, the pairs of satellites in the substructure need to pick out a particular scale.  

The types of substructure detectable in our analysis for any set is limited by the number of pairs required to identify a significant abundance of substructure on a particular scale, which depends on the total number of satellites in the set and on the radial number density profile and line-of-sight velocity distribution of that set. For example, the clustering signal of the first bin of the spatial two-point correlation function of the MW-A set is the result of 9 pairs of galaxies. If the MW-A set instead had 15 pairs of galaxies separated from 12.5~kpc to 25~kpc, while maintaing the same radial density profile, the increased clustering signal of the first bin would have indicated the presence of substructure at a significance of $3\sigma$. These 15 galaxy pairs could be part of a single substructure, such as a group or disk (for example, seven galaxies in four connected tetrahedrons), or in sets of streams ranging from a single stream of 16 galaxies to a set of 9 ``streams" (24 galaxies in six sets of three galaxies and three sets of two galaxies). 

A subset of the Milky Way satellite galaxies that make up the disk of satellites has been shown to occupy a similar orbital plane, which has been used as evidence to argue that the disk of satellites is rotationally supported \citep{2008ApJ...680..287M}. While we find qualitative agreement with the spatial clustering expected for the disk of satellites, our analysis found the Milky Way satellite galaxies are randomly distributed in line-of-sight velocity space, which is at odds with a disk of satellites exhibiting coherent motion. It is possible, however, that this disagreement is the result of the limited dimensionality of the line-of-sight velocity clustering results missing a correlation that would be present in a 3D velocity clustering analysis, since the argument for rotational support is based on the 3D velocities of the subset of satellites considered. Recently, the Andromeda galaxy was proposed to have a subset of satellites in a disk-like structure that has coherent rotational motion in radial velocity space \citep{2013Natur.493...62I}. We will examine the clustering of the Andromeda satellites in a future work.

There are several differences between the clustering results of the Milky Way satellite galaxies and the ``luminous" subhalo sets. The ``luminous" subhalo sets are more strongly clustered, on small scales in each space, than are the Milky Way satellite galaxies, as shown in Fig.~\ref{fig:main_pts} by comparing the full-area two-point correlation function of each set in each panel. The difference in clustering strength in configuration space is due to the difference in the mean radial number density profile of each set, which can be seen by comparing the two-point correlation functions of the mock sets that represent the clustering due to the underlying density profiles, illustrated by the various dotted curves in the top row. In addition to the greater degree of clustering in 4D phase-space, each of the ``luminous" subhalo sets are also clustered over a broader range of scales than the Milky Way satellites. Further, two of the ``luminous" subhalo sets, the LBA-A set and the EFmod-Arot set, each contain substructure in their 4D phase-space distribution. 

The line-of-sight velocity space results contain the most notable difference in clustering properties between the satellites and subhaloes. The Milky Way satellite galaxies are randomly distributed in line-of-sight velocity space, whereas the ``satellites" of each of the ``luminous" subhalo models are not only clustered, but clustered as a result of substructure. The significant detection of substructure in the dark subhalo set in line-of-sight velocity space indicates that this important difference between all three ``luminous" subhalo models and the observed satellite galaxies is likely to persist for semi-analytic models beyond the three studied here. The differences in the clustering results of the subhalo sets compared to the observed Milky Way satellites in all three spaces cannot be attributed to a difference in merger history, as the VL-II host Dark Matter halo was selected to have had no recent major mergers similar to the Milky Way's own quiescent recent merger history.

As noted above, the ``luminous" subhalo sets are more centrally concentrated in the Milky Way-analog dark halo than is the dark subhalo distribution, ``VL-A". Stronger central concentrations of subhalo sets have been found previously \citep{2009MNRAS.400.1593M, 2011MNRAS.417.1260F}. Selecting a ``luminous" sample injects a spatial bias by preferentially selecting subhaloes nearer to the center of the host Milky Way-analog dark halo. The recipes for defining the ``luminous" subsets are likely to be selecting the densest subhaloes in order to allow for the survival of the subhaloes to the present epoch in the central regions of the host dark halo. The strong clustering of the densest subhaloes is reminiscent of the `biased galaxy formation', whereby {\it n}-sigma density fluctuations are {\it a priori} clustered more strongly than 1-sigma density fluctuations on the same mass scale, with amplitude of the two-point correlation function scaling as the square root of $n$  \citep{1984ApJ...284L...9K}. 

The discussion presented so far does not include the results of the sets that are restricted to reside in the SDSS DR7 footprint, where there appears to be better agreement between the Milky Way satellite galaxies and the ``luminous" subhalo footprint sets, in terms of relative two-point correlation amplitude. There are, however, two critical differences in the clustering results of the satellite galaxies and subhalo footprint sets. The Milky Way satellite galaxies are consistent with a random distribution to within $2\sigma$ on all scales in each space, whereas the ``luminous" subhalo sets are clustered in each space on some scale. Each of the ``luminous" subhalo models also contain substructure within their distribution in at least one space: the LBA model in 4D phase-space, the EF model in line-of-sight velocity space, and the EFmod model in configuration and line-of-sight velocity space. While diminishing the overall clustering signal amplitude, the limited survey volume of the SDSS footprint region does not prevent detection of substructure in the ``luminous" subhalo sets' distributions on scales where the Milky Way satellite galaxies are randomly distributed, which underscores the same notable difference in the full-area clustering results.

None of the three semi-analytic models we considered here provides a good match to the Milky Way satellite galaxies in terms of clustering in the three spaces. The LBA model provides the best fit with the Milky Way satellites when comparing the overall two-point correlation function amplitudes, although the cause of the clustering signal for these two sets differ in a meaningful way: the LBA subhaloes are clustered due to substructure on several scales in line-of-sight velocity space and four-dimensional phase-space while the Milky Way satellites are not. The LBA semi-analytic model is also the least successful model of the three semi-analytic models considered, when accounting for the various small-scale issues of Cold Dark Matter theory. For example, the LBA model provides the poorest match of the ``luminous" subhalo sets we considered to the projected mass function and the luminosity function of the Milky Way satellite galaxies \citep{2008ApJ...679.1260M, 2009ApJ...696.2179K, 2011MNRAS.415L..40B, 2012MNRAS.422.1203B}. The EF and EFmod semi-analytic models are more physically motivated than the LBA model, but provide poorer matches to the clustering properties of the Milky Way satellites in terms of overall clustering strength in each space.    The most substantial difference in clustering properties between the Milky Way satellite galaxies and the EF and EFmod sets is difference in the line-of-sight velocity space results, where the Milky Way satellite galaxies are randomly distributed, while the EF and EFmod sets contain substructure in their line-of-sight velocity space distribution.

\section{\label{sec:Summary} Summary} 

We have analyzed the clustering properties of the Milky Way satellite galaxies and of several sets of subhaloes from the Via Lactea II Cold Dark Matter simulation in each of configuration space, line-of-sight velocity space, and four-dimensional phase space, using a two-point correlation function statistic. The clustering signal of each data set in each space was compared to the clustering signal of a mock set of ``galaxies" representing the contribution of the smooth underlying density profile and line-of-sight velocity distribution to each data set's two-point correlation function. A detection of substructure was defined as clustering of a data set in excess of the contribution from the underlying density profile and/or line-of-sight velocity distribution.

The clustering results of the Milky Way satellite galaxies and the ``luminous" subhalo sets contain several differences. The ``luminous" subhalo sets are more strongly clustered than are the Milky Way satellites in all three spaces and over a broader range of scales in four-dimensional phase-space. There is no significant evidence of substructure in any of the three spaces for the set of Milky Way satellite galaxies, while each of the ``luminous" subhalo models contain substructure within their line-of-sight velocity space distributions and, for two models, within their four-dimensional phase-space distribution. The ``luminous" subhalo sets do not contain substructure within their configuration space distributions. The greater degree of spatial clustering of the ``luminous" subhalo sets compared to the Milky Way satellite galaxies is due to the stronger central concentrations of the ``luminous" subhalo sets. The most notable difference between the ``luminous" subhalo models and the Milky Way satellite galaxies results is in line-of-sight velocity space, where the Milky Way satellite galaxies are randomly distributed, while the ``satellites" of each of the ``luminous" subhalo models are clustered due to substructure over several scales at greater than $3\sigma$ significance. The inconsistencies between the Milky Way satellite galaxies and the Via Lactea II subhalo sets for all clustering methods suggest a possible new `small-scale' tension between Cold Dark Matter theory and the observed satellites. The Via Lactea II simulation is, however, a single simulation and a more robust result would come from a comparison of the Milky Way satellite galaxy population to a statistical sample of Cold Dark Matter simulations. We leave such testing for future work.

 \acknowledgments
 B.B. and R.F.G.W. acknowledge the support of the US National
Science Foundation grant IIA-1124403 and R.F.G.W. thanks the
Aspen Center for Physics, supported by NSF grant PHY-1066293, for
hospitality while this work was completed.

\bibliography{Clustering_Bibv3}

\begin{thebibliography}{63}
\expandafter\ifx\csname natexlab\endcsname\relax\def\natexlab#1{#1}\fi

\bibitem[{{Abazajian} {et~al.}(2009){Abazajian}, {Adelman-McCarthy},
  {Ag{\"u}eros}, {Allam}, {Allende Prieto}, {An}, {Anderson}, {Anderson},
  {Annis}, {Bahcall}, \& et~al.}]{2009ApJS..182..543A}
{Abazajian}, K.~N., {Adelman-McCarthy}, J.~K., {Ag{\"u}eros}, M.~A., {et~al.}
  2009, ApJs, 182, 543, 543

\bibitem[{{Armandroff} {et~al.}(1995){Armandroff}, {Olszewski}, \&
  {Pryor}}]{1995AJ....110.2131A}
{Armandroff}, T.~E., {Olszewski}, E.~W., \& {Pryor}, C. 1995, AJ, 110, 2131,
  2131

\bibitem[{{Bellazzini} {et~al.}(2005){Bellazzini}, {Gennari}, \&
  {Ferraro}}]{2005MNRAS.360..185B}
{Bellazzini}, M., {Gennari}, N., \& {Ferraro}, F.~R. 2005, MNRAS, 360, 185, 185

\bibitem[{{Bellazzini} {et~al.}(2004){Bellazzini}, {Gennari}, {Ferraro}, \&
  {Sollima}}]{2004MNRAS.354..708B}
{Bellazzini}, M., {Gennari}, N., {Ferraro}, F.~R., \& {Sollima}, A. 2004,
  MNRAS, 354, 708, 708

\bibitem[{{Belokurov} {et~al.}(2006){Belokurov}, {Zucker}, {Evans},
  {Wilkinson}, {Irwin}, {Hodgkin}, {Bramich}, {Irwin}, {Gilmore}, {Willman},
  {Vidrih}, {Newberg}, {Wyse}, {Fellhauer}, {Hewett}, {Cole}, {Bell}, {Beers},
  {Rockosi}, {Yanny}, {Grebel}, {Schneider}, {Lupton}, {Barentine},
  {Brewington}, {Brinkmann}, {Harvanek}, {Kleinman}, {Krzesinski}, {Long},
  {Nitta}, {Smith}, \& {Snedden}}]{2006ApJ...647L.111B}
{Belokurov}, V., {Zucker}, D.~B., {Evans}, N.~W., {et~al.} 2006, APJL, 647,
  L111, L111

\bibitem[{{Belokurov} {et~al.}(2007){Belokurov}, {Zucker}, {Evans}, {Kleyna},
  {Koposov}, {Hodgkin}, {Irwin}, {Gilmore}, {Wilkinson}, {Fellhauer},
  {Bramich}, {Hewett}, {Vidrih}, {De Jong}, {Smith}, {Rix}, {Bell}, {Wyse},
  {Newberg}, {Mayeur}, {Yanny}, {Rockosi}, {Gnedin}, {Schneider}, {Beers},
  {Barentine}, {Brewington}, {Brinkmann}, {Harvanek}, {Kleinman}, {Krzesinski},
  {Long}, {Nitta}, \& {Snedden}}]{2007ApJ...654..897B}
---. 2007, ApJ, 654, 897, 897

\bibitem[{{Belokurov} {et~al.}(2008){Belokurov}, {Walker}, {Evans}, {Faria},
  {Gilmore}, {Irwin}, {Koposov}, {Mateo}, {Olszewski}, \&
  {Zucker}}]{2008ApJ...686L..83B}
{Belokurov}, V., {Walker}, M.~G., {Evans}, N.~W., {et~al.} 2008, APJL, 686,
  L83, L83

\bibitem[{{Belokurov} {et~al.}(2009){Belokurov}, {Walker}, {Evans}, {Gilmore},
  {Irwin}, {Mateo}, {Mayer}, {Olszewski}, {Bechtold}, \&
  {Pickering}}]{2009MNRAS.397.1748B}
---. 2009, MNRAS, 397, 1748, 1748

\bibitem[{{Belokurov} {et~al.}(2010){Belokurov}, {Walker}, {Evans}, {Gilmore},
  {Irwin}, {Just}, {Koposov}, {Mateo}, {Olszewski}, {Watkins}, \&
  {Wyrzykowski}}]{2010ApJ...712L.103B}
---. 2010, APJL, 712, L103, L103

\bibitem[{{Blumenthal} {et~al.}(1984){Blumenthal}, {Faber}, {Primack}, \&
  {Rees}}]{1984Natur.311..517B}
{Blumenthal}, G.~R., {Faber}, S.~M., {Primack}, J.~R., \& {Rees}, M.~J. 1984,
  Nature, 311, 517, 517

\bibitem[{{Boylan-Kolchin} {et~al.}(2011){Boylan-Kolchin}, {Bullock}, \&
  {Kaplinghat}}]{2011MNRAS.415L..40B}
{Boylan-Kolchin}, M., {Bullock}, J.~S., \& {Kaplinghat}, M. 2011, \mnras, 415,
  L40, L40

\bibitem[{{Boylan-Kolchin} {et~al.}(2012){Boylan-Kolchin}, {Bullock}, \&
  {Kaplinghat}}]{2012MNRAS.422.1203B}
---. 2012, \mnras, 422, 1203, 1203

\bibitem[{{Carrera} {et~al.}(2002){Carrera}, {Aparicio},
  {Mart{\'{\i}}nez-Delgado}, \& {Alonso-Garc{\'{\i}}a}}]{2002AJ....123.3199C}
{Carrera}, R., {Aparicio}, A., {Mart{\'{\i}}nez-Delgado}, D., \&
  {Alonso-Garc{\'{\i}}a}, J. 2002, AJ, 123, 3199, 3199

\bibitem[{{Cooper} {et~al.}(2010){Cooper}, {Cole}, {Frenk}, \&
  {Helmi}}]{2010arXiv1011.1926C}
{Cooper}, A.~P., {Cole}, S., {Frenk}, C.~S., \& {Helmi}, A. 2010, ArXiv
  e-prints, arXiv:1011.1926

\bibitem[{{Diemand} {et~al.}(2007){Diemand}, {Kuhlen}, \&
  {Madau}}]{2007ApJ...667..859D}
{Diemand}, J., {Kuhlen}, M., \& {Madau}, P. 2007, ApJ, 667, 859, 859

\bibitem[{{Diemand} {et~al.}(2008){Diemand}, {Kuhlen}, {Madau}, {Zemp},
  {Moore}, {Potter}, \& {Stadel}}]{2008Natur.454..735D}
{Diemand}, J., {Kuhlen}, M., {Madau}, P., {et~al.} 2008, Nature, 454, 735, 735

\bibitem[{{D'Onghia} \& {Lake}(2008)}]{2008ApJ...686L..61D}
{D'Onghia}, E., \& {Lake}, G. 2008, APJL, 686, L61, L61

\bibitem[{{Efstathiou}(1992)}]{1992MNRAS.256P..43E}
{Efstathiou}, G. 1992, MNRAS, 256, 43P, 43P

\bibitem[{{Font} {et~al.}(2011){Font}, {Benson}, {Bower}, {Frenk}, {Cooper},
  {De Lucia}, {Helly}, {Helmi}, {Li}, {McCarthy}, {Navarro}, {Springel},
  {Starkenburg}, {Wang}, \& {White}}]{2011MNRAS.417.1260F}
{Font}, A.~S., {Benson}, A.~J., {Bower}, R.~G., {et~al.} 2011, \mnras, 417,
  1260, 1260

\bibitem[{{Geha} {et~al.}(2009){Geha}, {Willman}, {Simon}, {Strigari}, {Kirby},
  {Law}, \& {Strader}}]{2009ApJ...692.1464G}
{Geha}, M., {Willman}, B., {Simon}, J.~D., {et~al.} 2009, ApJ, 692, 1464, 1464

\bibitem[{{Giavalisco} {et~al.}(1994){Giavalisco}, {Steidel}, \&
  {Szalay}}]{1994ApJ...425L...5G}
{Giavalisco}, M., {Steidel}, C.~C., \& {Szalay}, A.~S. 1994, APJL, 425, L5, L5

\bibitem[{{Harris} \& {Zaritsky}(2006)}]{2006AJ....131.2514H}
{Harris}, J., \& {Zaritsky}, D. 2006, AJ, 131, 2514, 2514

\bibitem[{{Hilditch} {et~al.}(2005){Hilditch}, {Howarth}, \&
  {Harries}}]{2005MNRAS.357..304H}
{Hilditch}, R.~W., {Howarth}, I.~D., \& {Harries}, T.~J. 2005, MNRAS, 357, 304,
  304

\bibitem[{{Ibata} {et~al.}(1995){Ibata}, {Gilmore}, \&
  {Irwin}}]{1995MNRAS.277..781I}
{Ibata}, R.~A., {Gilmore}, G., \& {Irwin}, M.~J. 1995, MNRAS, 277, 781, 781

\bibitem[{{Ibata} {et~al.}(2013){Ibata}, {Lewis}, {Conn}, {Irwin},
  {McConnachie}, {Chapman}, {Collins}, {Fardal}, {Ferguson}, {Ibata}, {Mackey},
  {Martin}, {Navarro}, {Rich}, {Valls-Gabaud}, \&
  {Widrow}}]{2013Natur.493...62I}
{Ibata}, R.~A., {Lewis}, G.~F., {Conn}, A.~R., {et~al.} 2013, \nat, 493, 62, 62

\bibitem[{{Irwin} {et~al.}(2007){Irwin}, {Belokurov}, {Evans}, {Ryan-Weber},
  {de Jong}, {Koposov}, {Zucker}, {Hodgkin}, {Gilmore}, {Prema}, {Hebb},
  {Begum}, {Fellhauer}, {Hewett}, {Kennicutt}, {Wilkinson}, {Bramich},
  {Vidrih}, {Rix}, {Beers}, {Barentine}, {Brewington}, {Harvanek},
  {Krzesinski}, {Long}, {Nitta}, \& {Snedden}}]{2007ApJ...656L..13I}
{Irwin}, M.~J., {Belokurov}, V., {Evans}, N.~W., {et~al.} 2007, APJL, 656, L13,
  L13

\bibitem[{{Kaiser}(1984)}]{1984ApJ...284L...9K}
{Kaiser}, N. 1984, APJL, 284, L9, L9

\bibitem[{{Kinemuchi} {et~al.}(2008){Kinemuchi}, {Harris}, {Smith},
  {Silbermann}, {Snyder}, {La Cluyz{\'e}}, \& {Clark}}]{2008AJ....136.1921K}
{Kinemuchi}, K., {Harris}, H.~C., {Smith}, H.~A., {et~al.} 2008, AJ, 136, 1921,
  1921

\bibitem[{{Koch} {et~al.}(2007){Koch}, {Kleyna}, {Wilkinson}, {Grebel},
  {Gilmore}, {Evans}, {Wyse}, \& {Harbeck}}]{2007AJ....134..566K}
{Koch}, A., {Kleyna}, J.~T., {Wilkinson}, M.~I., {et~al.} 2007, AJ, 134, 566,
  566

\bibitem[{{Koch} {et~al.}(2009){Koch}, {Wilkinson}, {Kleyna}, {Irwin},
  {Zucker}, {Belokurov}, {Gilmore}, {Fellhauer}, \&
  {Evans}}]{2009ApJ...690..453K}
{Koch}, A., {Wilkinson}, M.~I., {Kleyna}, J.~T., {et~al.} 2009, ApJ, 690, 453,
  453

\bibitem[{{Koposov} {et~al.}(2008){Koposov}, {Belokurov}, {Evans}, {Hewett},
  {Irwin}, {Gilmore}, {Zucker}, {Rix}, {Fellhauer}, {Bell}, \&
  {Glushkova}}]{2008ApJ...686..279K}
{Koposov}, S., {Belokurov}, V., {Evans}, N.~W., {et~al.} 2008, ApJ, 686, 279,
  279

\bibitem[{{Koposov} {et~al.}(2009){Koposov}, {Yoo}, {Rix}, {Weinberg},
  {Macci{\`o}}, \& {Escud{\'e}}}]{2009ApJ...696.2179K}
{Koposov}, S.~E., {Yoo}, J., {Rix}, H., {et~al.} 2009, ApJ, 696, 2179, 2179

\bibitem[{{Koposov} {et~al.}(2011){Koposov}, {Gilmore}, {Walker}, {Belokurov},
  {Wyn Evans}, {Fellhauer}, {Gieren}, {Geisler}, {Monaco}, {Norris}, {Okamoto},
  {Pe{\~n}arrubia}, {Wilkinson}, {Wyse}, \& {Zucker}}]{2011ApJ...736..146K}
{Koposov}, S.~E., {Gilmore}, G., {Walker}, M.~G., {et~al.} 2011, \apj, 736,
  146, 146

\bibitem[{{Kravtsov} {et~al.}(2004){Kravtsov}, {Gnedin}, \&
  {Klypin}}]{2004ApJ...609..482K}
{Kravtsov}, A.~V., {Gnedin}, O.~Y., \& {Klypin}, A.~A. 2004, \apj, 609, 482,
  482

\bibitem[{{Kroupa} {et~al.}(2005){Kroupa}, {Theis}, \&
  {Boily}}]{2005A&A...431..517K}
{Kroupa}, P., {Theis}, C., \& {Boily}, C.~M. 2005, AAP, 431, 517, 517

\bibitem[{{Kunkel} \& {Demers}(1976)}]{1976RGOB..182..241K}
{Kunkel}, W.~E., \& {Demers}, S. 1976, in Royal Greenwich Observatory Bulletin,
  Vol. 182, The Galaxy and the Local Group, ed. {R.~J.~Dickens, J.~E.~Perry,
  F.~G.~Smith, \& I.~R.~King}, 241--+

\bibitem[{{Landy} \& {Szalay}(1993)}]{1993ApJ...412...64L}
{Landy}, S.~D., \& {Szalay}, A.~S. 1993, ApJ, 412, 64, 64

\bibitem[{{Li} \& {Helmi}(2008)}]{2008MNRAS.385.1365L}
{Li}, Y., \& {Helmi}, A. 2008, MNRAS, 385, 1365, 1365

\bibitem[{{Lynden-Bell}(1976)}]{1976MNRAS.174..695L}
{Lynden-Bell}, D. 1976, MNRAS, 174, 695, 695

\bibitem[{{Lynden-Bell} \& {Lynden-Bell}(1995)}]{1995MNRAS.275..429L}
{Lynden-Bell}, D., \& {Lynden-Bell}, R.~M. 1995, MNRAS, 275, 429, 429

\bibitem[{{Madau} {et~al.}(2008){Madau}, {Diemand}, \&
  {Kuhlen}}]{2008ApJ...679.1260M}
{Madau}, P., {Diemand}, J., \& {Kuhlen}, M. 2008, ApJ, 679, 1260, 1260

\bibitem[{{Martin} {et~al.}(2008){Martin}, {de Jong}, \&
  {Rix}}]{2008ApJ...684.1075M}
{Martin}, N.~F., {de Jong}, J.~T.~A., \& {Rix}, H. 2008, ApJ, 684, 1075, 1075

\bibitem[{{Martin} {et~al.}(2007){Martin}, {Ibata}, {Chapman}, {Irwin}, \&
  {Lewis}}]{2007MNRAS.380..281M}
{Martin}, N.~F., {Ibata}, R.~A., {Chapman}, S.~C., {Irwin}, M., \& {Lewis},
  G.~F. 2007, MNRAS, 380, 281, 281

\bibitem[{{Mateo} {et~al.}(2008){Mateo}, {Olszewski}, \&
  {Walker}}]{2008ApJ...675..201M}
{Mateo}, M., {Olszewski}, E.~W., \& {Walker}, M.~G. 2008, ApJ, 675, 201, 201

\bibitem[{{Mateo}(1998)}]{1998ARA&A..36..435M}
{Mateo}, M.~L. 1998, ARAA, 36, 435, 435

\bibitem[{{Metz} {et~al.}(2007){Metz}, {Kroupa}, \&
  {Jerjen}}]{2007MNRAS.374.1125M}
{Metz}, M., {Kroupa}, P., \& {Jerjen}, H. 2007, MNRAS, 374, 1125, 1125

\bibitem[{{Metz} {et~al.}(2009){Metz}, {Kroupa}, \&
  {Jerjen}}]{2009MNRAS.394.2223M}
---. 2009, MNRAS, 394, 2223, 2223

\bibitem[{{Metz} {et~al.}(2008){Metz}, {Kroupa}, \&
  {Libeskind}}]{2008ApJ...680..287M}
{Metz}, M., {Kroupa}, P., \& {Libeskind}, N.~I. 2008, ApJ, 680, 287, 287

\bibitem[{{Mu{\~n}oz} {et~al.}(2009){Mu{\~n}oz}, {Madau}, {Loeb}, \&
  {Diemand}}]{2009MNRAS.400.1593M}
{Mu{\~n}oz}, J.~A., {Madau}, P., {Loeb}, A., \& {Diemand}, J. 2009, MNRAS, 400,
  1593, 1593

\bibitem[{{Riess} {et~al.}(2011){Riess}, {Macri}, {Casertano}, {Lampeitl},
  {Ferguson}, {Filippenko}, {Jha}, {Li}, \& {Chornock}}]{2011ApJ...730..119R}
{Riess}, A.~G., {Macri}, L., {Casertano}, S., {et~al.} 2011, ApJ, 730, 119, 119

\bibitem[{{Sakamoto} \& {Hasegawa}(2006)}]{2006ApJ...653L..29S}
{Sakamoto}, T., \& {Hasegawa}, T. 2006, APJL, 653, L29, L29

\bibitem[{{Simon} \& {Geha}(2007)}]{2007ApJ...670..313S}
{Simon}, J.~D., \& {Geha}, M. 2007, ApJ, 670, 313, 313

\bibitem[{{Starkenburg} {et~al.}(2009){Starkenburg}, {Helmi}, {Morrison},
  {Harding}, {van Woerden}, {Mateo}, {Olszewski}, {Sivarani}, {Norris},
  {Freeman}, {Shectman}, {Dohm-Palmer}, {Frey}, \&
  {Oravetz}}]{2009ApJ...698..567S}
{Starkenburg}, E., {Helmi}, A., {Morrison}, H.~L., {et~al.} 2009, ApJ, 698,
  567, 567

\bibitem[{{van den Bergh}(1999)}]{1999A&ARv...9..273V}
{van den Bergh}, S. 1999, AAPR, 9, 273, 273

\bibitem[{{van der Marel} {et~al.}(2002){van der Marel}, {Alves}, {Hardy}, \&
  {Suntzeff}}]{2002AJ....124.2639V}
{van der Marel}, R.~P., {Alves}, D.~R., {Hardy}, E., \& {Suntzeff}, N.~B. 2002,
  AJ, 124, 2639, 2639

\bibitem[{{Walker} {et~al.}(2006){Walker}, {Mateo}, {Olszewski}, {Bernstein},
  {Wang}, \& {Woodroofe}}]{2006AJ....131.2114W}
{Walker}, M.~G., {Mateo}, M., {Olszewski}, E.~W., {et~al.} 2006, AJ, 131, 2114,
  2114

\bibitem[{{Walsh} {et~al.}(2007){Walsh}, {Jerjen}, \&
  {Willman}}]{2007ApJ...662L..83W}
{Walsh}, S.~M., {Jerjen}, H., \& {Willman}, B. 2007, APJL, 662, L83, L83

\bibitem[{{White} \& {Rees}(1978)}]{1978MNRAS.183..341W}
{White}, S.~D.~M., \& {Rees}, M.~J. 1978, MNRAS, 183, 341, 341

\bibitem[{{Willman} {et~al.}(2005{\natexlab{a}}){Willman}, {Blanton}, {West},
  {Dalcanton}, {Hogg}, {Schneider}, {Wherry}, {Yanny}, \&
  {Brinkmann}}]{2005AJ....129.2692W}
{Willman}, B., {Blanton}, M.~R., {West}, A.~A., {et~al.} 2005{\natexlab{a}},
  AJ, 129, 2692, 2692

\bibitem[{{Willman} {et~al.}(2005{\natexlab{b}}){Willman}, {Dalcanton},
  {Martinez-Delgado}, {West}, {Blanton}, {Hogg}, {Barentine}, {Brewington},
  {Harvanek}, {Kleinman}, {Krzesinski}, {Long}, {Neilsen}, {Nitta}, \&
  {Snedden}}]{2005ApJ...626L..85W}
{Willman}, B., {Dalcanton}, J.~J., {Martinez-Delgado}, D., {et~al.}
  2005{\natexlab{b}}, APJL, 626, L85, L85

\bibitem[{{York} {et~al.}(2000){York}, {Adelman}, {Anderson}, {Anderson},
  {Annis}, {Bahcall}, {Bakken}, {Barkhouser}, {Bastian}, {Berman}, {Boroski},
  {Bracker}, {Briegel}, {Briggs}, {Brinkmann}, {Brunner}, {Burles}, {Carey},
  {Carr}, {Castander}, {Chen}, {Colestock}, {Connolly}, {Crocker}, {Csabai},
  {Czarapata}, {Davis}, {Doi}, {Dombeck}, {Eisenstein}, {Ellman}, {Elms},
  {Evans}, {Fan}, {Federwitz}, {Fiscelli}, {Friedman}, {Frieman}, {Fukugita},
  {Gillespie}, {Gunn}, {Gurbani}, {de Haas}, {Haldeman}, {Harris}, {Hayes},
  {Heckman}, {Hennessy}, {Hindsley}, {Holm}, {Holmgren}, {Huang}, {Hull},
  {Husby}, {Ichikawa}, {Ichikawa}, {Ivezi{\'c}}, {Kent}, {Kim}, {Kinney},
  {Klaene}, {Kleinman}, {Kleinman}, {Knapp}, {Korienek}, {Kron}, {Kunszt},
  {Lamb}, {Lee}, {Leger}, {Limmongkol}, {Lindenmeyer}, {Long}, {Loomis},
  {Loveday}, {Lucinio}, {Lupton}, {MacKinnon}, {Mannery}, {Mantsch}, {Margon},
  {McGehee}, {McKay}, {Meiksin}, {Merelli}, {Monet}, {Munn}, {Narayanan},
  {Nash}, {Neilsen}, {Neswold}, {Newberg}, {Nichol}, {Nicinski}, {Nonino},
  {Okada}, {Okamura}, {Ostriker}, {Owen}, {Pauls}, {Peoples}, {Peterson},
  {Petravick}, {Pier}, {Pope}, {Pordes}, {Prosapio}, {Rechenmacher}, {Quinn},
  {Richards}, {Richmond}, {Rivetta}, {Rockosi}, {Ruthmansdorfer}, {Sandford},
  {Schlegel}, {Schneider}, {Sekiguchi}, {Sergey}, {Shimasaku}, {Siegmund},
  {Smee}, {Smith}, {Snedden}, {Stone}, {Stoughton}, {Strauss}, {Stubbs},
  {SubbaRao}, {Szalay}, {Szapudi}, {Szokoly}, {Thakar}, {Tremonti}, {Tucker},
  {Uomoto}, {Vanden Berk}, {Vogeley}, {Waddell}, {Wang}, {Watanabe},
  {Weinberg}, {Yanny}, \& {Yasuda}}]{2000AJ....120.1579Y}
{York}, D.~G., {Adelman}, J., {Anderson}, Jr., J.~E., {et~al.} 2000, AJ, 120,
  1579, 1579

\bibitem[{{Zucker} {et~al.}(2006{\natexlab{a}}){Zucker}, {Belokurov}, {Evans},
  {Kleyna}, {Irwin}, {Wilkinson}, {Fellhauer}, {Bramich}, {Gilmore}, {Newberg},
  {Yanny}, {Smith}, {Hewett}, {Bell}, {Rix}, {Gnedin}, {Vidrih}, {Wyse},
  {Willman}, {Grebel}, {Schneider}, {Beers}, {Kniazev}, {Barentine},
  {Brewington}, {Brinkmann}, {Harvanek}, {Kleinman}, {Krzesinski}, {Long},
  {Nitta}, \& {Snedden}}]{2006ApJ...650L..41Z}
{Zucker}, D.~B., {Belokurov}, V., {Evans}, N.~W., {et~al.} 2006{\natexlab{a}},
  APJL, 650, L41, L41

\bibitem[{{Zucker} {et~al.}(2006{\natexlab{b}}){Zucker}, {Belokurov}, {Evans},
  {Wilkinson}, {Irwin}, {Sivarani}, {Hodgkin}, {Bramich}, {Irwin}, {Gilmore},
  {Willman}, {Vidrih}, {Fellhauer}, {Hewett}, {Beers}, {Bell}, {Grebel},
  {Schneider}, {Newberg}, {Wyse}, {Rockosi}, {Yanny}, {Lupton}, {Smith},
  {Barentine}, {Brewington}, {Brinkmann}, {Harvanek}, {Kleinman}, {Krzesinski},
  {Long}, {Nitta}, \& {Snedden}}]{2006ApJ...643L.103Z}
---. 2006{\natexlab{b}}, APJL, 643, L103, L103

\end{thebibliography}

\begin{deluxetable}{lrrrrc}
\tablecaption{Milky Way Satellite Galaxies\label{tab:Gallist}}
\tabletypesize{\scriptsize}
\tablewidth{0pt}
\tablehead{
\colhead{Galaxy} & \colhead{$\alpha (^{\circ})$} & \colhead{$\delta (^{\circ})$} & \colhead{$r_{helio} (kpc)$\tablenotemark{a}} & \colhead{$v_{helio} (km s^{-1})$\tablenotemark{b}} & \colhead{References}
}
\startdata
Leo I 			& 152.11	& 12.31	& 254 $(\pm 19)$	& 282.9$(\pm 0.5)$	& 1, 2, 3 \\ 
Leo II 			& 168.37 	& 22.15 	& 233 $(\pm 15)$	&79.1 $(\pm 0.6)$    & 1, 4, 5 \\  
Sextans 			& 153.26 	& -1.62 	& 86 $(\pm 4)$		& 227 $(\pm 3.0)$    & 1 \\
Canes Venatici I 	& 202.02 	& 33.56 	& 218 $(\pm 10)$	& 30.9 $(\pm 0.6)$    &  6, 7 \\
Hercules 		& 247.77 	& 12.79 	& 131 $(\pm 12)$ 	& 45 $(\pm 1.1)$	& 6, 7 \\
Bootes I 			& 210.02 	& 14.51 	&  62 $(\pm 3)$ 	& 99 $(\pm 2.1)$  & 6, 8 \\
Leo IV 			& 173.24 	& -0.52 	& 160 $(\pm 15)$	& 132.3 $(\pm 1.4)$    &  6, 7\\
Leo V 			& 172.79 	& 2.22 	& 180 $(\pm 10)$ 	& 173 $(\pm 3.1)$  	& 9 \\
Ursa Major I 		& 158.70 	& 51.94 	& 96.8 $(\pm 4)$	& -55.3 $(\pm 1.4)$	&  6, 7 \\
Canes Venatici II	& 194.29 	& 34.33 	& 160 $(\pm 5)$ 	& -128.9 $(\pm 1.2)$	&  6, 7 \\
Coma Bernices 	& 186.74 	& 23.92 	& 44 $(\pm 4)$ 		& 98.1 $(\pm 0.9)$    & 6, 7 \\
Bootes II 		&  209.53 	& 12.85 	& 42 $(\pm  8)$ 	& -117 $(\pm 5.2)$	& 6, 12\\
Ursa Major II 		& 132.87 	& 63.14 	& 30 $(\pm 5)$		& -116.5$(\pm 1.9)$	& 6, 7 \\
Segue 1 		& 151.76 	& 16.07 	& 23 $(\pm 2)$ 		& 206 $(\pm 1.3)$	&  6, 13 \\ 
Willman 1		& 162.34 	& 51.05 	& 38 $(\pm 7)$ 		& -12.3 $(\pm 2.5)$	& 6, 8 \\
Ursa Minor   	& 227.30 	& 67.22 	& 76 $(\pm 4.0)$	& -247.4  $(\pm 1.0)$	& 1, 15, 16 \\
\tableline
Fornax \T		& 40.00 	& -34.45 	& 138 $(\pm 8.0)$ 	&  53.3$(\pm 0.8)$	& 1, 14 \\
Sculptor 			& 15.04	& -33.71 	& 79 $(\pm 4.0)$ 	& 108 $(\pm 3.0)$	&  1 \\
Carina 			& 100.40 	& -50.97 	& 101$(\pm 5.0)$	& 224 $(\pm 3.0)$	& 1  \\
Draco 			& 260.08 	& 57.91 	&  82.4 $(\pm 5.8)$	& -293.3$(\pm 1.0)$   & 1, 16, 17 \\
Sagittarius 		&  283.76 	& -30.48 	&  24 $(\pm 2.0)$	& 140 $(\pm 2.0)$  & 1, 10  \\
Segue 2  		& 34.82	& 20.18	&  35 $(\pm 1.6)$ 	& -39.2 $(\pm 2.5)$   & 11 \\
LMC			& 80.89 	& -69.76 	& 49.8 $(\pm 1.6)$	& 262.2 $(\pm 3.4)$	& 1, 18, 19 \\
SMC  			& 13.18 	& -72.83 	& 60.6 $(\pm 3.8)$	& 145.6 $(\pm 0.6)$  & 1, 20, 21
\enddata
\tablecomments{The galaxies listed in the upper section make up the ``MW-B" set. All galaxies listed here are in the ``MW-A" set.}
\tablenotetext{a}{Galactocentric distances are determined using $R_{\odot} = 8~kpc$.}
\tablenotetext{b}{Galactic standard of rest line-of-sight velocity, $v_{GSR}$, is calculated using $v_{GSR} = v_{helio} + 9\cos(l)\cos(b) + 232\sin(l).*\cos(b) + 7\sin(b)$.}
\tablerefs{1 -  \citet{1998ARA&A..36..435M}, 2 - \citet{2004MNRAS.354..708B}, 3 - \citet{2008ApJ...675..201M}, 4 - \citet{2005MNRAS.360..185B}, 5 - \citet{2007AJ....134..566K}, 6 - \citet{2008ApJ...684.1075M}, 7 - \citet{2007ApJ...670..313S}, 8 - \citet{2007MNRAS.380..281M}, 9 - \citet{2008ApJ...686L..83B}, 10 - \citet{1995MNRAS.277..781I}, 11 - \citet{2009MNRAS.397.1748B}, 12 - \citet{2009ApJ...690..453K}, 13 - \citet{2009ApJ...692.1464G}, 14 - \citet{2006AJ....131.2114W}, 15 - \citet{2002AJ....123.3199C}, 16 - \citet{1995AJ....110.2131A}, 17 - \citet{2008AJ....136.1921K}, 18 - \citet{2011ApJ...730..119R}, 19 - \citet{2002AJ....124.2639V}, 20 - \citet{2005MNRAS.357..304H}, 21 - \citet{2006AJ....131.2514H}}
\end{deluxetable}	

\clearpage

\begin{deluxetable}{l | c | c | c | c | c | c | c | c}
\tablecaption{Data Set Details and Results\label{tab:Cluster_nums}}
\tablewidth{0pt}
\tabletypesize{\scriptsize}
\tablehead{
\colhead{Data Set} & \colhead{N} & \colhead{$n_r$} & \colhead{$n_v$\tablenotemark{a}} & \colhead{$r_0$(kpc)} & \colhead{$r_c$(kpc)} &\colhead{$r(kpc) > r_{sm}$} &\colhead{$ |\delta \overrightarrow{v_{los}}|(km s^{-1}) > |\delta \overrightarrow{v_{los}}|_{sm}$} &\colhead{$\theta > \theta_{sm}$}
}
\startdata
MW-A 	& 24  & 3 & G & 126	& 149 	& N/A & N/A & N/A \\ \hline
MW-B 	& 16  & 3 & G & 44	 	& 44	 	& N/A  & N/A & N/A  \\ \hline
  &   &  & &  & & & $ 1(4.6)$, $2$-$3(>7)$, $4(6.7)$, & \vspace{-0.13cm} \\ 
VL-A 	& 7739 & 7 &  5,5 & 125 & 157  & $1$-$8(\geq6.6)$ & & $1(4.5)$, $14(3.2)$, $15(>7)$  \vspace{-0.13cm}  \\  
& &  &  & & & &$5(5.3)$, $6(4.2)$, $7(4.7)$ &  \\ \hline

&   &  & &  & & & $1(3.3)$, $2(4.5)$ , $3(5.1)$, & \\
VL-Arot 	& 7739  & 7 & 5,5 & 125 	& 157 	& $1$-$8 (\geq6.4)$ & $4(5.8)$, $5(5.7)$, $6(3.3)$, & $1(2.8)$, $13-15(>7)$ \\
 & &  &  & & & & $7(5.5)$, $8(3.8)$  \\ \hline
 
 &   &  & &  & & & & $1(5.1)$, $2$-$3(>7)$, $4(6.6)$, \vspace{-0.13cm} \\
VL-B 	& 1992  & 7 & 5,5 &  80 	& 126	 & $1$-$7 (>7)$ & $2$-$3(>7)$, $4(6.3)$, $5(3.7)$ & \vspace{-0.13cm} \\
 & &  &  & & & & & $5(>7)$, $6(6.4)$, $7(3.7)$  \\ \hline
  &   &  & &  & & & &$1(2.6)$, $2(2.7)$, $3(3)$, \vspace{-0.13cm} \\
  &   &  & &  & & & $1$-$5(>7)$, $6(5.6)$, $7(5.9)$,  & \vspace{-0.13cm} \\
VL-Brot 	& 1967  & 7 & 5,5 & 79 &  126 	& $1$-$7 (\geq7)$ & &$4(3.1)$, $5(4.6)$, $6(5.6)$,  \vspace{-0.13cm}  \\
 & &  &  & & & &$8(2.9)$ &  \vspace{-0.13cm} \\ 
 & &  &  & & & & &  $7(7)$, $8$-$9(>7)$, $10(5.2)$ \\ \hline
 
  &   &  & &  & & & &  $9(2.7)$, $11(3.2)$, $12(2.6)$,  \vspace{-0.13cm}\\
LBA-A 	& 65  & 3 & 5 & 131 	& 152 	& $12 (3.4)$& $3(3)$, $8(3.1)$, $16(2.8)$ & \vspace{-0.13cm} \\
 & &  &  & & & & &  $13(3)$ \\ \hline
LBA-Arot 	& 65  & 3 & 6 & 131 	& 152 	& $12 (3.3)$ &  $3(2.9)$ & N/A  \\ \hline
LBA-B 	& 21  & 3 & 5 & 30	 	& 31	 	& N/A  & N/A  & $8(2.6)$  \\ \hline
LBA-Brot 	& 23  & 3 & 6 & 28	 	& 31 	& N/A  & N/A  & N/A \\ \hline
  &   &  & &  & & & $2(5.2)$, $3(3.2)$, $15(3.1)$, & \vspace{-0.13cm} \\
EF-A		& 102  & 4 & G & 119 	& 140 	& N/A &   & N/A \vspace{-0.13cm}  \\
& &  &  & & & & $16(3.5)$ & \\ \hline
EF-Arot 	& 102 & 4 & 5 & 119 	& 140 	& N/A & $1(4.2)$, $2(7)$, $8(4.1)$ & $12(3.1)$   \\ \hline
EF-B 	& 30  & 4 & G & 69 		& 114	& N/A & $2(2.7)$ & N/A \\ \hline
EF-Brot 	& 29  & 4 & 5 & 81 		& 106	 & N/A  & N/A  & N/A \\ \hline
EFmod-A 	& 120  & 6 & G & 115 	& 126 	& N/A & $3(5.5)$, $4(2.9)$ & N/A \\ \hline
EFmod-Arot & 120  & 6 & G & 114 	& 126 	& N/A & N/A & $9(4)$ \\ \hline
EFmod-B 	& 28  & 6 & G & 81 		& 106 	& N/A & N/A  & N/A \\ \hline
EFmod-Brot & 32  & 6 & G & 69	& 106 	& $2(2.7)$ & $8(3)$ & N/A 
\enddata
\tablecomments{Column 1: List of all data sets, Column 2: Number of Set Members, Column 3: Degree of polynomial for radial number density profile, Column 4: Degree of polynomial for line-of-sight velocity distribution, Column 5: Correlation length, Column 6: Critical length ($r_c$ where $\xi(r_c) = 0$), Column 7-9: Bins and significance level in parenthesis for data set clustering signal in excess of  clustering signal due to smooth underling density profile and/or line-of-sight velocity distribution. (Column 7: Configuration space, Column 8: Line-of-sight velocity space, Column 9: Four-dimensional phase-space)}
\tablenotetext{a}{G - for a Gaussian distribution. Two entries for separate positive and negative line-of-sight distribution velocity fits (pos,neg).}
\end{deluxetable}

\end{document}